\documentclass{svjour3}     

\usepackage[utf8]{inputenc}
\usepackage{color}
\usepackage{multirow}
\usepackage{amssymb}
\usepackage{paralist}
\usepackage{graphicx}
\usepackage{subcaption}
\usepackage{tcolorbox}
\usepackage{framed}
\usepackage[framemethod=default]{mdframed}
\newmdenv[linecolor=black,backgroundcolor=gray!30]{resultframe}
\usepackage{numprint}

\usepackage{natbib}

\usepackage{url}
\npthousandsep{,}
\usepackage[labelformat=simple]{subcaption}

\newcommand{\hone}{\textbf{$H_{apk}$}}
\newcommand{\htwo}{\textbf{$H_{sc}$}}
\newcommand{\htree}{\textbf{$H_{gh}$}}

\title{An Empirical Study on Quality of Android Applications written in Kotlin language}

\def\makeheadbox{{%
\hbox to0pt{\vbox{\baselineskip=10dd\hrule\hbox
to\hsize{\vrule\kern3pt\vbox{\kern3pt
\hbox{\bfseries Published in Empirical Software Engineering, Springer, 2019}
\kern3pt}\hfil\kern3pt\vrule}\hrule}%
\hss}}}

\date{}
\author{Bruno Góis Mateus   \and  Matias Martinez }

\institute{B. Gois Mateus \at
            Université Polytechnique Hauts-de-France, France\\
              UVHC - Campus Mont Houy - 59313 Valenciennes, France \\
              \email{Bruno.GoisMateus@etu.uphf.fr}\\
        \and
        M. Martinez \at
            Université Polytechnique Hauts-de-France, France\\
              UVHC - Campus Mont Houy - 59313 Valenciennes, France \\
            \email{Matias.Martinez@uphf.fr}
}

\newcommand{\inred}[1]{{\color{red}{#1}}}
\newcommand{\mcomment}[1]{{\color{red}{#1}}}
\newcommand{\rcomment}[1]{{\color{blue}{#1}}}

\begin{document}

\maketitle

\begin{abstract}

{\bf Context:}
During the last years, developers of mobile applications have the possibility to use new paradigms and tools for developing mobile applications.
For instance, since 2017, Android developers have the official support to write Android applications using Kotlin language.
Kotlin is programming language fully
interoperable with Java that combines object-oriented and functional features.

{\bf Objective:}
The goal of this paper is twofold.
First, it aims to study the degree of adoption of Kotlin language on the development of open-source Android applications and to measure the amount of Kotlin code inside those applications.
Secondly, it aims to measure the quality of Android applications that are written using Kotlin and to compare it with the quality of Android applications written using Java.

{\bf Method:}
We first defined a method to detect Kotlin applications from a dataset of open-source Android applications.
Then, we analyzed those applications to detect instances of code smells and computed an estimation of the quality of the applications.
Finally,  we studied how the introduction of Kotlin code impacts on the quality of an Android application.

{\bf Results:}
Our experiment found that 11.26\% of applications from a dataset with \numprint{2167} open-source applications have been written (partially or fully) using Kotlin language. 
We found that the introduction of Kotlin code increases the quality, in terms of the presence of 10 different code smells studied, 4 object-oriented and 6 Android, of the majority of the Android applications initially written in Java.

\end{abstract}

\section{Introduction}
\label{sec:introduction}

In 2017, smartphone companies shipped a total of 1.46 billion devices.
The vast majority of them (85\%) run one platform, Android from Google, which is much more adopted than the second most used platform,  iOS from Apple,  with the 14.7\% of the worldwide smartphone volume \citep{mobileshare}.

For developing a mobile application capable of running on devices with Android, 
Google provides an official IDE (Integrated Development Environment), named Android Studio,\footnote{\url{https://developer.android.com/studio/}} and a SDK (Software Development Toolkit).\footnote{\url{https://developer.android.com/about/}} 
Android allows running native applications originally written in Java.
The SDK compiles the Java code into Dalvik bytecode, which is then packaged on an \emph{apk}. 
Then, developers submit those apks to applications stores, such as the official, named Google Play Store.\footnote{\url{https://play.google.com}}
Android users can install applications into their Android devices by downloading those apks directly from the applications stores.
Those Android devices include a virtual machine capable of running apks.

During the last years, different development approaches and frameworks have emerged to ease the development of mobiles applications \citep{Nagappan2016,Martinez:2017:TQI}.
For instances, approaches such as PhoneGap/Cordova\footnote{\url{https://cordova.apache.org}} from Adobe, Xamarin from Microsoft\footnote{\url{https://www.xamarin.com}} and React-Native from Facebook\footnote{\url{https://facebook.github.io/react-native}} aim at facilitating the development of multi-platform mobile applications by allowing developers to write applications using a non-native programming language and then to obtain a version of each platform (Android and iOS).

Meanwhile, Google and Apple continue evolving their development toolkits for building native applications with the goal of avoiding mobile developers migrate to such third-party development frameworks, as Xamarin Visual Studio\footnote{\url{https://visualstudio.microsoft.com/xamarin/}} from their competitor Microsoft.
In June 2014 Apple released Swift, a modern, multi-paradigm language that combines imperative, object-oriented and functional programming for developing iOS applications.

In  2017 Google announced that the Kotlin programming language\footnote{\url{https://kotlinlang.org/}} as an officially supported language for Android development. 
Kotlin is a pragmatic programming language that runs on the Java virtual machine and Android. It combines object-oriented and functional features, and it is fully interoperable with Java. Consequently, it is possible to mix Kotlin and Java code in the same application, to call Koltin code from Java code and the opposite as well.
Android official blog\footnote{\url{https://android-developers.googleblog.com/}} states that, by the end of 2017, Kotlin has been used in more than 17\% of projects in Android Studio 3.0 \citep{kotlinshare}. 

The goal of this paper is twofold.
First,  to study the adoption of Kotlin language on Android applications and the amount of Kotlin code written on those applications.
Secondly, to measure the quality of Android applications that are written using Kotlin.

\begin{tcolorbox}
Our motivation is to know whether the Android applications written using Kotlin language have better quality than the applications written using the traditional approach for developing native Android applications.
\end{tcolorbox}

Several previous studies investigated the quality of mobile applications, focusing on presence of \emph{code smells}, aka \emph{anti-patterns} \citep{reimann2014tool,Hecht2015,Hecht2015a,Palomba2017}, energy consumption \citep{Morales2016,Morales2017,Cruz2017,Carette2017,Saborido2018}, performance \citep{Hecht2016,Carette2017, Saborido2018}, and hybrid applications \citep{Malavolta2015EndUsers,Malavolta2015Hybrid}.
However, to our best knowledge, neither the adoption of Kotlin language for developing Android applications nor the quality of Android applications written using Kotlin language were studied.

To carry out our experiment, we first build a dataset of open-source Android applications.
Then, we analyze the code of those applications to compute the amount of Kotlin code included in each application.
We also study how the amount of Kotlin and Java code evolve along the history of an Android application.
After that, we focus on the quality of Android applications. 
As proposed by \cite{Hecht2015}, we measure the quality of an Android application regarding the presence of code smells.
We first carry out an experiment for detecting instances of ten code smells proposed by the literature \citep{Hecht2015,hecht_thesis}. 
We compare the presence of code smells between two sets of Android applications: one that includes applications written using Koltin and the other with applications written using Java. 
Finally, 
we study how the introduction of Kotlin code impacts on the quality of Android applications initially written in Java.

The research of this paper is guided by the following research questions.

\newcommand{\rqadoption}{What is the degree of adoption of Kotlin in mobile development in our dataset of open-source applications?}
\newcommand{\rqproportionkotlincode}{What is the proportion of Kotlin code in mobile applications?}
\newcommand{\rqevolutionproportion}{How does code evolve along the history of an Android application after introducing Kotlin code?}
\newcommand{\rqdifferencewithjava}{Regarding the studied code smells, is there a difference between Kotlin and Java Android apps in terms of code smells presence?}
\newcommand{\rqscorekotlin}{How frequent does the introduction of Kotlin positively impact on the quality of the versions of an Android application?}

\begin{itemize}
\item \textbf{RQ 1:} \emph{\rqadoption}

We inspected three different datasets of Android applications (F-Droid,\footnote{\url{https://f-droid.org/}} AndroidTimeMachine\footnote{\url{https://androidtimemachine.github.io}} and Androzoo\footnote{\url{http://androzoo.uni.lu}}) and we filtered applications that were built, at least partially, using Kotlin.

{\bf \textit{Finding 1:}} The \numprint{11.26}\%  of the open-source Android  applications from our dataset contain Kotlin code (i.e., 244 out of 2167); 

\item \textbf{RQ 2:} \emph{\rqproportionkotlincode}

We analyzed every filtered application and we computed the proportion of code written in Kotlin language.

{\bf \textit{Finding 2:}} The 33.61\% of the apps that include Kotlin (i.e., 82 out of 244) are entirely written in Kotlin, i.e., those do not include Java code. 

\item \textbf{RQ 3:}  \emph{\rqevolutionproportion}

We computed the amount of Kotlin code for each commit of an Android application to analyze the evolution trends of the Kotlin code along the history.

{\bf \textit{Finding 3:}}  For the 63.9\% of the applications written in Kotlin (totally or partially), the amount of Kotlin code increases over the evolution whereas the amount of Java code is reduced. 

\item \textbf{RQ 4:} \emph{\rqdifferencewithjava}

We replicated the experiment done by \cite{Habchi2017} which compares iOS and Android apps, to compare code smells found in two Android apps datasets: one with Kotlin code, other without any line of Kotlin.

{\bf \textit{Finding 4:}} For 3 out of 4 object-oriented code smells affect more Kotlin than Java applications, but for all object-oriented smells Java applications have in median more entities affected with statistical relevance.

\item \textbf{RQ 5:} \emph{\rqscorekotlin}

We first computed the quality of each version of Kotlin applications based on the quality models from \cite{Hecht2015}. Then, we measured the impact over the quality of introducing Kotlin code in applications initially written in Java.

{\bf \textit{Finding 5:}} The introduction of Kotlin code in Android applications initially written in Java increases the quality of, at least, the 50\% of the apps.
\end{itemize}   

The contributions of this paper are:

\begin{itemize}
\item A methodology for detecting applications written in Kotlin; 
\item A dataset of 244 open-source Android applications written partially or totally with Kotlin;
\item A study about the amount of Java and Koltin code along the evolution of Android applications;
\item A study that compares the presence of code smells in Kotlin and Java applications;
\item A study about the measurement of the impact over the software quality of introducing Kotlin code in Android applications.
\end{itemize}

The paper continues as follows.
Section \ref{sec:relatedwork} presents the related work.
Section \ref{sec:kotlinintro} describes Kotlin and the development of Android applications using Kotlin language.
Section \ref{sec:methodology} introduces the methodologies used to respond the research questions.
Section \ref{sec:results} presents the results and the answers to the research questions.
Section \ref{sec:threatsvalidity} presents the threats to the validity.
Section \ref{sec:discussion} presents the discussion about our findings.
Section \ref{sec:conclusion} concludes the paper.
All the data presented in this paper is publicly available in our appendix: \url{https://github.com/UPHF/kotlinandroid}. 
\section{Related Work}
\label{sec:relatedwork}

Kent Beck coined the term \emph{code smell} in the context of identifying
quality issues in code that can be refactored to improve the maintainability of a software~\citep{fowler1999}. 
Since there, the software engineering community has explored various associated dimensions that include proposing a catalog of smells, detecting smells using a variety of techniques, exploring the relationships among smells, and identifying the causes and effects of smells~\citep{Sharma2018}. However, according to~\cite{Aniche2017}, traditional code smells capture very general principles of good design. Moreover, they suggested that specific types of code smells are needed to capture ``bad practices'' on software systems based on a specific platform, architecture or technology. In this context, it possible to find work focused on smells specific to the usage of object-relational mapping frameworks~\citep{chen2014}, Android applications~\citep{verloop2013,reimann2014tool,Hecht2015a},  Cascading Style Sheets (CSS)~\citep{mazinanian2014} and Model View Controller (MVC) Architecture~\citep{Aniche2017}.

In this section, we discuss the relevant literature about
code smells detection and fixing in mobile applications and related work on software evolution.

\subsection{Detection of anti-patterns (code smells) on mobile app}

\cite{Mannan2016} compared the presence of well-known object-oriented code smells in 500 Android applications and 750 desktop applications in Java. They concluded that there is not a major difference between these two types of applications concerning the density of code smells. However, they observed that the distribution of code smells on Android applications is more diversified than for desktop applications.
\cite{Khalid2016} conducted a study about the relation of the presence of code smells and applications' rating.
They analyzed \numprint{10000} Android applications and their reviews at Google Play Store and found that there is  a correspondence between three categories of warnings and the complaints in the review-comments of end users. 

\cite{reimann2014tool} proposed a catalog of 30 quality smells dedicated to Android. These code smells are mainly originated from the good and bad practices documented online in the official Android documentation or by developers reporting their experiences on blogs.
\cite{Hecht2015a,Hecht2015} proposed a tooled approach, named Paprika, to identify object-oriented and Android-specific code smells from binaries (apk) of mobile applications. \cite{Palomba2017} proposed a detection tool, called aDoctor, to detect 15 Android code smells using static analysis code techniques. They tested aDoctor on a testbed of 18 Android applications and attained a detection precision close to 100\%.

While \cite{Mannan2016} and \cite{Khalid2016} focused on detection of object-oriented smells in mobile applications, \cite{reimann2014tool} proposed a catalog of smells dedicated to Android. Moreover, \cite{Hecht2015,Hecht2015a} and \cite{Palomba2017} proposed tools to identify Android smells.

To the best of our knowledge, nobody studied code smells on Android applications written with Kotlin.

\subsection{Code evolution}

Object-oriented metrics gained popularity to assess software quality, since their definition by \cite{Chidamber1994}.  

\cite{Li2017evo} conducted an empirical study on long spans in the lifetime of 8 typical open-source mobile applications to have a better understanding of the evolution of mobile applications. 
Their results indicated that a subset of Lehman’s laws~\citep{lehman1980programs} still applies to open-source mobile applications. 
Moreover, they found that the growth of mobile applications is non-smooth and the software instability increases with the addition of third-party libraries.

To the best of our knowledge, no work studied the evolution of Kotlin applications.

\subsection{Presence of code smells (Anti-patterns) throughout the software evolution}

To assess software quality other work focus on the presence of code smells.

\cite{Palomba2015} presented HIST (Historical Information for Smell deTection), an approach aimed at detecting five different code smells by exploiting information extracted from versioning systems. They compared their approach over a manually-built oracle of smells, identified in twenty Java open-source projects with traditional approaches, based on the analysis of a single project snapshot. They found that HIST can identify code smells that cannot be identified by competitive approaches solely based on code analysis of a single system’s snapshot.
Also, they confirmed that there is a potential to combine historical and structural information to achieve better smell detection.

\cite{Tufano2015} conducted a sizable empirical study over the change history of 200 open-source projects from different software ecosystems, including Android applications. They investigated when bad smells are introduced by developers and the circumstances and reasons behind their introduction. They found that most times code artifacts are affected by code smells since their creation. Moreover, they observed that the main activities whose developers tend to introduce smells are implementing new features and enhancing existing ones.

\cite{Hecht2015a,Hecht2015} proposed a tooled approach, called Paprika, to identify object-oriented and Android-specific anti-patterns from binaries of mobile applications, and to analyze applications' quality along their evolution. They considered 106 Android applications that differ both in internal attributes, such as their size and external attributes from the perspective of end users. They collected several versions (apks) of each application to form a total of \numprint{3568} versions to estimate software quality during their evolution. One of the main findings of their work is that mobile applications developers need to allocate more quality assurance efforts.  
\cite{Palomba2018} presented a large-scale empirical study on the diffuseness of code smells and their impact on code change- and fault-proneness. They analyzed a total of 395 releases of 30 open-source projects, considering 13 different code smells. 
They found that smells characterized by long and/or complex code (e.g., Complex Class) are highly diffused, and  smelly classes have a higher change- and fault-proneness than smell-free classes.

Therefore, \cite{Tufano2015} and \cite{Palomba2015} focused on code smells, but none of them investigated the presence of Android-specific smells during the software evolution. On the other hand, \cite{Hecht2015a,Hecht2015} proposed a tool capable of identifying both type of smells, object-oriented, and Android and they evaluated the evolution of software quality. 

However, none of these work studied the presence of smells on the mobile application written in Kotlin.

\subsection{Relation between code smells and programming languages in mobile applications}

\cite{Habchi2017} studied code smells in the iOS ecosystem, considering Swift and Objective-C languages and how it is compared with Android smells. They proposed a catalog of 6 iOS-specific code smells that they identified from developers’ feedbacks and the platform official. 
To identify those code smells they extended Paprika~\citep{Hecht2015,Hecht2015a}. Then, they analyzed 103 Objective-C applications and 176 Swift applications hosted on GitHub and discovered that code smells tend to appear with the same proportion or only a slight difference in Objective-C and Swift and that the applications written in Objective-C and Swift are very different concerning object-oriented metrics. Furthermore, they analyzed \numprint{1551} Android open-source applications from F-Droid and found that Android applications tend to contain more code smells than iOS applications in both languages, except for the SAK code smell, which appears in the same proportion for all languages. 

Although \cite{Habchi2017} focused on the relation between code smells and programming languages in the context of mobile applications, they did not consider applications written in Kotlin.

\subsection{Studies that measure the impact of code smells on runtime.}

Besides some work with a focus on identifying code smells, other authors have focused on better understanding the impact of code smells on runtime.

\cite{Hecht2016} conducted an empirical study with different versions of open-source Android applications to determine whether fixing Android anti-patterns have a significant impact on the user interface (number of delayed frames) and memory usage. They reported that correcting these Android code smells effectively improves the user interface and memory usage in a significant way. 
 
\cite{Cruz2017} studied the impact of eight best performance-based practices on the energy consumed by Android applications.
They carried out an empirical study with 6 applications from F-droid and manual refactoring was applied for each detected pattern in those applications. 
They found that fixing ViewHolder, DrawAllocation, WakeLock, ObsoleteLayout-Param, and Recycle improved energy efficiency. However, fixing UnusedResources and UselessParent did not provide any significant change in energy consumption while fixing Overdraw increase energy consumption by 2.2\%. 

\cite{Saborido2018} performed a comparison among the implementation HashMap provided by Java SDK and the different implementations of map data structure offered by Android API.
They analyzed \numprint{5713} Android applications available on GitHub and published on Google Play.
They found that SparseArray variants should be used instead of HashMap and ArrayMap when keys are primitive types because it is more efficient in terms of CPU time, memory and energy consumption,  disagreeing partially with the Android documentation. 

These studies focused on object-oriented and Android smells and how they impact on different runtime aspects. Nevertheless, these experiments did not consider Kotlin applications.

 \subsection{Approaches that automatically correct code smells}

Previous work from \cite{Mannan2016,Khalid2016,Li2017c,Palomba2017,Hecht2015a,Hecht2015,Habchi2017,Hecht2016,Cruz2017,Saborido2018} focused on studying and detecting code smells. Other worls focused on applying ``correction'' action to overcome the code smells. 
 
 \cite{Carette2017} proposed a tooled and reproducible approach, called Hot-Pepper, to automatically correct code smells and evaluate the impact on energy consumption. 
 They conducted an empirical study on five Android applications to assess the impact on energy consumption. Their results confirmed that Android anti-patterns have an impact on the energy consumption of applications.
 
\cite{Morales2016,Morales2017} introduced a novel approach for refactoring mobile applications while controlling for energy consumption, named EARMO, that uses evolutionary multi-objective techniques. They evaluated it on a benchmark of 20 free and open-source Android applications. The results showed that EARMO can propose solutions to remove a median of 84\% of anti-patterns, with a median execution time of less than a minute. Concerning the difference in energy consumption after refactoring, they observed that three applications improved their energy consumption with statistically significant results. 

\cite{Cruz2018} presented a tool capable of identifying and applying automatic refactor of five energy code smells: View Holder,  Draw Allocation, Wake Lock, Recycle and ObsoleteLayoutParam~\citep{Cruz2017}.  
They analyzed 140 free and open-source Android applications collected from F-droid.
Their experiment yielded a total of 222 refactorings  in 45 applications, which were submitted to the original repositories as PRs.  18 applications had successfully merged their PR.
These studies~\citep{Morales2016,Morales2017,Cruz2017,Cruz2018, Carette2017} applied refactoring at Java code level to remove code smells.

As far as we know, no studies focused on apply refactoring on Kotlin applications.

\section{Brief Introduction to Mobile Development using Kotlin Language}
\label{sec:kotlinintro}

In this section, we briefly describe Kotlin and  Android application development using Kotlin.

\subsection{Kotlin programming languages}
\label{kotlin}
Kotlin is a statically typed programming language that  reduces language verbosity relative to Java (e.g., semicolons are optional as a statement terminator `;'). Furthermore, Kotlin applications run on top of the Java Virtual Machine (JVM). The Kotlin compiler \emph{kotlinc} compiles Kotlin code to Java bytecode, which can be executed by that JVM. For this reason, Kotlin is interoperable with Java and other JVM languages as well (e.g., Scala or Groovy).
This interoperability can be seen in two manners. First, Kotlin developers can use libraries (e.g., jars) written in another language such as Java, and secondly, developers can create applications using Kotlin together with other JVM languages.

\subsection{New Features from Kotlin}
Kotlin and Java are modern programming languages that run on top of the JVM. 
Nevertheless, Kotlin presents some differences compared to Java. 
To our opinion, the most relevant is that, in addition of the object-oriented paradigm,  Kotlin introduces functional programming, which allows, for instance, the use of high order functions. 

According to the official documentation,\footnote{\url{https://kotlinlang.org/docs/reference/comparison-to-java.html}} the following list presents some features from Kotlin not present in Java:
\begin{inparaenum}[\it 1)]
    \item the combination of Lambda Expressions and Inline functions, 
    \item Extension functions,
    \item Null-safety,
    \item Smart casts,
    \item Properties,
    \item Primary constructors,
    \item First-class delegation,
    \item Type inference for variable and property types,
    \item Singletons,
    \item Declaration-site variance and Type projections,
    \item Range expressions,
    \item Operator overloading,
    \item Companion objects,
    \item Data classes,
    \item Separate interfaces for read-only and mutable collections, and
    \item Coroutines.
\end{inparaenum}

Moreover, according to the mentioned documentation, 
Kotlin fixes a series of issues that Java suffers from:
\begin{inparaenum}[\it 1)]
 \item Null references are controlled by the type system,
 \item No raw types,
 \item Arrays in Kotlin are invariant,
 \item Kotlin has proper function types, as opposed to Java's SAM-conversions,
 \item Use-site variance without wildcards, and
 \item Kotlin does not have checked exceptions.
\end{inparaenum}

\subsection{Kotlin for Mobile development}

To develop a mobile application using Kotlin, developers can use the same tools that Google provides for developing Android applications using Java.
Those tools are:
\begin{inparaenum}[\it a)]
\item the Software Development Kit (SDK), and 
\item the Official Android Integrated Development Environment (IDE).
\end{inparaenum}
For instance, Android Studio 3.0+ fully supports the development of Android applications using Kotlin code and provides features such as autocomplete, debugging, refactoring, and lint check.
Using Android Studio, a mobile developer can: 
\begin{inparaenum}[\it 1)]
\item start a new Android project for developing an app using Kotlin  from scratch, 
\item add new Kotlin files to an existing project already written in Java, or
\item converts existing Java code to Kotlin. 
\end{inparaenum}

Android applications' source code (written in Java, Kotlin or both) are compiled to Dalvik bytecode, which runs on top of an adapted JVM for Android devices named \emph{Dalvik} or \emph{Art} (according to the Android version). 
The Dalvik bytecode is stored into .dex files (those are similar to .class files for JVM). 
Moreover, Dalvik bytecode is packaged in a file named Application Package Kit (APK), which groups the bytecode of a mobile app, similarly to jar files for class files in Java. 
Android SDK provides a feature of inspecting these files using a tool named \emph{apkanalyzer}.\footnote{\url{https://developer.android.com/studio/build/apk-analyzer}} 

To generate an apk, every app project must have an \textit{AndroidManifest.xml} file at the root of the project source set.\footnote{\url{https://developer.android.com/guide/topics/manifest/manifest-intro}}
The manifest file describes essential information about an application and is consumed by the Android build tools, the Android operating system, and Google Play. 
The manifest must include: the app's package name,  components of the app (activities, services, broadcast receivers, and content providers), permissions that the app needs to access protected parts of the system or other applications, etc. 

\subsection{Advantages claimed by Kotlin community}
\label{sec:advantagesclaimed}
 
To the best of our knowledge, nobody has studied the advantages of using Kotlin. However, the community of Kotlin developers enumerates some features in favor of Kotlin, suggesting that mobile developers should migrate from Java to Kotlin. 
For example, \cite{kotlin_adv} suggests 17 reasons 
\begin{inparaenum}[\it 1)]
\item Java interoperability,
\item familiar syntax,
\item string interpolation,
\item type inference,
\item smart casts,
\item intuitive equals,
\item default arguments,
\item named arguments,
\item the \emph{when} expression,
\item properties,
\item the data class,
\item operator overloading,
\item destructuring declarations,
\item ranges,
\item extension Functions,
\item null safety, and
\item better lambdas.
\end{inparaenum}
To the best of our knowledge, no work has empirically validated those reasons in the context of mobile development.

\section{Methodology}
\label{sec:methodology}

In this section, we present the methodology used in this paper for studying Android applications written with Kotlin.
Section~\ref{sec:method:dataset}  describes the method to collect  Android applications from different sources and presents the resulting dataset of applications studied along this paper. 
Section \ref{sec:method:filteringapps}  presents the heuristics for classifying Kotlin applications  from the mentioned dataset, used to respond the RQ 1.
Section \ref{sec:method:proportion} presents the method to obtain the proportion of Kotlin code, used to respond the RQ 2.
Section \ref{sec:methodology:codeevolutionclassification} presents the different evolution trends of source code, used to respond the RQ 3. 
Section \ref{sec:method:analysisapps} describes the code smells from the bibliography, used to respond the RQ 4.
Finally, section \ref{sec:method:qualityscore} describes 
\begin{inparaenum}[\it 1)]
\item the technique to calculate the quality score of Android applications, and 
\item the process to measure the impact of introducing Kotlin code, used to respond to RQ 5. 
\end{inparaenum}

\subsection{Creation of a dataset of Kotlin applications}
\label{sec:method:dataset}

\subsubsection{Criteria for building our dataset of mobile applications}
As the goal of our experiment is to study the use and quality of Kotlin code in Android applications,
we decided to study mobile applications that have at least one version released in 2017 or later.
As Kotlin was announced as an official language for Android development in 2017,\footnote{\url{https://android-developers.googleblog.com/2017/05/android-announces-support-for-kotlin.html}} before that date, Android developers did not have support from Google for developing Android applications using Kotlin language.
Therefore, we considered that applications whose last versions date from 2016 or earlier could not give us much information about the use of Kotlin language in the Android domain.

Moreover, we need that our dataset of Android applications includes, for each application: 
\begin{inparaenum}[\it a)]
\item its source code hosted in a code repository (e.g., GitHub), and 
\item binary files (apk) of the released versions. 
\end{inparaenum}

\subsubsection{Selecting dataset of Android applications}
We created our dataset of mobile applications by combining three already defined datasets: F-droid, AndroidTimeMachine~\citep{Geiger2018:data} and AndroZoo~\citep{Allix2016}. To our knowledge, those are the three largest publicly available datasets of Android applications that contain applications recently released. 

Let us now to introduce each dataset and to explain the reasons for choosing them. 

\paragraph{F-droid} is a directory of open-source Android applications that contains  \numprint{1509} applications.\footnote{Last visit: 06/04/2018.}
A mobile application has one or more versions, each of them represented by an apk. 
At the main page of each application, F-Droid provides the links to download the last three versions of an application and a link for another page that contains a list of all the versions.

F-droid provides all the information we need, i.e., access to a code repository and apks of the different released versions of the applications.
However, the number of applications (\numprint{1509}) is low compared with other datasets.
For this reason, we decided to mine other Android datasets with the goal of including more applications in our study.

\paragraph{AndroidTimeMachine}\label{sec:atmdescription}
is a graph database of Android applications which are both accessible on GitHub and Google Play~\citep{Geiger2018:data}. To create this dataset, the authors defined and executed a 4-step process: \begin{inparaenum}[\it 1)] \item identifies open-source Android applications hosted on GitHub, \item extracts their package names, \item checks their availability on the Google Play store, and \item matches each GitHub repository to its corresponding app entry in the Google Play store~\citep{Geiger2018:data}.
\end{inparaenum}
In total, AndroidTimeMachine has \numprint{8,431} applications, and it is based on publicly-available GitHub mirror available in BigQuery.\footnote{\url{https://cloud.google.com/bigquery/public-data/github}} 
Using the Neo4j database is possible to retrieve for each application its source-code repository link and the application's package name. 
However, this dataset does not provide any apks from its applications.
For this reason, we decided to mine the missing apks on AndroZoo dataset. 

\paragraph{AndroZoo} is dataset of millions of Android applications collected from various data sources \citep{Allix2016}, including major market places \emph{Google Play}, \emph{Anzhi}, and \emph{AppChine}, as well as smaller directories mobile, \emph{AnGeeks}, \emph{Slideme}, \emph{ProAndroid}, \emph{HiApk}, and \emph{F-Droid}~\citep{Geiger2018}. 
In total, AndroZoo has \numprint{4390288} applications that correspond to a total of \numprint{7795372} apks (versions).\footnote{Last visit: 16/10/2018.} 
The list of available APKs is regularly updated on the AndroZoo website, along with metadata for each app as, the main package name, the size of the APK, the version, and the market where the app was downloaded from, etc ~\citep{Allix2016}. However, AndroZoo does not provide a link for the source-code repository of an application, even if an application is open-source.

\subsubsection
{Building our study dataset}
\label{sec:method:fdroid}
Now, we present the different steps to build the dataset used in this study.

\paragraph{Step 1: Mining F-Droid.} 

Using the upload version date included in F-Droid,
we retrieved 926 applications from F-Droid that fulfill our selection criterion.\footnote{Experiment executed the June 4th, 2018.}
The total number of versions (i.e., apks) found corresponding to those applications was \numprint{13094}. 
We could download \numprint{11675} apks (89\%). 
The rest of the apks files (11\%) were not available.

\paragraph{Step 2: Mining AndrodTimeMachine.} 
The version of  AndroidTimeMachine  presented by ~\cite{Geiger2018:data}  contains~\numprint{8431} applications and it was executed in October 2017. As their infrastructure is publicly available,\footnote{AndroidTimeMachine resources: \url{https://androidtimemachine.github.io/dataset/} and \url{https://github.com/AndroidTimeMachine/open_source_android_apps}}
we re-executed it on October 2018 with the goal of including more recent applications.

Once we retrieved the list of applications, 
differently of the original work from~\cite{Geiger2018:data}, 
we carried out a new step to keep applications whose code repositories have only one Android manifest file. We added this new step because we discovered that the AndroidTimeMachine match algorithm produced wrong results (i.e., applications linked with wrong repositories) when a repository has more that one AndroidManifest.xml files.
In our appendix we list those applications that suffer the mentioned problem.\footnote{\url{https://github.com/UPHF/kotlinandroid/blob/master/docs/multiple_manifest_problem.md}.}

Table~\ref{tab:comp:steps} shows the comparison between the numbers found by us resulting from the re-execution of each step and the numbers of the original dataset by \cite{Geiger2018:data}. 
In conclusion, we retained  \numprint{2156} applications from AndroidTimeMachine, all of them fulfill our selection criterion.

\begin{table}
\centering
\caption{Steps executed to build a recent version of AndroidTimeMachine}
\begin{tabular}{|l|c|c|}
\hline
\multirow{2}{*}{Steps} & \multicolumn{2}{|c|}{AndroidTimeMachine}\\ 
\cline{2-3}
& Original & Updated \\\hline
Finding Android Manifest Files & \numprint{378610} & \numprint{358518} \\\hline
Extracting package Names & \numprint{112153} & \numprint{114152} \\\hline
Discarding repositories with more than one manifest file & - & \numprint{49570} \\\hline
Filtering applications on Google Play & \numprint{9478} & \numprint{3664} \\\hline
Matching of Google Play pages to GitHub repositories & \numprint{8431} & \numprint{3664} \\\hline
Filtering applications from 2017 or later & - & \numprint{2156} \\\hline
\end{tabular}
\label{tab:comp:steps}
\end{table} 
\paragraph{Step 3: Combining AndrodTimeMachine with AndroZoo:} 
Since AndroidTimeMachine does not provide apks from the applications, 
we retrieved from AndroZoo the apks corresponding to each application of AndroidTimeMachine.
To obtain those apks, we first extracted the package name located on the Android manifest, and then we queried the AndroZoo HTTP API.
In total, \numprint{1531} applications from AndrodTimeMachine were present in AndroZoo. 
To assure that these applications were correctly linked with repositories, we manually checked, for each application, its repository page and its page on Google Play Store. Then we found 54 applications linked with wrong repositories. We removed them, remaining \numprint{1477} applications.\footnote{In the appendix we list that applications that suffer the mentioned problem: \url{https://github.com/UPHF/kotlinandroid/blob/master/docs/wrong_match.md}.}

\paragraph{Step 4: Combining F-Droid and AndroZoo.} 
Finally, we kept \numprint{1241} applications from AndroZoo. The rest, \numprint{236}, were already discovered from F-Droid during the step 1.

\paragraph{The resulting dataset.} Our final dataset, named FAMAZOA (\underline{F}-droid \underline{A}ndroidtime\underline{M}achine \underline{A}ndroZoo \underline{O}pen-source \underline{A}pplications) corresponds to:  

$$ FAMAZOA =  F\-droid \cup ( AndroZoo \cap AndroidTimeMachine)$$

\begin{figure}
    \center
    \includegraphics[width=0.8\textwidth]{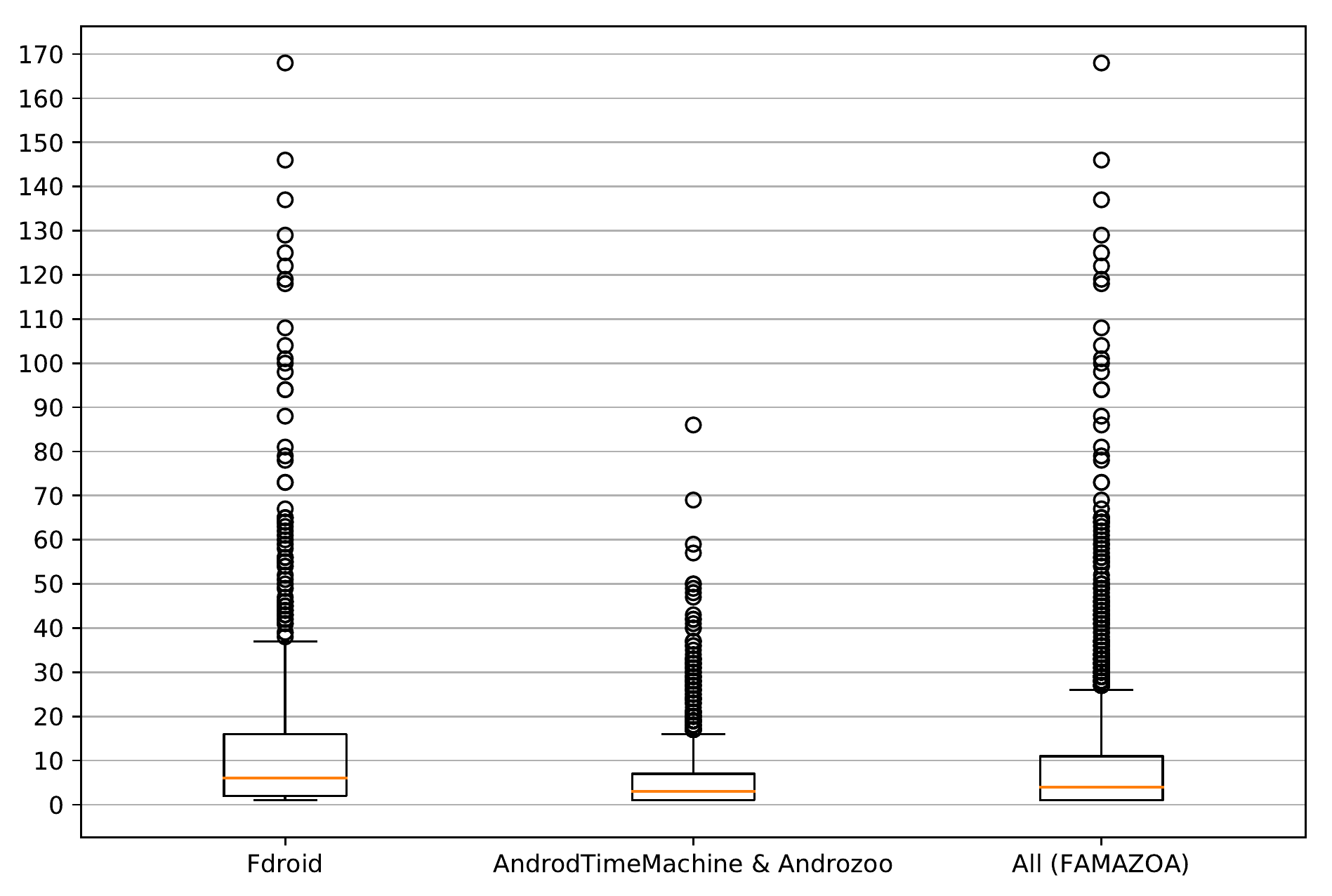}
    \caption{Distribution of number of version (apk) per application.
    }
    \label{fig:box:datasets}
\end{figure}

In total, FAMAZOA has \numprint{2167} applications (926 + \numprint{1241})  and \numprint{19838} apks. 
Figure~\ref{fig:box:datasets} shows the distribution of apks (released versions) per applications.
From F-droid we downloaded in median 6 apks, 
and from AndroidZoo we downloaded in media 3 apk per application.
In summary, FAMAZOA has in median 4 apk per application.
The complete list of applications from FAMAZOA is publicly available in our appendix.\footnote{FAMAZOA dataset: \url{https://github.com/UPHF/kotlinandroid/blob/master/docs/final_dataset.md}}

\subsection{Detecting Kotlin applications}
\label{sec:method:filteringapps}

For responding our  RQ$_1$ (\textit{\rqadoption}), 
we built a process to classify both applications and apks (retrieved in Section \ref{sec:method:dataset}) in three categories of applications:
\begin{inparaenum}[\it 1)]
\item written with Java (not include any line of Kotlin code),
\item written partially with Kotlin, and
\item written totally with Kotlin, we call these applications as `pure Kotlin'.
\end{inparaenum}
Note that we only focused on applications' source code and bytecode, discarding third-party libraries,  which neither their source code nor bytecode (jars) are included in the applications' code repositories.

\begin{figure}
    \includegraphics[width=\textwidth]{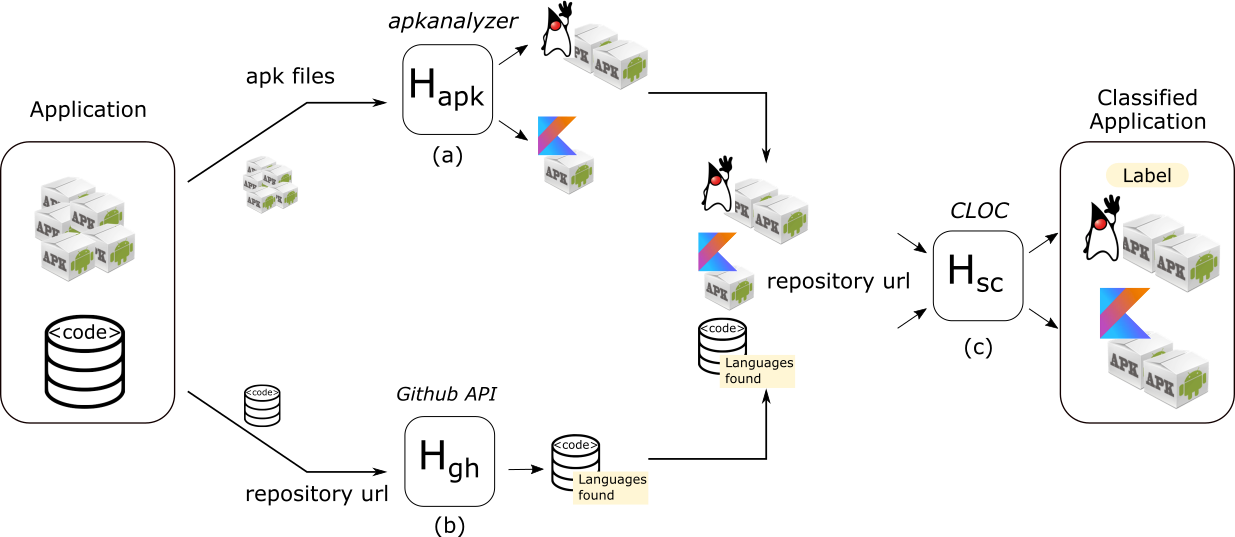}
    \caption{Our pipeline to classify Android applications  (Section \ref{sec:method:filteringapps}.). We take as input an application that corresponds to its apks and its source-code repository url and then classify it between Java, Kotlin or Pure Kotlin in three steps. 
    First, in step (a), heuristic \hone{} considers the content of apk. 
    Second, in step (b), heuristic \htree{} uses the GitHub API to get information about the programming languages used if the application is hosted there.
    Third, in step (c), heuristic \htwo{}  analyzes the applications' commit history to verify the presence of Kotlin source code. (d) Finally, we classify (label) as Java, Kotlin or Pure Kotlin.}
    \label{fig:dataset_pipeline}
\end{figure} 
For each application from FAMAZOA, which has a link to the code repository and a set of apks, we applied three heuristics to classify an application and its apk. Figure~\ref{fig:dataset_pipeline} shows this classification process.

We  first applied an heuristics \hone{}, Figure~\ref{fig:dataset_pipeline}(a), that consists of looking for a folder called \textbf{kotlin} inside the apk file. Having such folder indicates the presence of  Kotlin code inside the application.  
To automatize this task,  a tool named \textit{apkanalyzer} included in the Android SDK. 
Using this heuristic, we first classified each version (apk) of an application.
Then, if at least one apk of an application is classified as Kotlin, the heuristic classified the application as `Kotlin'. 
Otherwise, it classifies as `Java'.
The \hone{}  provides a cheap and fast approach to get an initial guess about the presence of Kotlin code.

At the same time, we applied our second heuristic \htree{}, Figure~\ref{fig:dataset_pipeline}(b), which relies on the GitHub API:  for each application hosted on GitHub, we queried the GitHub API to retrieve the amount of code (expressed in bytes) from the most recent version (i.e., the HEAD) grouped by programming language. We classified the app as `Kotlin' if Kotlin language is present in the response of the API. The input of this heuristic is a URL to a GitHub repository.

Finally, once we retrieved a set of candidate Kotlin applications using \hone{} and \htree{}, i.e., $H_{apk} \cup H_{gh}$,
 we applied the heuristic \htwo{} over them, Figure~\ref{fig:dataset_pipeline}(c), 
to assert the presence of Kotlin code and to measure how much Kotlin an application has.  
The heuristic \htwo{} inspected every commit of the source code repository of an application.
An application was classified as `Kotlin' if the heuristic found at least one commit that has Kotlin code,  and as `Pure Kotlin' if all commits contained only Kotlin code.
To carry out this task, the heuristic used the tool CLOC\footnote{\url{http://cloc.sourceforge.net/}} which returns a list with the languages used in an application and the amount of code regarding no-blank lines. 
\htwo{} is time-consuming because it requires to analyze the source code of each commit of an application.
Therefore, to execute \htwo{} for every application from our dataset is prohibitive.

Note that, differently of \htwo{}, heuristic \htree{} only focuses on the most recent version hosted on GitHub (the API only retrieves that information). Consequently, it cannot detect applications that:
\begin{inparaenum}[\it a)]
\item do not contain Kotlin code in the most recent version (last commit), but
\item contain Kotlin code in older versions.
\end{inparaenum}

\subsection{Analyzing the proportion of Kotlin code}
\label{sec:method:proportion}

For responding the RQ 2 (\textit{\rqproportionkotlincode}),
we first retrieved the code of the last version of each application classified as `Kotlin' or `Pure Kotlin' using the heuristics presented in Section \ref{sec:method:filteringapps}.
For each application under analysis, we retrieved the corresponding code repository.
By construction, our dataset (build from F-Droid and AndroidTimeMachine dataset) only include open-source applications which code repositories are publicly available (Section \ref{sec:method:dataset}).

Once we retrieved all the repositories, 
for each repository (associated to one application), we executed the CLOC over the most recent version (i.e., the last commit), then we calculated the proportion of Kotlin code (excluding blanks and comments) concerning the total code. 
Note that we discarded analyzing files that did not contain Java or Kotlin code, such as XML, CSS, JavaScript, and others.

\subsection{Analyzing the code evolution of Android applications}
\label{sec:methodology:codeevolutionclassification}

For responding the RQ 3 (\textit{\rqevolutionproportion}), 
we inspected the source code repositories to analyze the evolution trends of Kotlin code along the histories of the applications. 
The applications that we considered in this experiment are those applications classified as Kotlin or pure Kotlin (Section \ref{sec:method:filteringapps}).

For each application, we visited each commit from its code repository in chronological order (i.e., starting from the oldest one) for calculating the amount of code (also using CLOC) of the version related to each commit. 

In this experiment, we focused on analyzing the evolution trend of two particular languages: Java (i.e., the traditional used for developing Android applications) and Kotlin.

We defined 12 cases that represent different evolution trends of Kotlin and Java code.
They are:
\begin{enumerate}[\it ET 1: ]
\item  Kotlin is the initial language and the amount of Kotlin grows along the history.
\item  Kotlin code replaces all Java code between two consecutive versions.
\item  Kotlin code replaces some Java in consecutive versions (i.e., amount of Java code drops), but after the drop, the amount of Java continues growing.
\item Kotlin increases together with Java.
\item  Kotlin grows and Java decreases and last version of the app has both languages.
\item  Kotlin grows and Java decreases until the amount of Java code is 0. 
\item  Kotlin grows and Java remains constant. 
\item  Kotlin is constant and Java changes.
\item Kotlin and Java remain constant.
\item  Kotlin introduced in the app but lately disappears (the amount is 0).
\item Java replaces Kotlin code.
\item Other.
\end{enumerate}

Note that it could exist more evolution trends that we have not included in the previous list.
We included only those we have observed during our experiment and are particularly interesting for this paper.

To classify each Kotlin application according to evolution trend, 
we first plotted the amount of code (lines of code) of Kotlin and Java for each commit. Figure \ref{fig:evolutions_trends1} shows some of such plots.
Then, we manually selected the evolution trend that is most representative (i.e., that better fits) to the code evolution of that application. 
We classified an application with a given evolution trend only when both authors of the paper fully agree on the classification.
Otherwise, we classified the applications without consensus as \emph{Other}. 
Moreover, we make publicly available in our appendix all the plots and the resulting classifications for further analysis of our experiments.\footnote{\url{https://github.com/UPHF/kotlinandroid/tree/master/docs/evolution/}}

\subsection{Analyzing the difference between Kotlin and Java applications and the quality of Android application}
\label{sec:method:analysisapps}

For responding the RQ 4 (\textit{\rqdifferencewithjava}),
we run a code smells detection tool over Android applications from the dataset presented in Section \ref{sec:method:filteringapps}. 
In this paper we focused on two kinds of smells: object-oriented and Android-related code smells.

\subsubsection{Selection code smells detecting tool}
\label{sec:method:selection}

To detect code smells from Android applications we selected the tool named Paprika~\citep{Hecht2015,Hecht2015a}.\footnote{\url{https://github.com/GeoffreyHecht/paprika}}
The reasons of selecting Paprika were:
\begin{inparaenum}[\it a)]
\item it can detect object-oriented and Android specific code smells;
\item it was designed for detecting code smells on Android applications without requiring code source: the input of Paprika is the apk of one Android application;
\item as it works at JVM bytecode level (i.e., an apk contains bytecode), it can analyze Android application written on Java and/or Kotlin;
\item it is open-source and hosted on GitHub;
\item previous work had extensively used Paprika for analyzing mobiles applications \citep{Hecht2015,Hecht2015a,Hecht2016,Habchi2017,Carette2017,Grano2017}; 
\item the tool implementation was deeply validated, including an experiment done together with Android developers \citep{hecht_thesis}.
\end{inparaenum}

\begin{table}
\centering
\caption{Paprika supported code-smells. The column `Considered' shows the 10 code smells studied in our work (\checkmark) and the 7  not studied ($X$).}
\begin{tabular}{|p{1cm}|l|c|c|}
\hline
Type& Code smell name &Entity& Considered \\
\hline
\hline
\multirow{4}{1.8em}{Object-Oriented} 
& Blob Class (BLOB) &Classes& \checkmark\\
& Swiss Army Knife (SAK) &Interface &\checkmark \\
& Complex Class (CC) &Classes& \checkmark \\
& Long Method (LM) &Method& \checkmark \\
\hline
\multirow{12}{2em}{Android-Specific} 
& Hashmap Usage (HMU) &Class&$X$\\
& Unsupported Hardware Acceleration (UHA) &Class&$X$ \\
& Leaking Inner Class (LIC) &Inner classes& $X$ \\
& Member Ignoring Method (MIM) &Methods& $X$ \\
& Internal Getter/Setter (IGS) &Methods& $X$\\
& No Low Memory Resolver (NLMR) &Activities& \checkmark \\
& Heavy ASynctask (HAS) &Async Tasks& \checkmark \\
& Heavy Service Start (HSS) &Services& \checkmark \\
& Heavy Broadcast Receiver (HBR) &Broadcast Receivers& \checkmark \\
& Init OnDraw (IOD) &View& \checkmark \\
& Invalidate Without Rect (IWR) &View& $X$\\
& UI Overdraw (UIO) &View& \checkmark \\
& Bitmap Format Usage (BFU) &-& $X$\\\hline
\end{tabular}
\label{tab:paprika_smells}
\end{table}

Table \ref{tab:paprika_smells} shows the code smells that Paprika is able to identify:  4 are object-oriented, and 13 are Android code smells. It also presents the entities that are related to each code smell. 
Some entities are related to object-oriented smells (Classes, Methods), others to Android smells (Activities, Async Task).\footnote{Version of Paprika used: commit 5ebd34 \url{https://github.com/GeoffreyHecht/paprika/commit/5ebd349ed3067914386e8c6a05e87ff161f9edd1}}

The input of Paprika is an apk (i.e., a version of an Android application) and returns a list with all instances of the smells found in that apk.
Moreover, 
for each instance of some smells (incl. BLOB, CC, and HSS),
Paprika outputs a fuzzy value (between 0 and 1) calculated using fuzzy logic~\citep{zadeh1974}.
A fuzzy value represents the degree of truth of the detected instance \citep{Habchi2017,hecht_thesis}.

To detect occurrences of code smells, Paprika uses metrics associated with entities. 
For example, the code smell \emph{Long Method} (LM) is an object-oriented smell, and it is related to \emph{methods}:
an instance of LM is a method which the number of instructions is higher to a given threshold.

As output, Paprika produces 
\begin{inparaenum}[\it a)]
\item a list of smells found, and
\item the metrics associated with the entities used for detecting smells, for example, the number of methods, activities, services.
\end{inparaenum}
Those metrics are later used to calculate the software quality score (Section \ref{sec:method:qualityscore}).

\subsubsection{Code Smells considered in our study}
\label{sec:method:smells}

We now describe the code smells that we study in this paper. 
Table \ref{tab:paprika_smells} shows them with a \checkmark in column ``Considered''.

Firstly, we considered the four object-oriented smells (BLOB, SAK, and CC, related to classes, LM related to methods) because they can also exist on Kotlin applications. 
Let us briefly describe each of them.
A Blob class (BLOB), also know as God class, is a class with a large number of attributes and/or methods~\citep{brown1998}. 
A Swiss army knife (SAK) is an interface with a large number of methods \citep{hecht_thesis}.
A complex class (CC) is a class containing complex methods. These classes are hard to understand and maintain and need to be refactored~\citep{fowler1999}. On Paprika, the class complexity is calculated by summing the internal methods complexities and the complexity of a method is calculated using McCabe's Cyclomatic Complexity~\citep{mccabe1976,Hecht2015}.
Long methods (LM) have much more lines than other methods, becoming complex, hard to understand and maintain.

Secondly, we considered 6 Android platform related code smells that Paprika can detect, and we discarded 7.
Those Android smells we considered are:
\begin{inparaenum}[\it 1)]
\item NLMR (related with activities),
\item HAS (async taks), 
\item HSS (async taks),
\item HBR (broadcast receivers),
\item UIO (views), and 
\item IOD (views)
\end{inparaenum}

Let us briefly describe each of them.
\emph{No Low Memory Resolver} (NLMR)~\citep{Hecht2015a} occurs when activities do not have the  method \textbf{onLowMemory()} overrided and if this method is not implemented by an activity, the Android system could kill a process related with this activity to free memory, and consequently, it could cause an abnormal termination of programs~\citep{smell_catalogue}. 

\emph{Heavy ASynctask} (HAS)~\citep{hecht_thesis}, \emph{Heavy Service Start} (HSS)~\citep{hecht_thesis} and \emph{Heavy BroadcastReceiver} (HBR)~\citep{Hecht2015} are similar: they occur when heavy operations are executed at the main thread in different Android components, Async Task, Service and BroadcastReceiver, respectively~\citep{blog_has,blog_hss,blog_hbr}. 

 \emph{UI Overdraw} (UIO)~\citep{Hecht2015} and \emph{Init OnDraw} (IOD)~\citep{hecht_thesis} are related to custom views.
The smell UIO  produces overdraw views because of missing methods invocations, such as \emph{clipRect} and \emph{quickReject}~\citep{google_dev_uio}, which could avoid the overdraw.
IOD happens when new objects are created inside onDraw method that could be executed many times by second, resulting in many allocations of new objects~\citep{google_dev_iod}.

\subsubsection{Code Smells ignored in our study}
\label{sec:smells:ignored}

We ignored 7 Android-related code smells that Paprika can identify.
Table \ref{tab:paprika_smells} shows them with a ``$X$'' in column ``Considered''.
These smells are MIM, LIC, IGS, BUF, HMU, UHA and IWR.
In the remainder of this section, we describe them and explain why we decided to exclude them.

\emph{Member Ignoring Method} (MIM)~\citep{smell_catalogue} occurs when a method does not access any class's attribute.  In Android, it is recommended to use a static method instead, because static method invocations are about 15\%-20\% faster than a dynamic invocation~\citep{android_dev_mim}.
The smell \emph{Leaking Inner Class} (LIC)~\citep{Hecht2015a} occur when an application uses a non-static and anonymous inner class, since in Java this type of inner class holds a reference to the outer class, and consequently it could provoke a memory leak in Android systems~\citep{smell_catalogue,androidpatterns_lic}. 
We decided to discard MIM and LIC because Kotlin does not have static methods~\citep{kotlin_doc_static}.

\emph{Internal Getter/Setter} (IGS)~\citep{smell_catalogue} impacts on performance and energy consumption of applications \citep{Hecht2016,Morales2016,Morales2017,Kessentini2017,Grano2017,Palomba2017,Carette2017}. 
However, this code smell only impacts when an application runs on Android platforms 2.3 or less \citep{Cruz2018}.
We discarded the smell because the number of active Android devices that run those platform versions is smaller than 0.5\%.\footnote{\url{https://developer.android.com/about/dashboards/} Last visit: 06/11/2018}

\emph{Bitmap Format Usage} (BFU) is related with image format~\citep{Carette2017}.
We discard it because it is related to neither Kotlin nor Java code, i.e., the smell is independent of the programming language used.

\emph{HashMap Usage} (HMU)~\citep{Carette2017} occurs when developers use small \textbf{HashMap} instances instead of using \textbf{ArrayMap} or \textbf{SimpleArrayMap}, both provided by the Android framework~\citep{medium_hmu,android_dev_hmu}. 
However, the results found by~\cite{Saborido2018} show that 
\textbf{ArrayMap} is generally slower and less efficient regarding energy consumption than \textbf{HashMap}. Moreover, they showed that when the keys used are primitive types, developers should adopt \textbf{SparseArray} variants, because they are more efficient concerning CPU time, memory and energy consumption. 
We discarded this smell because of:
\begin{inparaenum}[\it a)]
\item the mentioned finding from \cite{Saborido2018},  and 
\item the Paprika's mechanism used to identify HMU occurrence does not take into account the key's type.
\end{inparaenum}

Finally, we discarded 2 smells related to custom views.
\emph{Unsupported Hardware Acceleration} (UHA) \citep{hecht_thesis} occurs when developers call a method that is not hardware accelerated, so it runs on the CPU instead of GPU, impacting on performance and energy consumption \citep{google_dev_uha_iwr}.
We discard it because the occurrences of this smell depend neither on the developer nor programming languages.
The smell \emph{Invalidate Without Rect} (IWR) \citep{hecht_thesis} appears  when the \emph{onDraw} method is not implemented properly, resulting in overdraw views \citep{google_dev_uha_iwr}. When developers do not specify the rectangle area that should be updated, the whole view is redraw, even some area that is not visible, resulting in performance problems. \cite{google_dev_uha_iwr} indicated that developers should call the method \textit{invalidate(Rect dirty)}, specifying the area to be drawn, to avoid this smell.
However, this method was deprecated in API 28 and since API 21 its calls is ignored. Consequently, we discarded this smell.

\subsubsection{Analyzing the difference between Kotlin and Java applications in terms of presence of 4 object-oriented and 6 Android code smells}
\label{sec:method:diff}

First, we analyzed all applications from FAMAZOA to
identify the object-oriented and Android smells listed in Section~\ref{sec:method:smells}.
We run Paprika for all version of each application (i.e., apks).
Then, we computed the percentage of Java and pure Kotlin applications affected by each code smell. 
An application $a$ is affected by a code smells $s$ if $a$ has at least one instance of $s$. 
This approach splits applications into two groups: 
\begin{inparaenum}[\it a)]
\item affected by $s$, and
 not affected by $s$.
\end{inparaenum}

We also calculated a metric that computes the ratio between the number of instances of one code smell and the number of concerned entities related to this smell. The goal of this metric is to quantify the importance of the difference in proportions of the smells, as done by \cite{Habchi2017}. 
Thus, for each application $a$, the ratio of a smell $s$  \citep{Habchi2017} is defined as:
\begin{equation} 
\label{eq:1}
ratio_s(a) = \frac{fuzzy\_value_s(a)}{number\_of\_entities_s(a)} 
\end{equation}

where $fuzzy\_values_s(a)$ is the sum of the fuzzy values (which vary from 0 to 1) of the detected instances of the smell $s$ in the app $a$, and $number\_of\_entities_s(a)$ is the number of the entities concerned by the smell $s$ in the app $a$. Section \ref{sec:method:selection} describes relationship between the smells and the entities concerned by them.

We computed the Cliff's $\delta$~\citep{romano2006} as well, which indicates the magnitude of the effect size~\citep{cliff2014} of the treatment on the dependent variable. 
In our study, we used the Cliff's $\delta$ to determine for each smell $s$ which group of applications, pure Kotlin or Java applications, have more entities affected by $s$, and whether the effect size found presents a significance difference.

The Cliff's Delta estimator can be obtained with Equation~\ref{eq:cliff}.

\begin{equation} 
\label{eq:cliff}
\delta =  \frac{\#(x_1 > x_2) - \#(x_1 < x_2)}{n_1*n_2}
\end{equation}

In this expression, $x_1$ and $x_2$ are scores within group 1 (Kotlin applications) and group 2 (Java applications), and $n_1$ and $n_2$ are the sizes of the sample groups. The cardinality symbol \# indicates counting. This statistic estimates the probability that a value selected from one of the groups, in our case a ratio$_s$, is greater than a value selected from the other group, minus the reverse probability~\citep{Macbeth2011}.

According to \cite{romano2006}, the effect size is small for $0.147 \leq d < 0.33$, medium for $0.33 \leq d < 0.474$, and large for $d \geq 0.474$.
We opted for the Cliff's $\delta$ test since it is suitable for non-normal distributions. Moreover, Cliff's $\delta$ is also recommended for comparing samples of different sizes \citep{Macbeth2011}.

\subsection{Calculating quality scores of Android applications}
\label{sec:method:qualityscore}
For responding to the last RQ 5 (\textit{\rqscorekotlin}),
we used the technique presented by \cite{Hecht2015} for scoring each version (apk) from a mobile application. 
The score serves as an estimation of the mobile app quality in a particular version (apk) and is based on the consistency between the application's size and the number of detected code smells.

\subsubsection{Defining a quality model}
To compute the software quality score based on one type of code smell $s$, the technique from \cite{Hecht2015a} first builds an estimation model using linear regression, which represents the relationship between the number of code smells of type $s$  
and
one metric which represents the size of the application in terms of the numbers of entities associated with $s$.
For example, for the smell BLOB, the metric of size that we consider is the total number of classes, 
and for the smell LM the metric is the number of methods.
Table \ref{tab:paprika_smells} presents the relation between the types of smells and entities.
We built a quality model for each code smell that we consider in this study (Section \ref{sec:method:smells}).

To obtain the quality score for one application (apk), 
the technique takes as input the number of code smells and a value of the size of that app, and produces as output a score.
A higher positive score implies better quality. 
As described by \cite{Hecht2015}, the software quality score of an application at a particular version is computed as the additive inverse of the residual. Consequently, a larger positive residual value suggests worst software quality because it means the apk has more smells with respect to its size than the norm (i.e., linear regression),  whereas a larger negative residual value implies better quality because of the lower number of smells. 

\subsubsection{Training a quality model}
We created a quality score model, i.e., a linear regression, for each code smell that we considered in Section \ref{sec:method:smells}.
We trained the linear model using a dataset defined by \cite{Hecht2015} which contains \numprint{3568} Android versions (apk) extracted from the Google Play store between June 2013 and June 2014.
We selected this dataset for training the model for two main reasons:
\begin{inparaenum}[\it 1)]
\item it was previously used in a similar experiment for training quality models \cite{Hecht2015,hecht_thesis}; and
\item its applications do not include Kotlin code. 
The trained model created from this dataset represents the quality of applications build previously to Kotlin was released. Thus, we used it as baseline to measure if Kotlin applications have more (or less) quality than applications written before Kotlin was released. 
\end{inparaenum}

To create a quality model, we first run Paprika over the dataset previously mentioned.
The output of Paprika is the training set. 
Each element of the training dataset (a row) corresponds to a single apk $a$ and has the following information: 
\begin{inparaenum}[\it a)]
\item the number of instances of a smell $s$  in application $a$, and 
\item the value associated to the entity of smell $s$ in $a$.
\end{inparaenum}
For example, for smell Long Method (LM), we compute the linear regression between: 
\begin{inparaenum}[\it 1)]
\item the number of instances of smell LM,\footnote{\cite{hecht_thesis} considers that a method is ``long" (LM) if it has more than 17 instructions.} and
\item the total number of methods.
\end{inparaenum}

\subsubsection{Using the a quality model for measuring quality of Kotlin applications}

Once trained, we computed the quality scores (one per code smell) for each apk from our dataset of applications classified as `Kotlin' (Section \ref{sec:method:filteringapps}). 
Those apks conform to our \emph{test} dataset.
Note that the training dataset does not include any apk from the \emph{test} dataset. 
Thus, we discarded the possibility of having overfitting in our models.

\subsubsection{Measuring the impact on the quality of introducing Kotlin}

We measured the impact on the quality of introducing Kotlin code in one application as follows.
The applications that we studied were those that initially have 1 or more apks classified as Java and then it has 1 or more apks classified as Kotlin, using the heuristic presented in Section \ref{sec:method:filteringapps}. 
For each of those applications and for each  and for each code smell $s$,
we first compared the quality scores of $s$ between the apk that introduces Kotlin code and the previous apk (i.e., which has Java code and no Kotlin).
Then, we compared the quality score between the last Java apk (i.e., the version just before the introduction of Kotlin code) and the most recent Kotlin version available.
These two comparisons have different goals: the first one aims at measuring the impact just after the introduction of Kotlin in an app; the second one aims at studying the impact after the application (that now includes Kotlin code) has evolved.

\subsubsection{Detecting changes in the quality evolution trends after introducing Kotlin}
\label{sec:methodology:changes_evolve_trend}

A quality evolution trend describes how the quality scores from the versions of an application $a$ change along its evolution. 
In this paper we studied whether the introduction of Kotlin code into an app $a$ produces a \emph{positive} change in the quality evolution trend.
We analyzed 5 major quality evolution trends defined by \cite{Hecht2015}: 
\begin{inparaenum}[\it A)]
\item Constant decline,
\item Constant rise,
\item Stability,
\item Sudden decline, and
\item Sudden rise.
\end{inparaenum}

For each application $a$, we classified manually the quality evolution trend \emph{before} and \emph{after} the introduction of Kotlin on $a$.
Then, we considered that the introduction of Kotlin produces a \emph{positive} change if:
\begin{inparaenum}[\it 1)]
\item the trend before the introduction is `Decline' or `Stability' (trends A, C or D); and
\item the trend after the introduction is exclusively `Rise' (trends B or E). 
\end{inparaenum}
Note that we discarded analyzing those applications whose trends (before or after) do not fit on any of the defined  trends.

\section{Results}
\label{sec:results}

\pgfmathtruncatemacro{\ukotlindown}{244}
\pgfmathtruncatemacro{\ujavadown}{1923}

\pgfmathtruncatemacro{\udataset}{\ukotlindown + \ujavadown}

\pgfmathsetmacro{\percukotlin}{(\ukotlindown / \udataset) * 100)}
\pgfmathsetmacro{\percujava}{(\ujavadown / \udataset) * 100)}

\newcommand{\kotlindownandrozoo}{464}
\newcommand{\kotlindownfdroid}{1126}
\pgfmathtruncatemacro{\kotlindown}{1590}

\newcommand{\javadownandrozoo}{7199}
\newcommand{\javadownfdroid}{11049}

\pgfmathtruncatemacro{\javadown}{18248}

\pgfmathtruncatemacro{\dataset}{19838}

\newcommand{\ukotlinanalyzed}{89}
\newcommand{\ujavaanalyzed}{785}

\newcommand{\kotlinanalyzed}{1016}
\newcommand{\javaanalyzed}{9763}

\pgfmathsetmacro{\perckotlin}{8.01}
\pgfmathsetmacro{\percjava}{91.98}

\begin{table}[t]
\centering
\caption{Classification of FAMAZOA's applications according to the programming language.}
\label{tab:summ}
\begin{tabular}{|l|r|r|r|}
\hline
Information & Total & Kotlin & Java \\
\hline
\hline
Unique apps &
\numprint{\udataset} & \ukotlindown & \ujavadown \\
\hline
Versions    & \numprint{\dataset} & \kotlindown & \javadown \\
\hline
\end{tabular}
\end{table}

\pgfkeys{pgf}

\subsection{RQ$_1$: \rqadoption}
\label{sec:results:rqone}

\begin{figure}
    \centering
    \begin{subfigure}[b]{0.45\textwidth}
        \includegraphics[width=\textwidth]{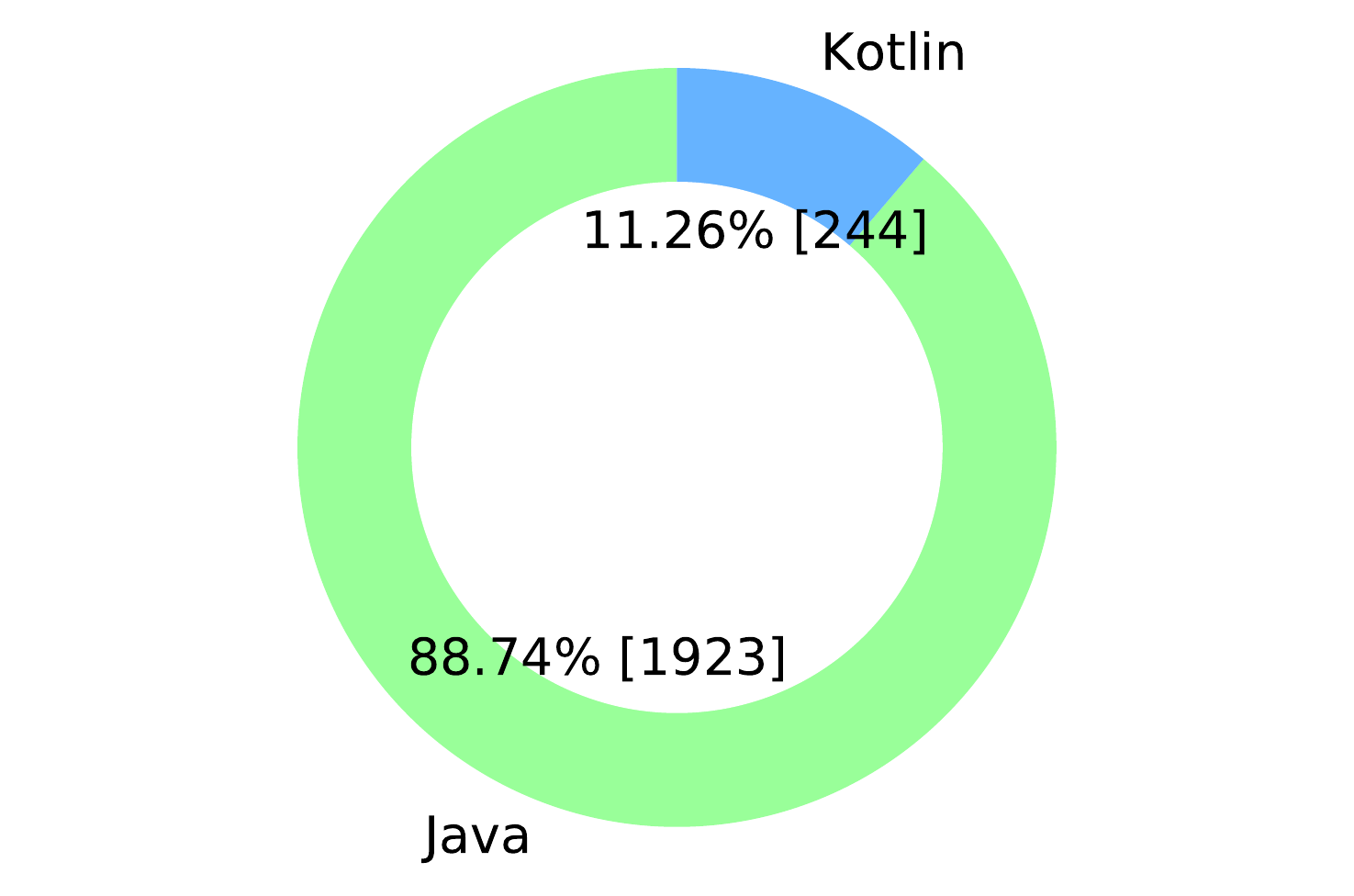}
        \caption{Unique applications}
        \label{fig:unique_cmp}
    \end{subfigure}
     \begin{subfigure}[b]{0.45\textwidth}
        \includegraphics[width=\textwidth]{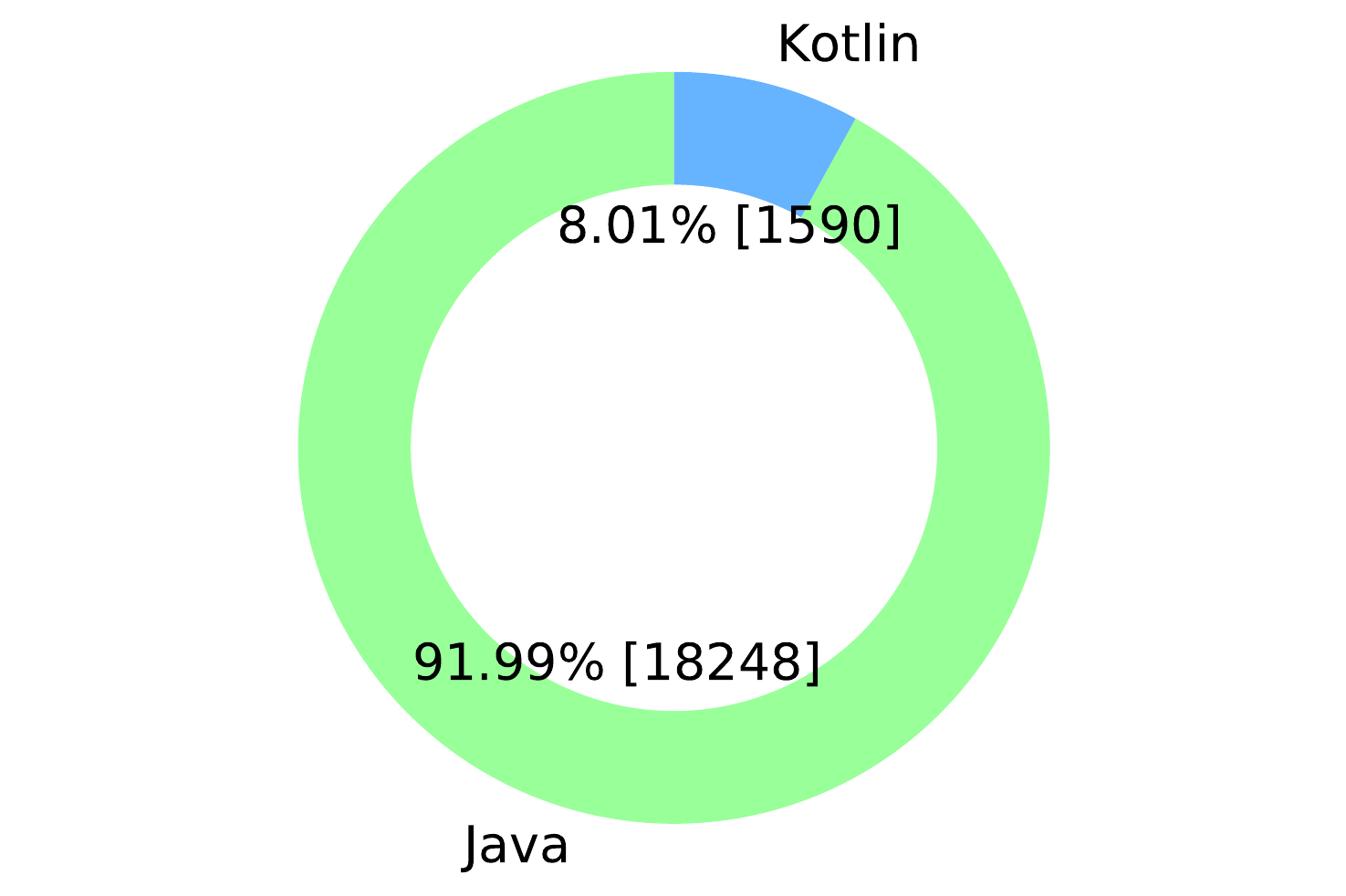}
        \caption{All versions (apk)}
        \label{fig:all_versions_cmp}
    \end{subfigure}
    \caption{Distributions between Kotlin and Java applications and versions.}
    \label{fig:kotlin_vs_java}
\end{figure}

Table~\ref{tab:summ} summarizes the classification of applications done using the methodology  presented in Section \ref{sec:method:filteringapps}.
FAMAZOA has \numprint{\udataset}~applications and using our heuristics we classified~\ukotlindown~(\pgfmathprintnumber[fixed, precision=2]{\percukotlin}\%) as `Kotlin' applications. Consequently, the remaining applications, \numprint{\ujavadown}~(\pgfmathprintnumber[fixed, precision=2]{\percujava}\%), were classified as `Java'. 
Figure~\ref{fig:unique_cmp} shows these percentages.  
Considering the number of versions (apk), we found 
\numprint{\kotlindown}~apks (\pgfmathprintnumber[fixed, precision=2]{\perckotlin}\%)
with Kotlin code and 
\numprint{\javadown}~(\pgfmathprintnumber[fixed, precision=2]{\percjava}\%)
without Kotlin code. 

Now, let us to explain how we arrived to detect 244 Kotlin applications from FAMAZOA. 
First, the heuristic \hone{} (apk analyzer, Section \ref{sec:method:filteringapps}) classified 265 applications as `Kotlin'.
Those applications have, at least, one apk classified as `Kotlin'.
For 76 of them, all apks are classified as `Kotlin'.

Then, \htree{} (GitHub API, Section \ref{sec:method:filteringapps}) classified 234 applications as `Kotlin'. 
193 of them, were also classified as `Kotlin' by \hone{}.
Up to here, both heuristics have classified 297 unique applications as `Kotlin'.

Finally, 
we applied \htwo{}(Source Code analysis Section \ref{sec:method:filteringapps}) on the repositories of those  297 applications, finding 244 that contain Kotlin code.
The list of Kotlin applications can be found in our appendix.\footnote{Applications classified as Kotlin: \url{https://github.com/UPHF/kotlinandroid/blob/master/docs/final_kotlin_dataset.md}}

\begin{tcolorbox}
{\bf Response to RQ 1}: \emph{\rqadoption}

We found that \ukotlindown~out of \numprint{\udataset}~applications from our dataset have, at least, one version released between the years 2017 and 2018 written (totally or partially) using the Kotlin language.
\end{tcolorbox}

\subsection{RQ$_2$. \rqproportionkotlincode}

\newcommand{\purekotlin}{82}
\newcommand{\morethaneighty}{145}
\newcommand{\lessthanten}{45}

\newcommand{\ukotlindownrepo}{244}
\pgfmathsetmacro{\percpurekotlin}{(\purekotlin / \ukotlindownrepo) * 100}

\pgfmathsetmacro{\percnotpurekotlin}{100-\percpurekotlin}

\pgfmathsetmacro{\percmorethaneighty}{(\morethaneighty / \ukotlindownrepo) * 100}
\pgfmathsetmacro{\perclessthanten}{(\lessthanten / \ukotlindownrepo) * 100}

\begin{figure}
    \centering
    \includegraphics[width=0.8\textwidth]{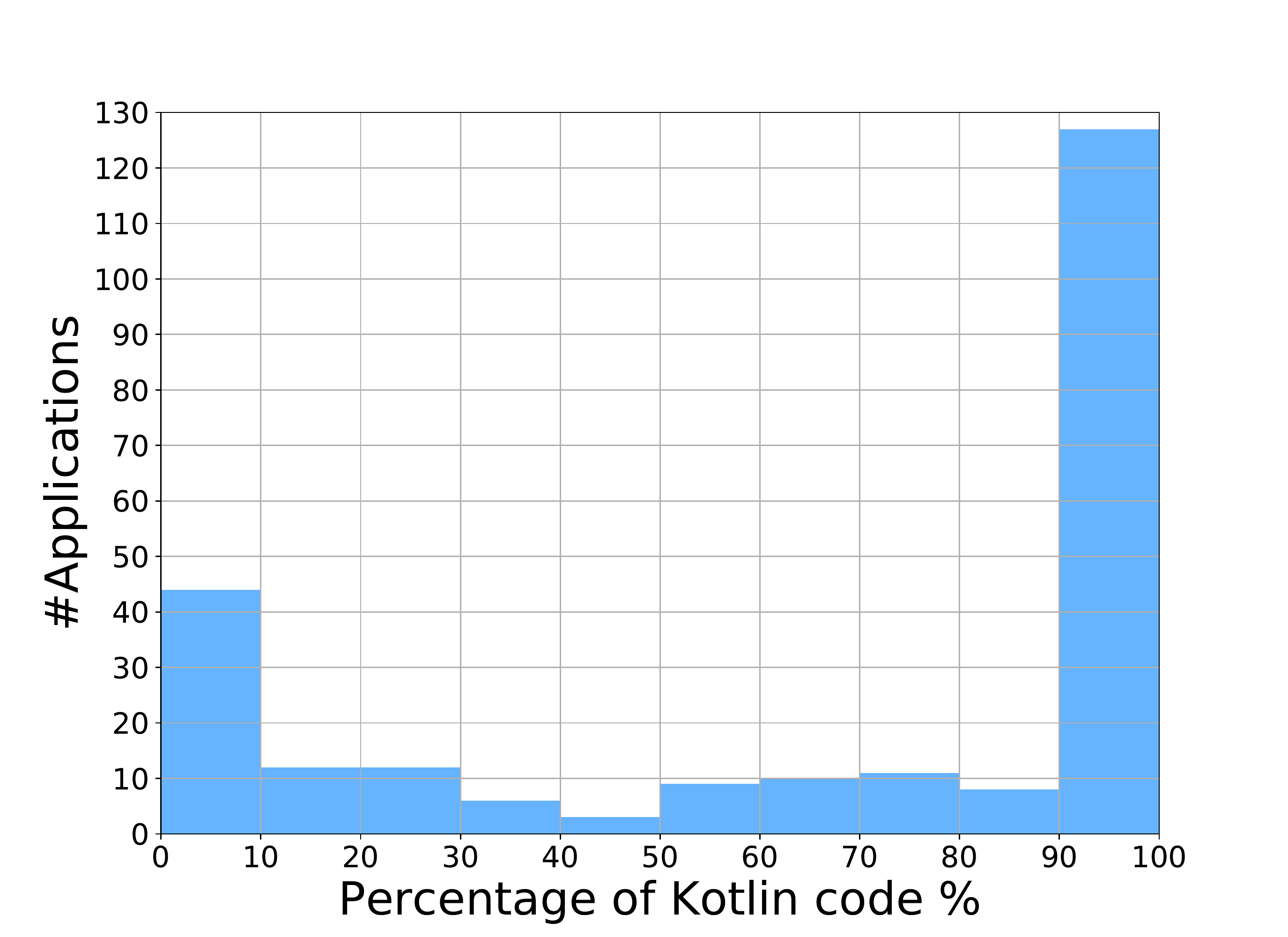}
    \caption{Distribution of applications according the percentage of Kotlin code. 59.43\% of Kotlin applications have more than 80\% of source code written in Kotlin.}
    \label{fig:histogram_kotlin_percentage}
\end{figure}  
For computing the percentage of Kotlin code, we executed the methodology presented in Section \ref{sec:method:proportion} over the   \ukotlindownrepo~applications that contain at least one commit with Kotlin code.

Figure~\ref{fig:histogram_kotlin_percentage} shows the distribution of Kotlin applications according to the percentage of Kotlin code. We found that 
\purekotlin~out of \ukotlindownrepo~(\pgfmathprintnumber[fixed, precision=2]{\percpurekotlin}\%)~applications have only Kotlin code.
The rest of the Kotlin applications (\pgfmathprintnumber[fixed, precision=2]{\percnotpurekotlin}\%)  are also written in Java. 
Furthermore, we found that \morethaneighty~out of \numprint{\ukotlindownrepo}~(\pgfmathprintnumber[fixed, precision=2]{\percmorethaneighty}\%)~applications have at least 80\% of Kotlin code and, on the contrary,  \lessthanten~out of \udataset~(\pgfmathprintnumber[fixed, precision=2]{\perclessthanten}\%)~applications, have less than 10\% of Kotlin code.

\begin{tcolorbox}
{\bf Response to RQ 2:}  \emph{\rqproportionkotlincode}

Considering the last version of each application, the majority of Kotlin applications (59.43\%) have at least 80\% of lines of code written in Kotlin.
\end{tcolorbox}

\subsection{RQ$_3$. \rqevolutionproportion}

\newcommand{\etone}{19}
\newcommand{\ettwo}{15}
\newcommand{\etthree}{4}
\newcommand{\etfour}{8}
\newcommand{\etfive}{52}
\newcommand{\etsix}{48}
\newcommand{\etseven}{41}
\newcommand{\eteight}{43}
\newcommand{\etnine}{7}
\newcommand{\etten}{3}
\newcommand{\eteleven}{2}
\newcommand{\ettwelve}{2}

\pgfmathsetmacro{\petone}{(\etone/ \ukotlindown) * 100)}
\pgfmathsetmacro{\pettwo}{(\ettwo/ \ukotlindown) * 100)}
\pgfmathsetmacro{\petthree}{(\etthree/ \ukotlindown) * 100)}
\pgfmathsetmacro{\petfour}{(\etfour/ \ukotlindown) * 100)}
\pgfmathsetmacro{\petfive}{(\etfive/ \ukotlindown) * 100)}
\pgfmathsetmacro{\petsix}{(\etsix/ \ukotlindown) * 100)}
\pgfmathsetmacro{\petseven}{(\etseven/ \ukotlindown) * 100)}
\pgfmathsetmacro{\peteight}{(\eteight/ \ukotlindown) * 100)}
\pgfmathsetmacro{\petnine}{(\etnine/ \ukotlindown) * 100)}
\pgfmathsetmacro{\petten}{(\etten/ \ukotlindown) * 100)}
\pgfmathsetmacro{\peteleven}{(\eteleven/ \ukotlindown) * 100)}
\pgfmathsetmacro{\pettwelve}{(\ettwelve/ \ukotlindown) * 100)}

\begin{table*}
\centering
\caption{Classification of Android applications according to the evolution trend of Kotlin and Java source code.}
\begin{tabular}{|l|l|r|r|}
\hline
\multicolumn{2}{|c|}{Source Code Evolution Trend}&\# Apps & \%\\
\hline
\hline
ET 1 &Kotlin is the initial language and the amount of Kotlin grows& \etone & \pgfmathprintnumber[fixed, precision=1]{\petone}\\
ET 2 &Kotlin code replaces all Java code& \ettwo & \pgfmathprintnumber[fixed, precision=1]{\pettwo}\\
ET 3 & Kotlin code replaces some Java then  Java continues growing& \etthree & \pgfmathprintnumber[fixed, precision=1]{\petthree}\\
ET 4 &Kotlin increase together with Java& \etfour & \pgfmathprintnumber[fixed, precision=1]{\petfour}\\
ET  5&Kotlin grows and Java decreases (but never is zero)& \etfive & \pgfmathprintnumber[fixed, precision=1]{\petfive}\\
ET 6 & Kotlin grows and Java decreases until the Java code is 0& \etsix & \pgfmathprintnumber[fixed, precision=1]{\petsix}\\
ET 7  & Kotlin grows and Java remains constant& \etseven & \pgfmathprintnumber[fixed, precision=1]{\petseven}\\
ET 8 &Kotlin is constant and Java changes & \eteight & \pgfmathprintnumber[fixed, precision=1]{\peteight}\\
ET 9  & Kotlin and Java remain constant & \etnine & \pgfmathprintnumber[fixed, precision=1]{\petnine}\\
ET 10 & Kotlin introduced but lately disappears& \etten & \pgfmathprintnumber[fixed, precision=1]{\petten}\\
ET 11 & Java replaces Kotlin code & \eteleven & \pgfmathprintnumber[fixed, precision=1]{\peteleven}\\
ET 12 & Other & \ettwelve & \pgfmathprintnumber[fixed, precision=1]{\pettwelve}\\
\hline
\hline
\multicolumn{2}{|l|}{Total applications} &\ukotlindown& 100\%\\

\hline

\hline
\end{tabular}
\label{tab:evolution_trend_kotlin}
\end{table*}

\begin{figure}
\captionsetup{justification=centering}
    \centering
 \begin{subfigure}[b]{0.32\textwidth}
        \includegraphics[width=\textwidth]{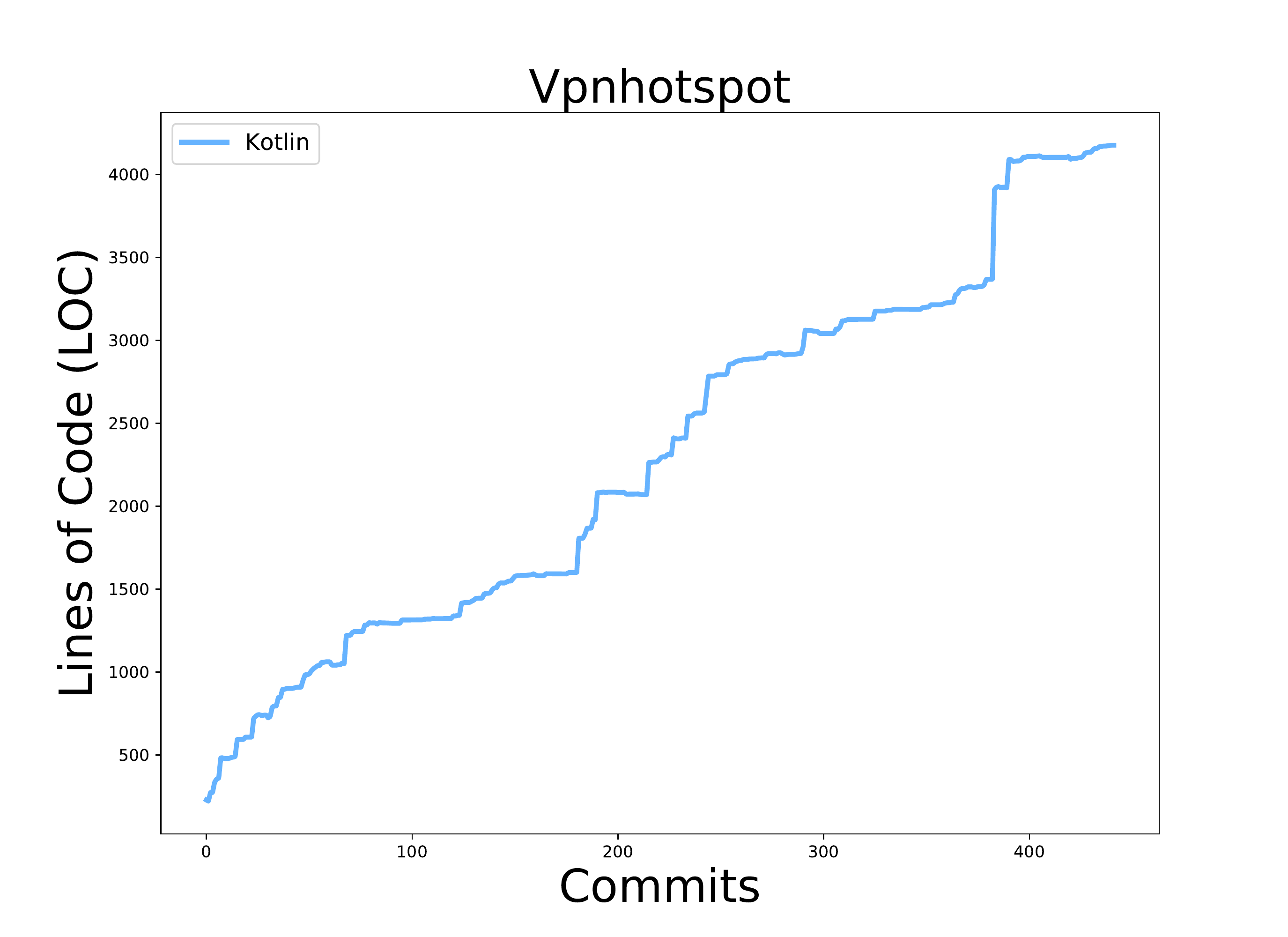}
        \caption{ET 1: Kotlin is the initial language.}
        \label{fig:et1}
    \end{subfigure}
    \begin{subfigure}[b]{0.32\textwidth}
         \includegraphics[width=\textwidth]{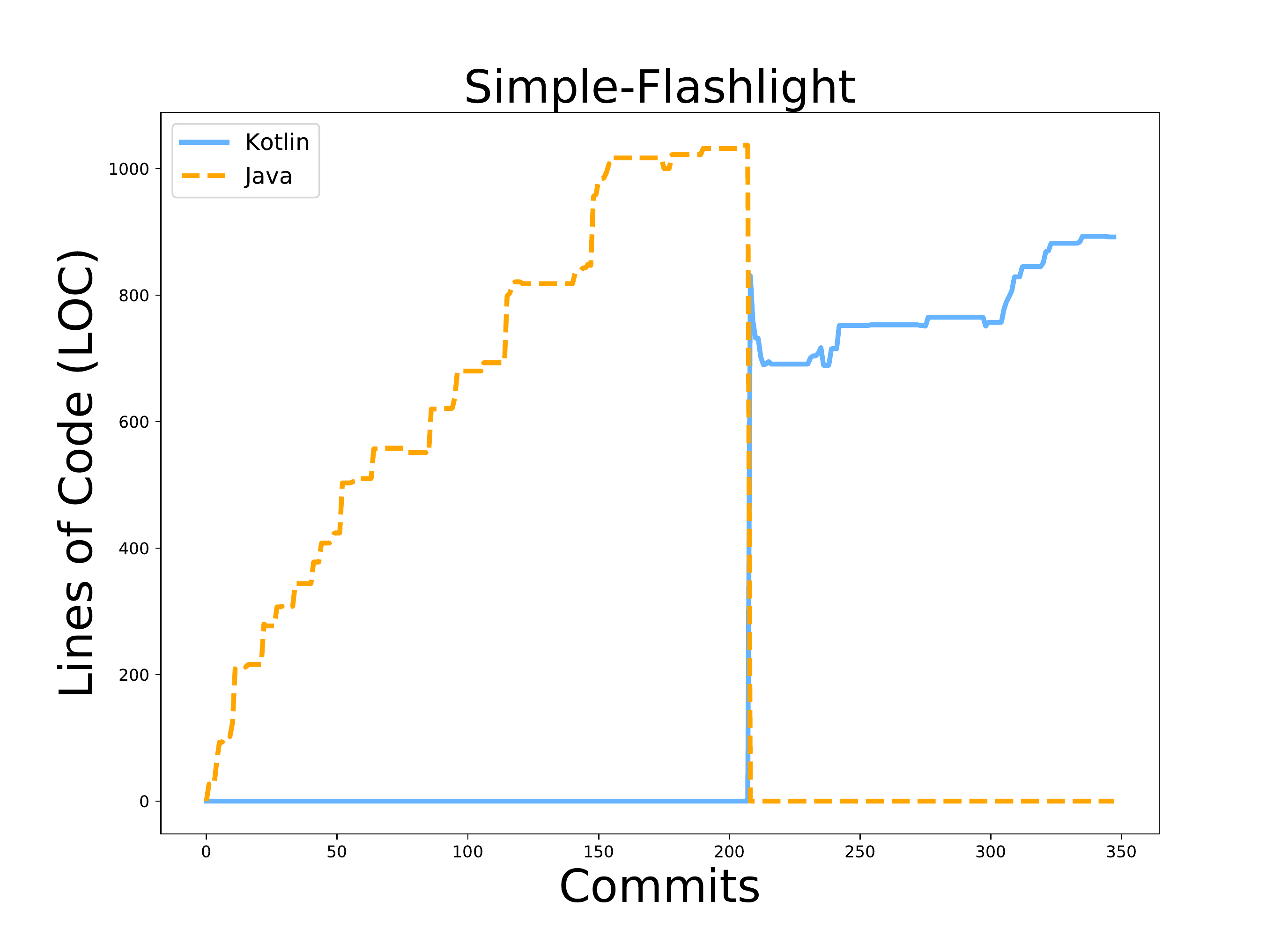}
           \caption{ET 2: Kotlin code replaces all Java code.}
      \label{fig:et2}
    \end{subfigure}
     \begin{subfigure}[b]{0.32\textwidth}
     \includegraphics[width=\textwidth]{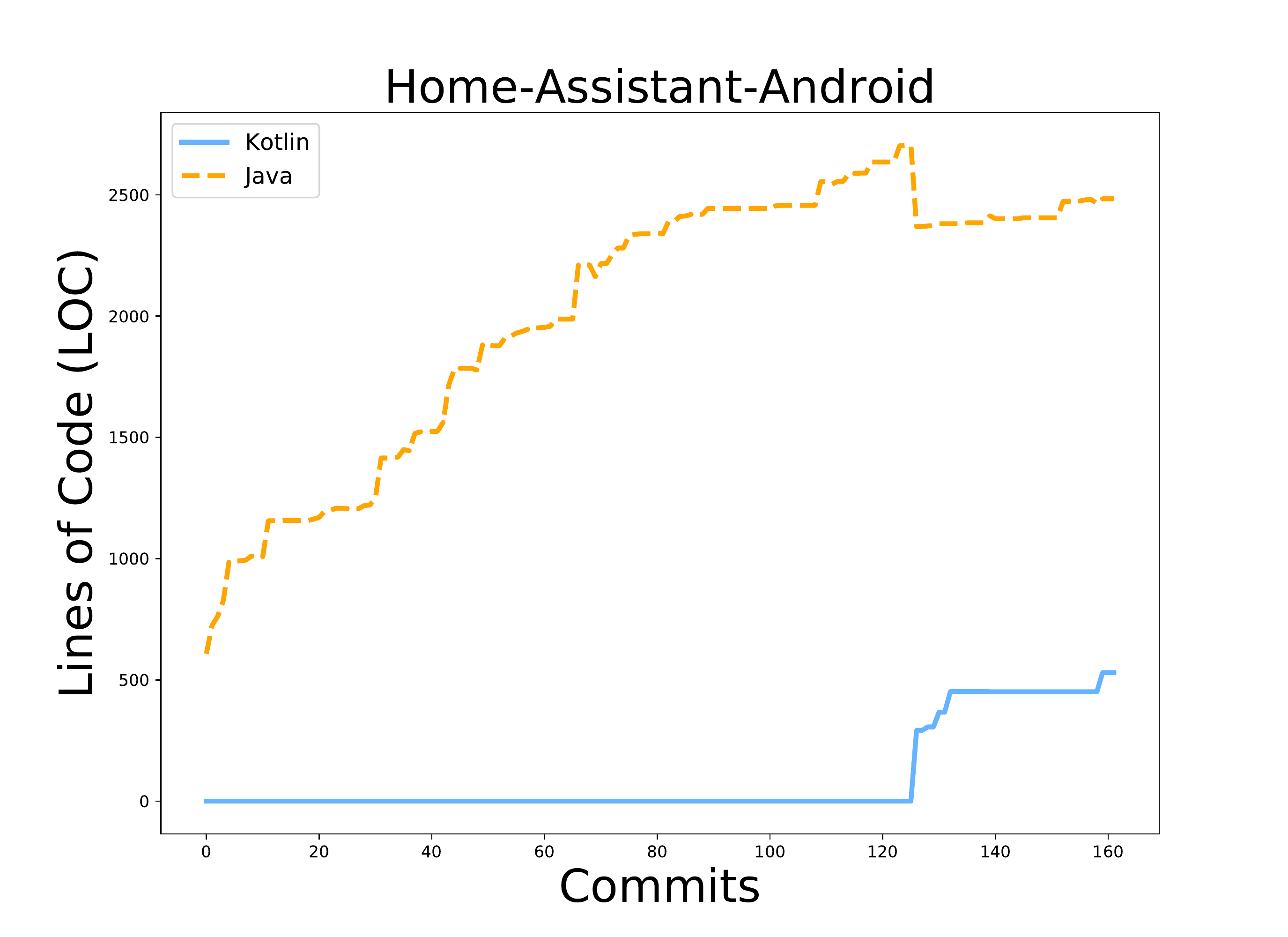}
        \caption{ET 3:  Kotlin code replaces some Java then Java continues growing.}
    \label{fig:et3}
    \end{subfigure}
    
    \begin{subfigure}[b]{0.32\textwidth}
       \includegraphics[width=\textwidth]{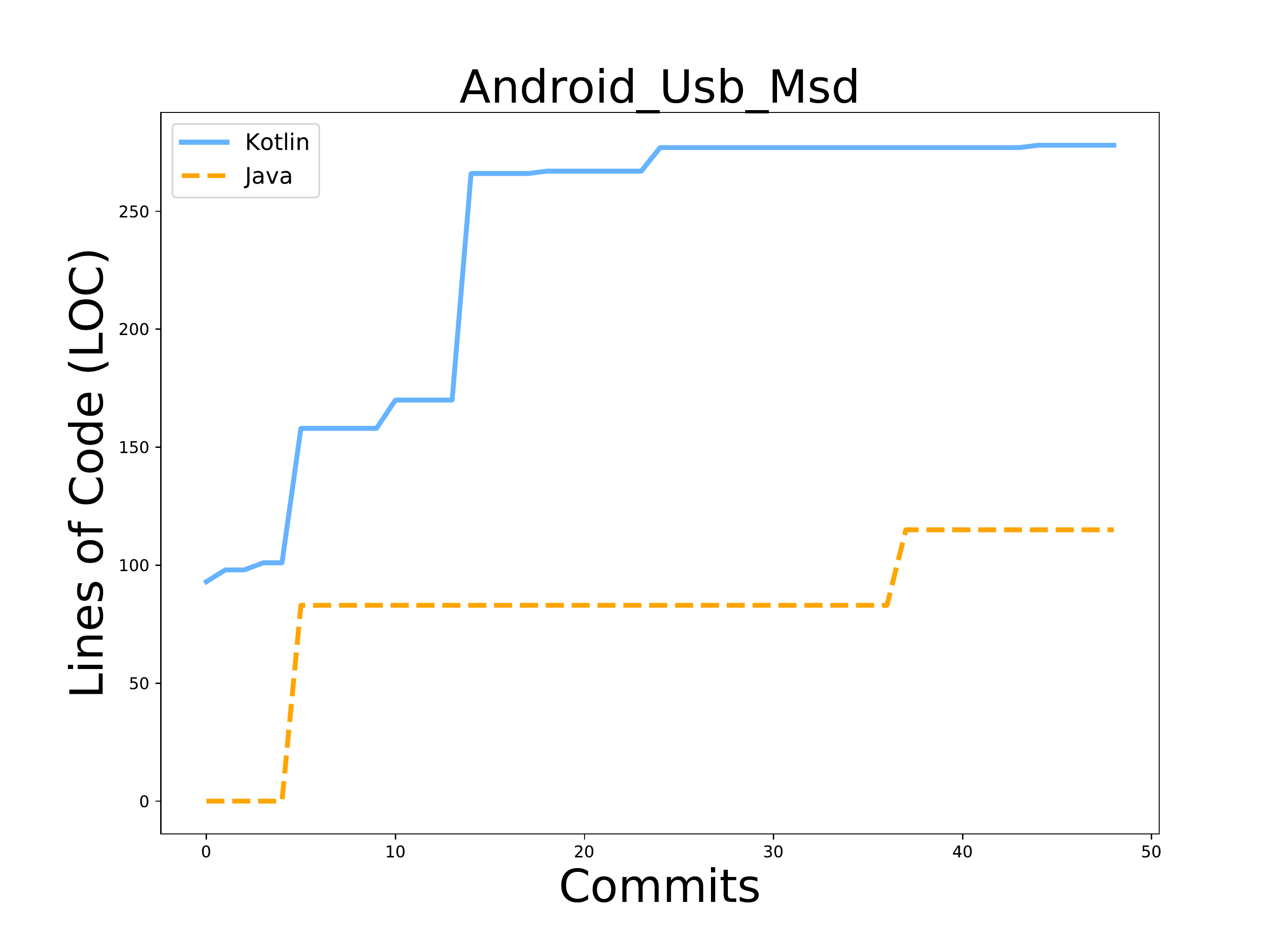}
        \caption{ET 4: Kotlin increase together with Java.}
      \label{fig:et4}
    \end{subfigure}
    \begin{subfigure}[b]{0.32\textwidth}
          \includegraphics[width=\textwidth]{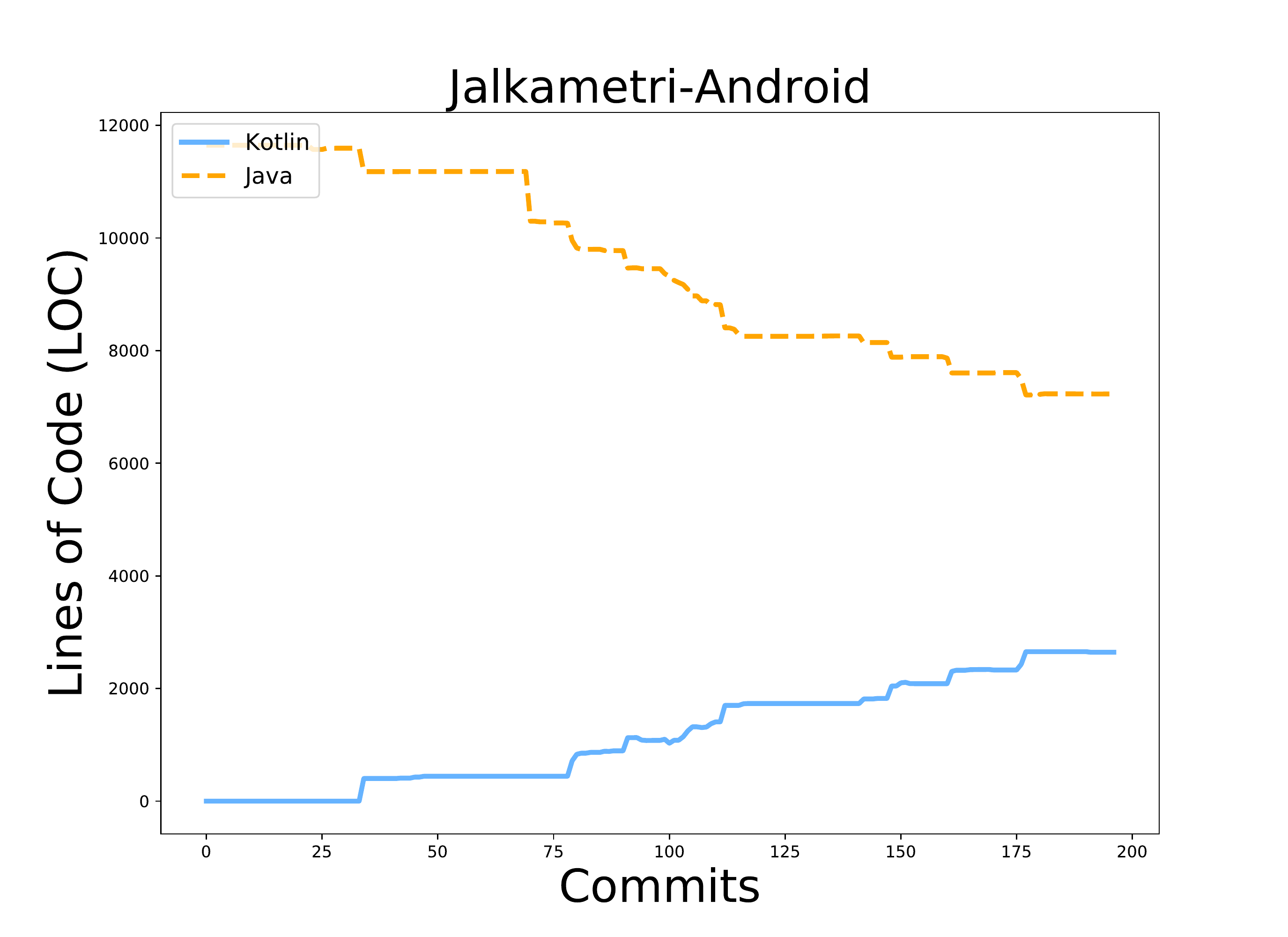}
        \caption{ET 5: Kotlin grows and Java decreases. Example 1.}
       \label{fig:et5}
    \end{subfigure}
      \begin{subfigure}[b]{0.32\textwidth}
         \includegraphics[width=\textwidth]{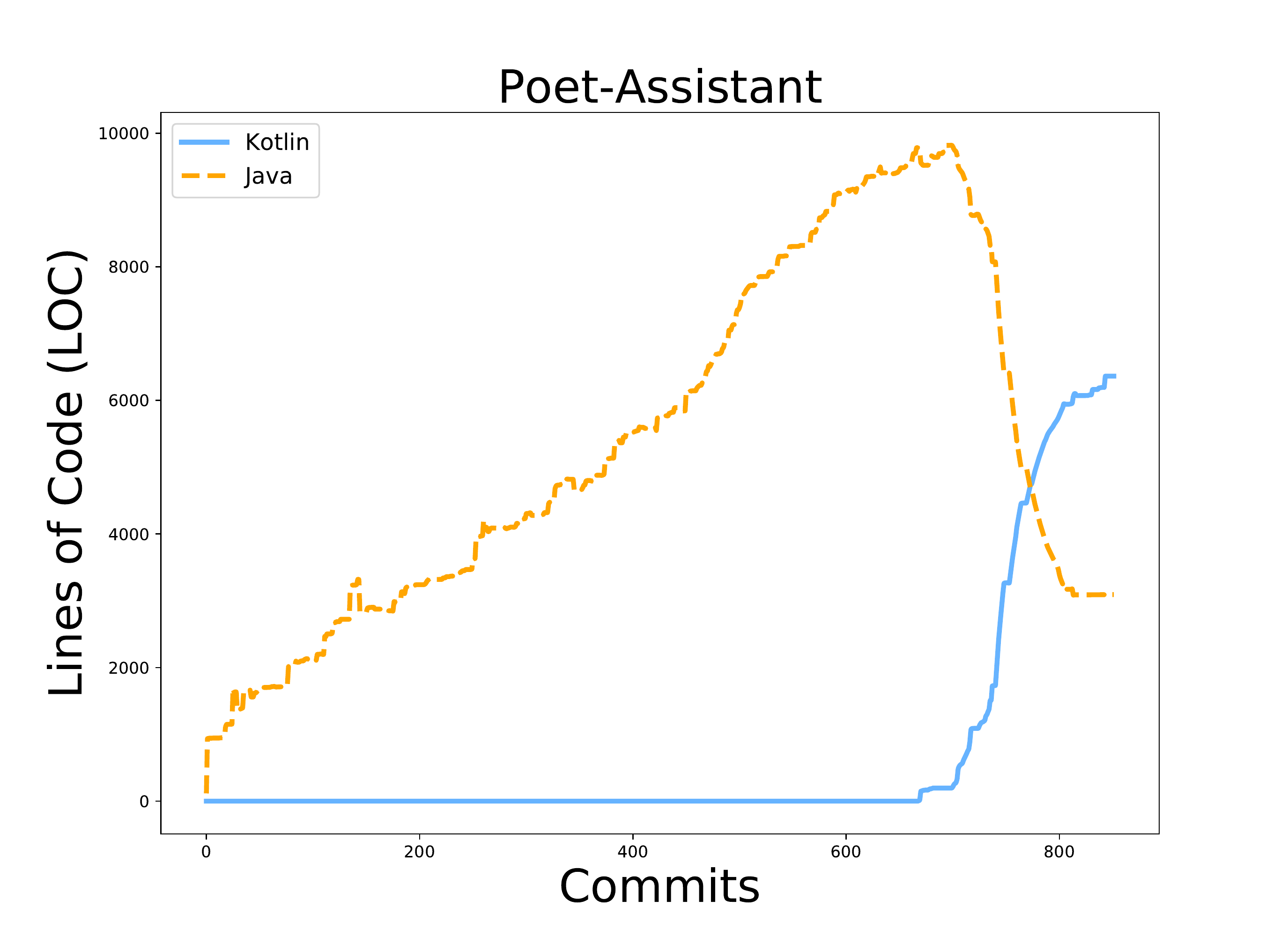}
        
        \caption{ET 5: Kotlin grows and Java decreases. Example 2.}
       \label{fig:et52}
    \end{subfigure}

    \begin{subfigure}[b]{0.32\textwidth}
      \includegraphics[width=\textwidth]{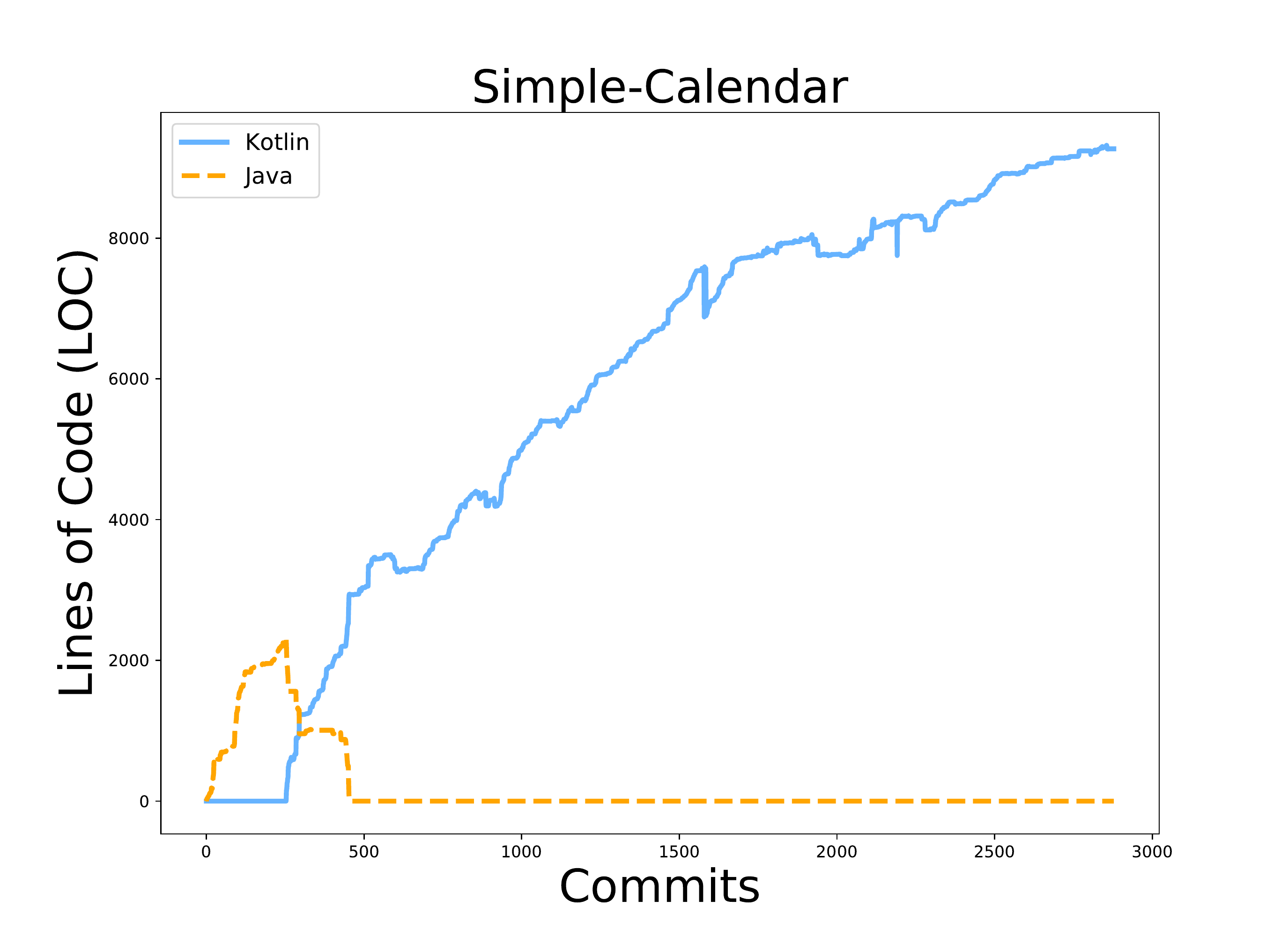} 
        \caption{ET 6: Kotlin grows and Java decrease until the Java code is 0.}
            \label{fig:et6}
    \end{subfigure}
     \begin{subfigure}[b]{0.32\textwidth}
        \includegraphics[width=\textwidth]{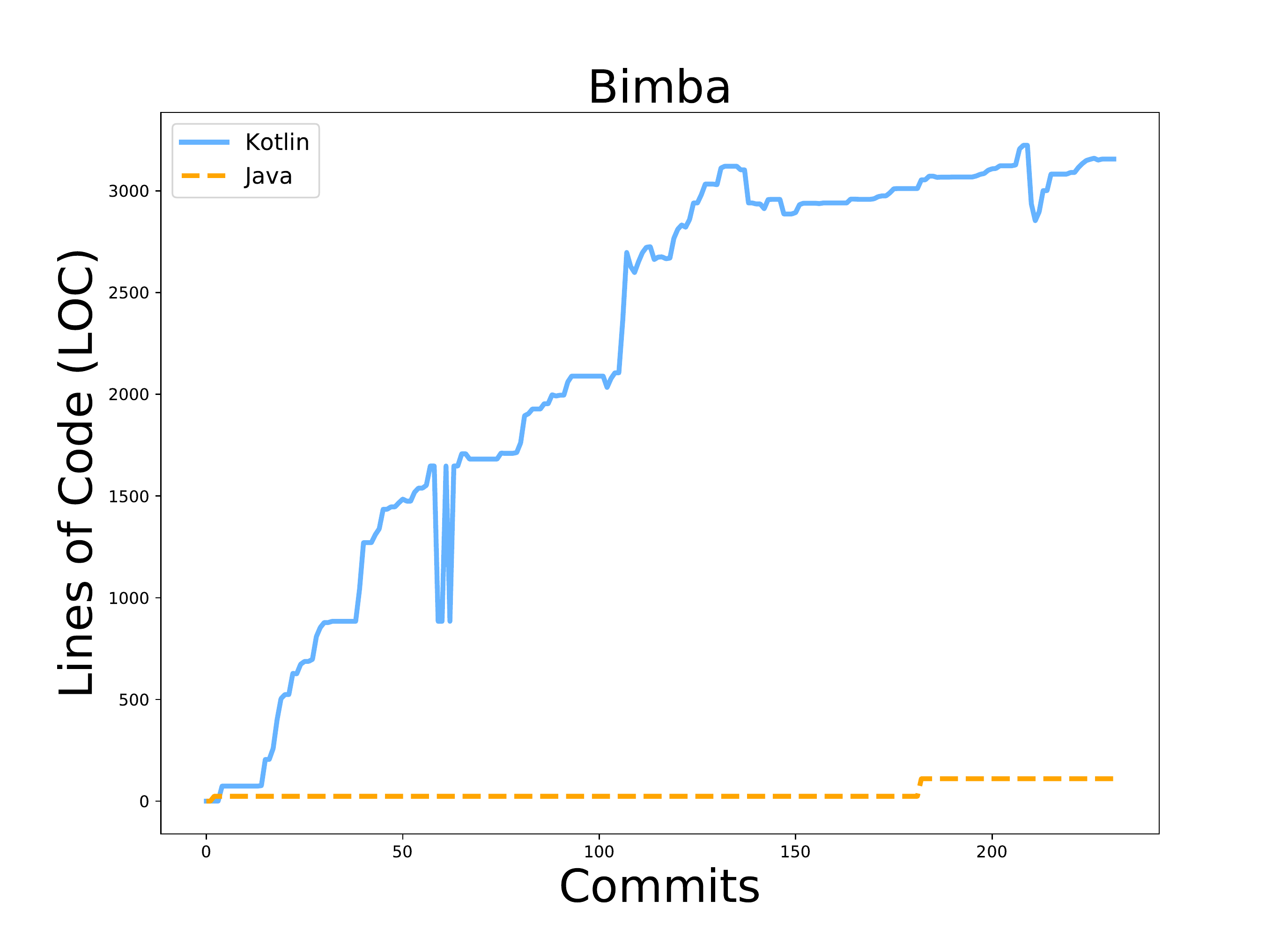}
        \caption{ET 7: Kotlin grows and Java remains constant.}
              \label{fig:et7}
    \end{subfigure}
    \begin{subfigure}[b]{0.32\textwidth}
           \includegraphics[width=\textwidth]{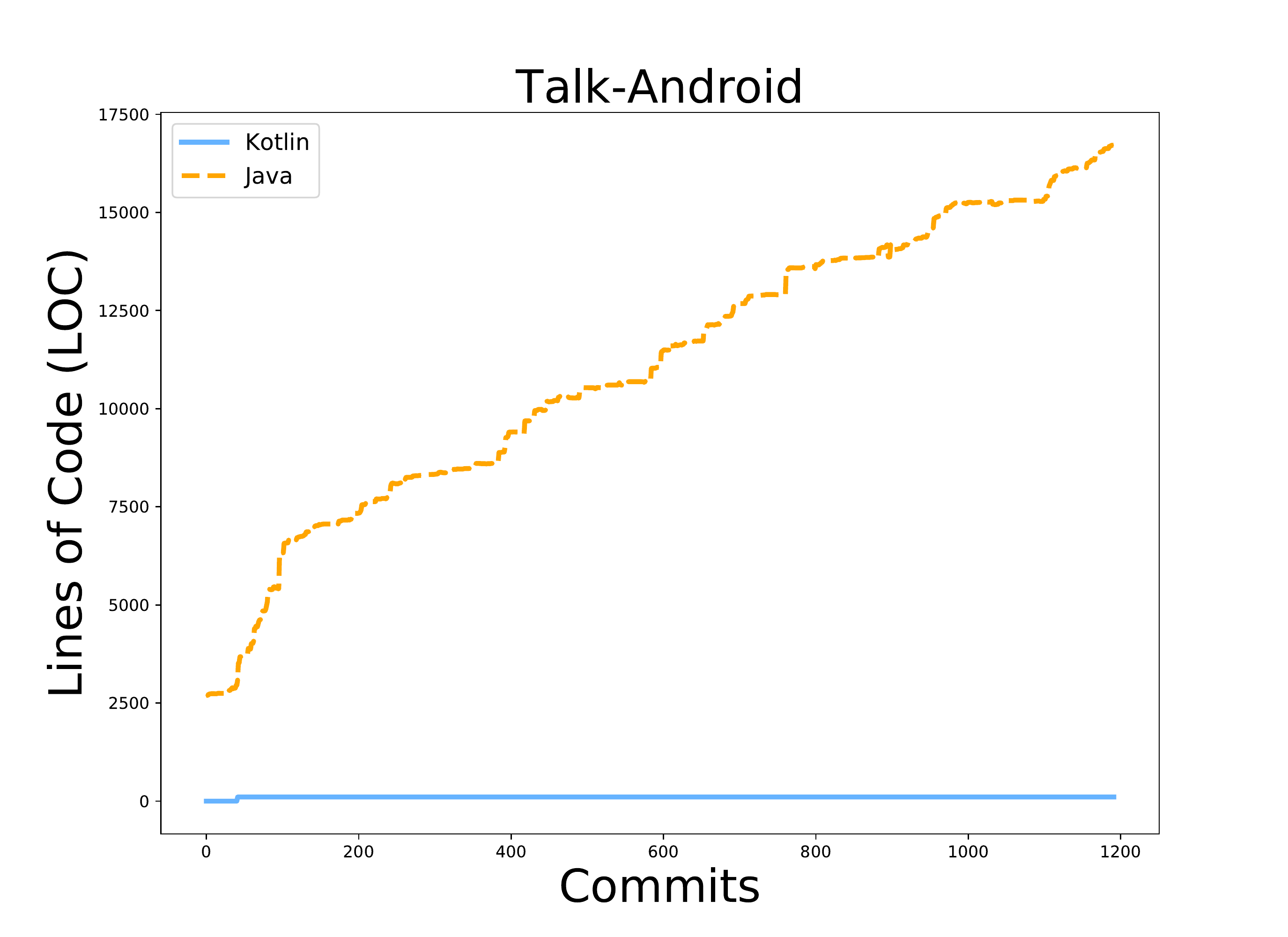}
        \caption{ET 8: Kotlin is constant and Java grows.}
       \label{fig:et8}
    \end{subfigure}
    
    \begin{subfigure}[b]{0.32\textwidth}
        \includegraphics[width=\textwidth]{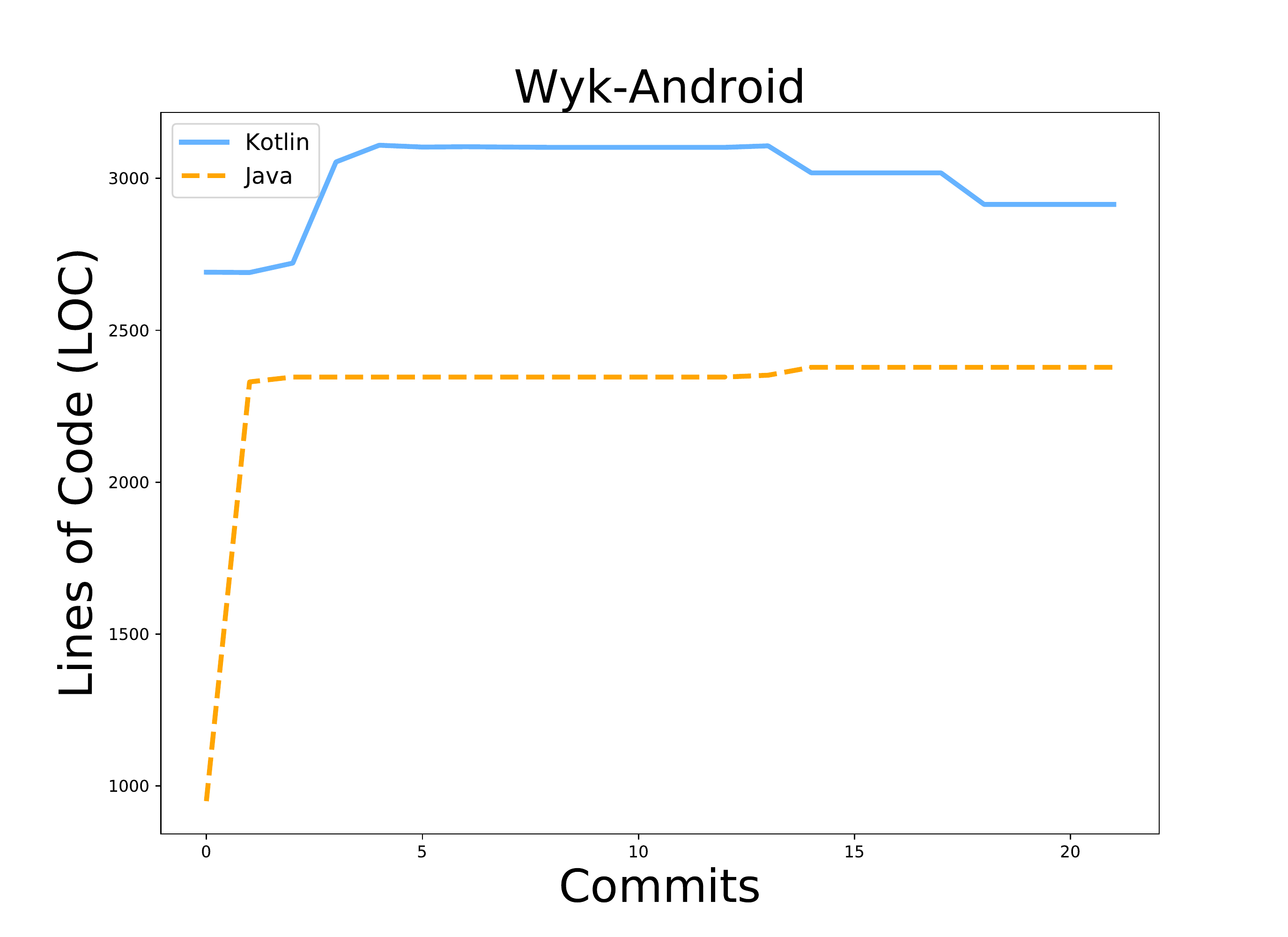}
        \caption{ET 9: Kotlin and Java remain constant.}
    \label{fig:et9}
    \end{subfigure}
    \begin{subfigure}[b]{0.32\textwidth}
        \includegraphics[width=\textwidth]{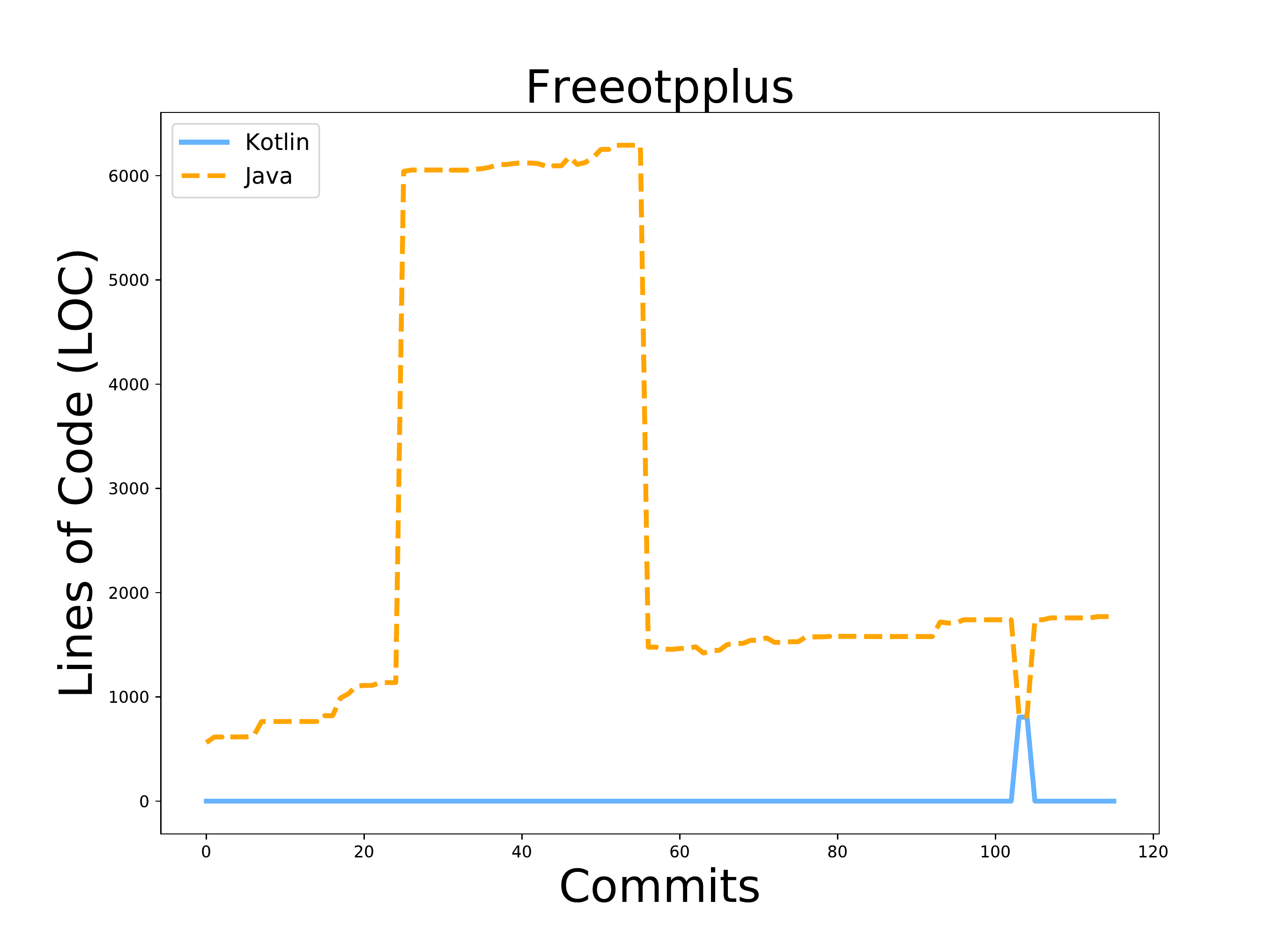}
        \caption{ET 10: Kotlin introduced but lately disappears.}
    \label{fig:et10}
    \end{subfigure}
    \begin{subfigure}[b]{0.32\textwidth}
         \includegraphics[width=\textwidth]{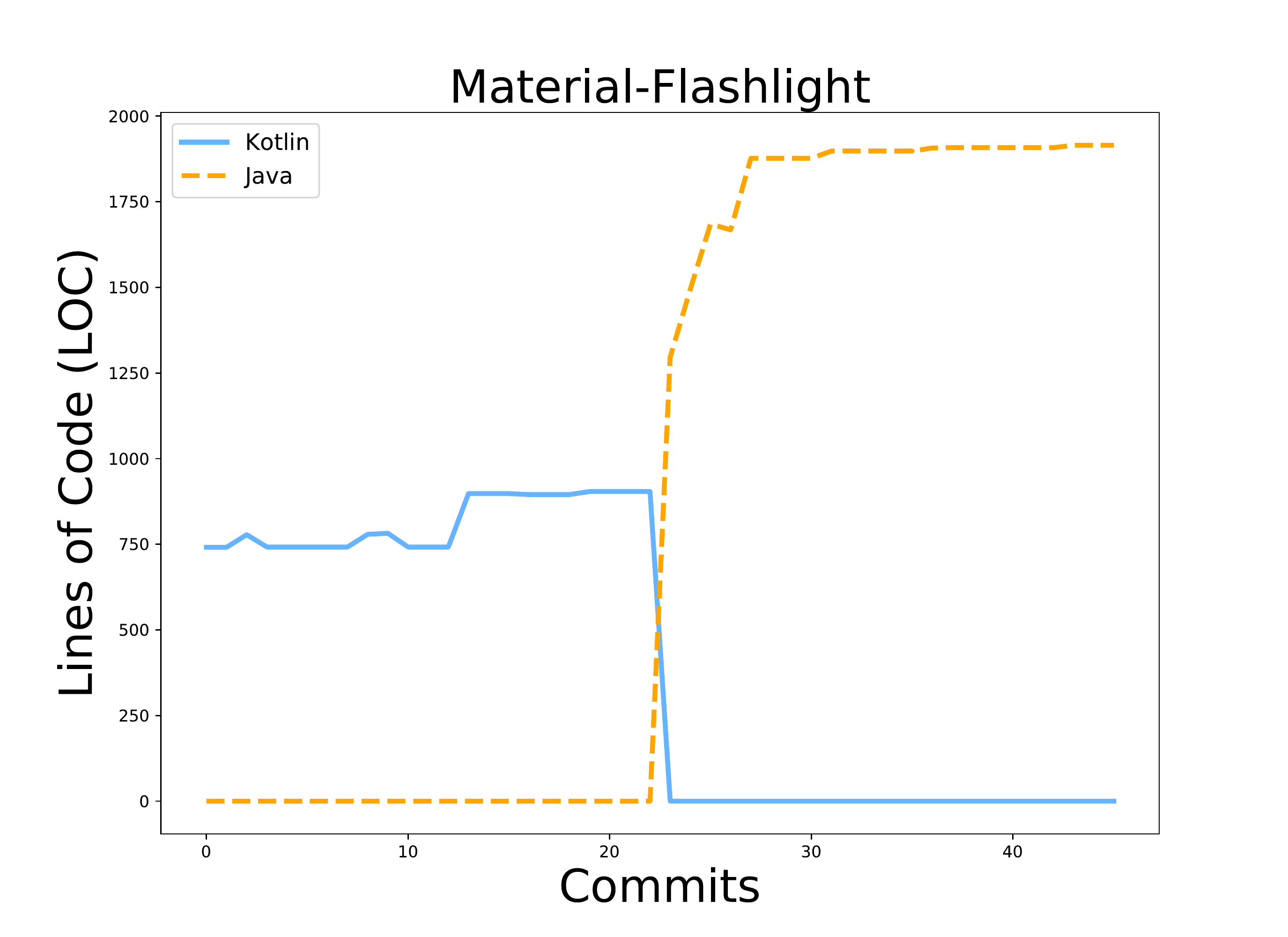}
           \caption{ET 11:Java replaces Kotlin code.}
      \label{fig:et11}
    \end{subfigure}
       
    \caption{Evolution trends of Kotlin and Java code. The figure presents one graphic per evolution trend described in Section \ref{sec:methodology:codeevolutionclassification}
    Each graphic presents the evolution of Kotlin and Java code along the history (i.e., commits) of one single application.  The x-axis corresponds to the commits and the y-axis corresponds to the amount of code (i.e., lines).
   }
    \label{fig:evolutions_trends1}
\end{figure}

We classified each Kotlin application according to the code evolution trends presented in Section \ref{sec:methodology:codeevolutionclassification}.
Table \ref{tab:evolution_trend_kotlin} shows the results and Figure \ref{fig:evolutions_trends1} displays, for each code evolution trend, the code evolution of one particular application as an example.

The most frequent code evolution trend we found is ET 5, with \etfive~out of \ukotlindown~(\pgfmathprintnumber[fixed, precision=1]{\petfive}\%) Kotlin applications. This evolution trend represents the cases that, 
after the first version (i.e., commit) that introduces Kotlin code, the amount of Kotlin code tends to grow, whereas the amount of Java code decreases. 
Sub-figure  \ref{fig:et5} shows the code evolution of app \emph{Jalkametri-Android}, which corresponds to that evolution trend.
Still, the last version of \emph{Jalkametri-Android} has more lines of code (LOC) of Java than Kotlin.
Another application classified as ET 5 is \emph{Poet-Assistant} (sub-figure \ref{fig:et52}).
However, unlike \emph{Jalkametri-Android}, the amount of Kotlin code in the last version is larger than the amount of Java code.
In the mentioned applications, the lines that represent the evolution of Kotlin code seem to be symmetric w.r.t those of Java. 
We suppose that in those cases some components of the application written in Java code were gradually migrated to Kotlin code.

A similar trend to ET 5 is ET 6: \etsix~Android applications (\pgfmathprintnumber[fixed, precision=1]{\petsix}\%) exhibited that trend. 
Unlike ET 5,  the amount of Java code in trend ET 6 gradually decreases until arriving at zero LOC. 
Since that moment, those applications do not contain Java code anymore. The sub-figure \ref{fig:et6} shows one application, \emph{Simple-Calendar}, whose first versions were written in Java. 
Then, the versions from commit 09ef99 to 206dfe, introduced Kotlin code and removed Java code. 
Finally, from commit eee184 the application is only composed by Kotlin code.

There are two trends ET 7 and 8 (with~\etseven~and~\eteight~applications, resp.) of which amount of code of a given language is almost constant since Kotlin is introduced. 
For example, the sub-figure \ref{fig:et8} shows the  code evolution of \emph{Talk-Android}, classified as ET 8.   
One commit (7f12) introduced a portion of Kotlin code (105 lines). Since then: 
\begin{inparaenum} [\it a)]
\item the amount of Kotlin code remains constant along the evolution (the last commit 724 has 106 LOC),
\item the amount of Java code constantly grows.
\end{inparaenum}
The sub-figure \ref{fig:et7} shows an inverted case (app \emph{Bimba}): the amount of Java code is constant while the amount of Kotlin code grows. 
There are also \etnine~applications whose amount of code written in both languages remain constant, such the app \emph{Wyk-Android} showed in sub-figure~\ref{fig:et9}. These applications represent the trend ET 9. 

The fifth most frequent evolution trend is ET 1, with~\etone~applications: those applications were initially written in Kotlin and did not include Java code in any version. 
Application \emph{Vpnhotspot} is one of them (sub-figure \ref{fig:et1}).
The evolution trend ET 2, with \ettwo~applications over~\ukotlindown~(\pgfmathprintnumber[fixed, precision=1]{\pettwo}\%), represents applications such \emph{Simple-Flashing},
sub-figure \ref{fig:et2}, whose are initially written in Java and then, in one commit (18b5c9), migrate the complete code base from Java to Kotlin. 
However, unlike with ET 5 and 6, in ET 2 no version shares Java and Kotlin code. 
On the other hand, we classified~\eteleven~applications with trend ET 11, i.e., applications whose evolution presents the opposite behavior when compared with ET 2, such application \emph{Material-Flashlight}, sub-figure~\ref{fig:et11}. These applications migrated from Kotlin to Java code.

Moreover, there were~\etfour~Android applications (\pgfmathprintnumber[fixed, precision=1]{\petfour}\%)  that correspond to trend ET 4: the amount of both Java and Kotlin grows.
The sub-figure \ref{fig:et4} shows the amount of code of the app \emph{Android USB MSD}: 
there, the introduction of code written in one language did not produce a decrease in the amount of code written in the other language. 

Another evolution trend is ET 3: when Kotlin code is introduced, the amount of Java code decreases (in similar proportions) but then the amount of Java starts growing again.
Sub-figure \ref{fig:et3} shows the code evolution of app \emph{Home-Assistant}, one of the~\etthree~applications (\pgfmathprintnumber[fixed, precision=1]{\petthree}\%) classified as ET 3.  Here, we suspect that developers migrated only a portion of code. 

Furthermore, there were~\etten~applications represented by ET 10: Kotlin code is introduced at some time but, at some version later, that code is removed. 
The sub-figure \ref{fig:et10} shows the code evolution of \emph{Freeotpplus} application.
There we can see that Kotlin is introduced in the commit 2dbc32, but later, after two commits, it is completely removed.

Finally, we assigned the trend ET 12 (Other) to  \ettwelve{} applications that we could not assign any of our previously defined evolution trends.
In our appendix, we presented the classification of the evolution trend discussed in this section.\footnote{Classification of the evolution trends. \url{https://github.com/UPHF/kotlinandroid/tree/master/docs/evolution}}

\begin{tcolorbox}
{\bf Response to RQ 3:} \emph{\rqevolutionproportion}

\pgfmathparse{\pettwo + \petfive + \petsix + \petseven}

For the \pgfmathprintnumber[fixed, precision=1]{\pgfmathresult}\% of the Kotlin applications,
the amount of Kotlin code increases along the Android application evolution and, at the same time, the amount of Java code decreases or remains constant (cases ET 2, 5,  6 and 7).
\pgfmathparse{\pettwo +\petsix}
For the \pgfmathprintnumber[fixed, precision=1]{\pgfmathresult}\% of applications, the Kotlin code replaces the totality of the Java code written on those applications (cases ET 2 and 6).

\end{tcolorbox}

\subsection{RQ$_4$: \emph{\rqdifferencewithjava}}
\label{sec:rquestionfour}

We applied the methodology presented in Section \ref{sec:method:analysisapps}.
We executed Paprika on all apks from FAMAZOA. 
Paprika successfully analyzed  \numprint{17725} apks from \numprint{2040} applications  (94\%) and threw an error over all apks from 127 applications.\footnote{In our appendix we list the reason of those Parpika failures \url{https://github.com/UPHF/kotlinandroid/blob/master/README.md#analyzing-apps-with-priprika}}

To respond to this research question, 
we compared the results found for Java applications and pure Kotlin applications. 
First, we focused on computing the number of affected \emph{applications}. Next, we focused on the number of affected \emph{entities}.

\subsubsection{Number of Affected Applications}

\newcommand{\kblob}{96.51}
\newcommand{\ksak}{75.71}
\newcommand{\kcc}{99.02}
\newcommand{\klm}{99.93}
\newcommand{\klic}{82.75}
\newcommand{\knlmr}{98.89}
\newcommand{\khas}{17.00}
\newcommand{\khss}{16.81}
\newcommand{\khbr}{40.36}
\newcommand{\kiod}{12.47}
\newcommand{\kuio}{51.55}

\newcommand{\pkblob}{95.12}
\newcommand{\pksak}{65.67}
\newcommand{\pkcc}{98.50}
\newcommand{\pklm}{99.80}
\newcommand{\pklic}{78.60}
\newcommand{\pknlmr}{99.30}
\newcommand{\pkhas}{09.55}
\newcommand{\pkhss}{17.31}
\newcommand{\pkhbr}{35.72}
\newcommand{\pkiod}{09.55}
\newcommand{\pkuio}{54.02}

\newcommand{\jblob}{93.53}
\newcommand{\jsak}{66.51}
\newcommand{\jcc}{96.89}
\newcommand{\jlm}{99.62}
\newcommand{\jlic}{87.27}
\newcommand{\jnlmr}{98.84}
\newcommand{\jhas}{22.61}
\newcommand{\jhss}{19.92}
\newcommand{\jhbr}{39.98}
\newcommand{\jiod}{06.50}
\newcommand{\juio}{45.33}

\begin{table}
\centering
\caption{Percentage of Android applications affected by code smell. An app $a$ is affected by a code smell $s$ if $a$ has at least one instance of $s$.}
\begin{tabular}{|l|c c c c|c c c c c c|} 
\hline
\multirow{5}{*}{Lang} &\multicolumn{10}{c|}{\% Affected applications by smells}\\
\cline{2-11}
&\multicolumn{4}{c|}{Object-oriented smells} & \multicolumn{6}{c|}{Android smells}\\
\cline{2-11}
& \tiny{Long}  & \tiny{Complex}  & \tiny{BLOB}  & \tiny{Swiss}   & \tiny{No Low} & \tiny{UI}  & \tiny{Heavy} & \tiny{Heavy} & \tiny{Heavy}  & \tiny{Init}  \\ 
& \tiny{Method} & \tiny{Class} & \tiny{class} & \tiny{Army} &  \tiny{Memory} & \tiny{Overdraw} & \tiny{Broadcast} & \tiny{Service} & \tiny{ASynctask} & \tiny{OnDraw} \\
&  &  &  & \tiny{Knife} & \tiny{Resolver} &  & \tiny{Receiver} & \tiny{Start} &  & \\

\cline{2-11}
&(LM) & (CC) & (BLOB) & (SAK) & (NLMR) & (UIO) & (HBR) & (HSS) & (HAS) & (IOD)\\
\hline
\hline
Pure Kotlin & \pklm & \pkcc & \pkblob & \pksak & \pknlmr & \pkuio  & \pkhbr & \pkhss & \pkhas & \pkiod \\
\hline
Java   & \jlm & \jcc & \jblob & \jsak & \jnlmr & \juio & \jhbr & \jhss & \jhas & \jiod  \\
\hline
\hline
K - J 
      & \pgfmathparse{\pklm - \jlm} \pgfmathprintnumber[fixed, precision=2]{\pgfmathresult}
      & \pgfmathparse{\pkcc - \jcc} \pgfmathprintnumber[fixed, precision=2]{\pgfmathresult}
      & \pgfmathparse{\pkblob - \jblob} \pgfmathprintnumber[fixed, precision=2]{\pgfmathresult}
      & \pgfmathparse{\pksak - \jsak} \pgfmathprintnumber[fixed, precision=2]{\pgfmathresult}
      & \pgfmathparse{\pknlmr - \jnlmr} \pgfmathprintnumber[fixed, precision=2]{\pgfmathresult} 
      & \pgfmathparse{\pkuio - \juio} \pgfmathprintnumber[fixed, precision=2]{\pgfmathresult}
      & \pgfmathparse{\pkhbr - \jhbr} \pgfmathprintnumber[fixed, precision=2]{\pgfmathresult}
      & \pgfmathparse{\pkhss - \jhss} \pgfmathprintnumber[fixed, precision=2]{\pgfmathresult}
      & \pgfmathparse{\pkhas - \jhas} \pgfmathprintnumber[fixed, precision=2]{\pgfmathresult}
      & \pgfmathparse{\pkiod - \jiod} \pgfmathprintnumber[fixed, precision=2]{\pgfmathresult} 
       \\ \hline 
\end{tabular}
\label{tab:apps_affected_smells}
\end{table} 
Table~\ref{tab:apps_affected_smells} shows, for each programming language and code smell,  the percentages of Android applications affected by such code smell, i.e., having one or more smell instance.

We found that 3 out of 4 (75\%) object-oriented smells affect more than 93\% of applications considering both languages, Kotlin and Java. LM is the most common smell, affecting approximately 99\% of the applications of both languages.
SAK is the least frequent smell, but it still affects the majority of applications, around 65\% of Kotlin applications and 66\% of Java applications. 
Therefore, our results agree with previous work~\citep{Hecht2015,Habchi2017} showing that LM, CC, BLOB and SAK respectively, are the most common OO smells in Android applications. 

Table~\ref{tab:apps_affected_smells} also shows differences between the percentages of applications written in Java and Kotlin (row K - J): 
for all OO smells (LM, CC, BLOB, and SAK), the differences are small: 0.18\%, 1.61\%, 1.59\% and -0.84\%, respectively. 
Our conclusion is twofold.
First, 3 out of 4 (75\%) of object-oriented code smells are more present in Kotlin applications.
However, they affect a similar proportion of both Kotlin and Java application.

Regarding Android smells,
we observed that NLMR is the most frequent smell, affecting the 99\% of Kotlin and 98\% Java applications.
Furthermore, the second most present Android smell, UIO, it affects 54\% and 45\% of Kotlin and Java applications, respectively. 
The third most present smell is HBR, that affects 35\% of Kotlin and 39\% of Java applications.
Other 3 smells (HSS, HAS, IOD) are present in, at most, 22\% of all applications.
Note that there are 3 Android smells that affect proportionally more Java applications. 
Moreover, the most significant difference between the proportion of applications affected by Android smells is 13.06\% from the HAS smell.

We observed a high number of applications affected by 
NLMR smell, because it is related to~\textit{Activities}, the main component of Android applications, which is present in almost every Android application. On the other hand, we found a smaller number of applications affected by the remaining smells. Since HBR, HSS and HAS are related with other Android components, \textit{BroadcastReceiver}, \textit{Service} and \textit{AsyncTask}, that are not essential for all Android applications and because UIO and IOD, that are smells related with \textit{View} component, affect only custom views implementation.  Nonetheless,  as the Android SDK provides several views implementations, most applications do not need custom views implementation.

\begin{tcolorbox}
{\bf Response to RQ 4}: \emph{\rqdifferencewithjava}

\underline{In terms of affected applications:} 
Considering all smells, 6 out of 10 (3 out of 4 object-oriented smells and 3 out of 6 Android smells) affect proportionally more applications with Kotlin code. However, for all object-oriented smells and 4 out of 6 Android smells, the difference between the percentage of affected Kotlin and Java applications is small (between 0.18 and 4.26\%). 
\end{tcolorbox}

\subsubsection{Number of Affected Entities}

Now, we study the proportion of entities affected by smells using the methodology presented in Section \ref{sec:method:diff}. 
Table~\ref{tab:smells_ratio} shows, for each smell and programming language, the median (med) of the ratios of smells  in the applications (Formula \ref{eq:1}) and the Cliff's $\delta$ effect sizes.

First, we observed that the most frequent smells (LM, CC, and BLOB) have a small median ratio.  
This means that, although they are present in most applications, only a few entities are affected by these smells. 

\newcommand{\kblobmed}{0.0163}
\newcommand{\kblobiqr}{0.0120}
\newcommand{\jblobmed}{0.0278}
\newcommand{\jblobiqr}{0.0265}

\newcommand{\ksakmed}{0.0008}
\newcommand{\ksakiqr}{0.0028}
\newcommand{\jsakmed}{0.0040}
\newcommand{\jsakiqr}{0.0085}

\newcommand{\kccmed}{0.0593}
\newcommand{\kcciqr}{0.0302}
\newcommand{\jccmed}{0.0781}
\newcommand{\jcciqr}{0.0530}

\newcommand{\klmmed}{0.0563}
\newcommand{\klmiqr}{0.0239}
\newcommand{\jlmmed}{0.0736}
\newcommand{\jlmiqr}{0.0407}

\newcommand{\knlmrmed}{0.3750}
\newcommand{\knlmriqr}{0.8000}
\newcommand{\jnlmrmed}{1.0000}
\newcommand{\jnlmriqr}{0.5000}

\newcommand{\kuiomed}{0.0769}
\newcommand{\kuioiqr}{0.2857}
\newcommand{\juiomed}{0.0000}
\newcommand{\juioiqr}{0.1818}

\newcommand{\khssmed}{0.0000}
\newcommand{\khssiqr}{0.0000}
\newcommand{\jhssmed}{0.0000}
\newcommand{\jhssiqr}{0.0000}

\newcommand{\khbrmed}{0.0000}
\newcommand{\khbriqr}{0.1115}
\newcommand{\jhbrmed}{0.0000}
\newcommand{\jhbriqr}{0.1666}

\newcommand{\khasmed}{0.0000}
\newcommand{\khasiqr}{0.0000}
\newcommand{\jhasmed}{0.0000}
\newcommand{\jhasiqr}{0.0000}

\newcommand{\kiodmed}{0.0000}
\newcommand{\kiodiqr}{0.0000}
\newcommand{\jiodmed}{0.0000}
\newcommand{\jiodiqr}{0.0000}

\begin{table}
\centering
\caption{Ratio comparison between Kotlin and Java.
The column `Cliff's $\delta$' shows the difference between the smell median ratio of pure Kotlin and Java applications: negative values mean that a smell affects fewer entities in Kotlin than in Java.
}
\begin{tabular}{|l | l | l | l | l |}
\hline
\multirow{2}{*}{Smell} & \multirow{2}{*}{Lang} & Median & \multirow{2}{*}{Cliff's  $\delta$} & Significance of\\ 
  &  & Ratio & & difference\\ 
\hline \hline
\multirow{2}{*}{LM} & Kotlin  & \klmmed &  \multirow{2}{*}{-0.3873}  & \multirow{ 2}{*}{Medium} \\
  & Java & \jlmmed &  & \\ \hline
\multirow{2}{*}{CC} & Kotlin  & \kccmed & 
\multirow{2}{*}{-0.3168} &  \multirow{2}{*}{Small} \\
  & Java & \jccmed &  &\\ \hline
\multirow{2}{*}{BLOB} & Kotlin  & \kblobmed  & \multirow{2}{*}{-0.4338} &  \multirow{2}{*}{Medium}\\
  & Java & \jblobmed & & \\ \hline
\multirow{2}{*}{SAK} & Kotlin  & \ksakmed  & \multirow{2}{*}{-0.2433} &  \multirow{2}{*}{Small} \\
  & Java & \jsakmed &  &\\
 \hline \hline
\multirow{2}{*}{NLMR} & Kotlin  & \knlmrmed &  \multirow{2}{*}{-0.2915} &  \multirow{2}{*}{Small} \\ 
 & Java & \jnlmrmed & &\\  \hline
\multirow{2}{*}{UIO} & Kotlin  & \kuiomed &  \multirow{2}{*}{0.1156} &  \multirow{2}{*}{Insignificant} \\ 
 & Java & \juiomed & &\\  \hline
\multirow{2}{*}{HBR} & Kotlin  & \khbrmed &  \multirow{2}{*}{-0.0699}&  \multirow{2}{*}{Insignificant} \\ 
 & Java & \jhbrmed &&\\  \hline
\multirow{2}{*}{HSS} & Kotlin  & \khssmed &  \multirow{2}{*}{-0.0240} &  \multirow{2}{*}{Insignificant}  \\
 & Java & \jhssmed  & & \\ \hline
\multirow{2}{*}{HAS} & Kotlin  & \khasmed &  \multirow{2}{*}{-0.1306}&  \multirow{2}{*}{Insignificant} \\
 & Java & \jhasmed & & \\ \hline
\multirow{2}{*}{IOD} & Kotlin  & \kiodmed &  \multirow{2}{*}{0.0341} &  \multirow{2}{*}{Insignificant} \\ 
 & Java & \jiodmed & &\\  \hline
\end{tabular}
\label{tab:smells_ratio}
\end{table} 
Moreover, the Cliff's $\delta$ values show that the difference between the smell median ratio of pure Kotlin and Java applications are statistically significant for all object-oriented smells: `small' difference for CC and SAK, and `medium' for  LM and BLOB. 
We conclude that although 3 out of 4 object-oriented smells affect more Kotlin applications, our results show that for 100\% of object-oriented smells, \emph{Java applications have in median more entities affected by them with statistical relevance.} 

Concerning Android smells, 
we observed that the median ratios for 4 out of 6 smells (HBR, HSS, HAS and IOD) are zero for Java and Kotlin applications, which is consistent with the Table~\ref{tab:apps_affected_smells}, since less than 50\% of applications for both languages are affected by these smells. 
Furthermore, Cliff's $\delta$ showed no significant difference for 5 out of 6 Android smells, including those previously mentioned. 
We conclude that very few entities are affected by these smells, even for HBR that affect more than 35\% of Android applications (see Table~\ref{tab:apps_affected_smells}). 
Concerning the smell NLMR, we observed, with statistical significance, that Java applications have more entities affected with a `small' significant difference. 
This result agrees with those from \cite{Habchi2017} and shows that NLMR affects all activities of most Java applications. 

\begin{tcolorbox}
{\bf Response to RQ 4 (cont.)}: \emph{\rqdifferencewithjava}

\underline{In terms of affected entities}:
Although 3 out of 4 (LM, CC, and BLOB) object-oriented smells affect more Kotlin applications, 
our results show that for all object-oriented smells (LM, CC, BLOB and SAK) Java applications have in median more entities affected by them with statistically relevance. Moreover, considering Android smells, Java applications have in median more entities affected by NLMR, the unique smell that presents statistical relevance.

\end{tcolorbox}

\subsection{RQ$_5$: \rqscorekotlin}

\newcommand{\kotlinimprovement}{56}

\newcommand{\lmfk}{28}
\newcommand{\lmlk}{28}
\pgfmathsetmacro{\plmfk}{(\lmfk / \kotlinimprovement)*100}
\pgfmathsetmacro{\plmlk}{(\lmlk / \kotlinimprovement)*100}
\newcommand{\positivelm}{10}
\pgfmathsetmacro{\ppositivelm}{(\positivelm / \lmlk)*100}

\newcommand{\ccfk}{35}
\newcommand{\cclk}{36}
\pgfmathsetmacro{\pccfk}{(\ccfk / \kotlinimprovement)*100}
\pgfmathsetmacro{\pcclk}{(\cclk / \kotlinimprovement)*100}
\newcommand{\positivecc}{13}
\pgfmathsetmacro{\ppositivecc}{(\positivecc / \cclk)*100}

\newcommand{\blobfk}{43}
\newcommand{\bloblk}{42}
\pgfmathsetmacro{\pblobfk}{(\blobfk / \kotlinimprovement)*100}
\pgfmathsetmacro{\pbloblk}{(\bloblk / \kotlinimprovement)*100}
\newcommand{\positiveblob}{18}
\pgfmathsetmacro{\ppositiveblob}{(\positiveblob / \bloblk)*100}

\newcommand{\sakfk}{44}
\newcommand{\saklk}{45}
\pgfmathsetmacro{\psakfk}{(\sakfk / \kotlinimprovement)*100}
\pgfmathsetmacro{\psaklk}{(\saklk / \kotlinimprovement)*100}
\newcommand{\positivesak}{17}
\pgfmathsetmacro{\ppositivesak}{(\positivesak / \saklk)*100}

\newcommand{\hbrfk}{45}
\newcommand{\hbrlk}{40}
\pgfmathsetmacro{\phbrfk}{(\hbrfk / \kotlinimprovement)*100}
\pgfmathsetmacro{\phbrlk}{(\hbrlk / \kotlinimprovement)*100}
\newcommand{\positivehbr}{7}
\pgfmathsetmacro{\ppositivehbr}{(\positivehbr / \hbrlk)*100}

\newcommand{\hasfk}{37}
\newcommand{\haslk}{30}
\pgfmathsetmacro{\phasfk}{(\hasfk / \kotlinimprovement)*100}
\pgfmathsetmacro{\phaslk}{(\haslk / \kotlinimprovement)*100}
\newcommand{\positivehas}{2}
\pgfmathsetmacro{\ppositivehas}{(\positivehas / \haslk)*100}

\newcommand{\hssfk}{45}
\newcommand{\hsslk}{42}
\pgfmathsetmacro{\phssfk}{(\hssfk / \kotlinimprovement)*100}
\pgfmathsetmacro{\phsslk}{(\hsslk / \kotlinimprovement)*100}
\newcommand{\positivehss}{6}
\pgfmathsetmacro{\ppositivehss}{(\positivehss / \hsslk)*100}

\newcommand{\iodfk}{36}
\newcommand{\iodlk}{36}
\pgfmathsetmacro{\piodfk}{(\iodfk / \kotlinimprovement)*100}
\pgfmathsetmacro{\piodlk}{(\iodlk / \kotlinimprovement)*100}
\newcommand{\positiveiod}{5}
\pgfmathsetmacro{\ppositiveiod}{(\positiveiod / \iodlk)*100}

\newcommand{\nlmrfk}{33}
\newcommand{\nlmrlk}{31}
\pgfmathsetmacro{\pnlmrfk}{(\nlmrfk / \kotlinimprovement)*100}
\pgfmathsetmacro{\pnlmrlk}{(\nlmrlk / \kotlinimprovement)*100}
\newcommand{\positivenlmr}{6}
\pgfmathsetmacro{\ppositivenlmr}{(\positivenlmr / \nlmrlk)*100}

\newcommand{\uiofk}{36}
\newcommand{\uiolk}{36}
\pgfmathsetmacro{\puiofk}{(\uiofk / \kotlinimprovement)*100}
\pgfmathsetmacro{\puiolk}{(\uiolk / \kotlinimprovement)*100}
\newcommand{\positiveuio}{8}
\pgfmathsetmacro{\ppositiveuio}{(\positiveuio / \uiolk)*100}

\begin{table*}
\centering
\caption{Changes on quality scores after introducing Kotlin. 
The table shows the number of applications that have improved a quality score associated with a smell. 
The column `First Kotlin' shows the comparison between the quality score of the last version without Kotlin and the first version with Kotlin.
The column `Last Kotlin' compares the quality score of the last version without Kotlin and the last version with Kotlin.
The column `Positive Change on Evolution Trend' counts the number of apps where the introduction of Kotlin produces a change in the quality evolution trend from `Decline' or `Stability' to `Rise'.
}
\begin{tabular}{|c|r|r|r|}
\hline
\multirow{2}{*}{Code Smell}& 
\multicolumn{2}{c|}{\# Apps Kotlin Improves Quality Score}& Positive Change\\
\cline{2-3}
& First Kotlin & \multicolumn{1}{c|}{Last Kotlin} & \multicolumn{1}{c|}{on Evolution Trend}\\
\hline 
\hline

LM & \lmfk (\pgfmathprintnumber[fixed, precision=2]{\plmfk}\%)
& \lmlk (\pgfmathprintnumber[fixed, precision=2]{\plmlk}\%)
& \positivelm/\lmlk (\pgfmathprintnumber[fixed, precision=2]{\ppositivelm}\%)\\

CC& \ccfk (\pgfmathprintnumber[fixed, precision=2]{\pccfk}\%)
& \cclk (\pgfmathprintnumber[fixed, precision=2]{\pcclk}\%)
& \positivecc/\cclk (\pgfmathprintnumber[fixed, precision=2]{\ppositivecc}\%)\\

BLOB& \blobfk (\pgfmathprintnumber[fixed, precision=2]{\pblobfk}\%)
& \bloblk (\pgfmathprintnumber[fixed, precision=2]{\pbloblk}\%)
& \positiveblob/\bloblk (\pgfmathprintnumber[fixed, precision=2]{\ppositiveblob}\%)\\

SAK& \sakfk (\pgfmathprintnumber[fixed, precision=2]{\psakfk}\%)
& \saklk (\pgfmathprintnumber[fixed, precision=2]{\psaklk}\%)
& \positivesak/\saklk (\pgfmathprintnumber[fixed, precision=2]{\ppositivesak}\%)\\
\hline
\hline
HBR& \hbrfk (\pgfmathprintnumber[fixed, precision=2]{\phbrfk}\%)
& \hbrlk (\pgfmathprintnumber[fixed, precision=2]{\phbrlk}\%)
& \positivehbr/\hbrlk (\pgfmathprintnumber[fixed, precision=2]{\ppositivehbr}\%)\\

HAS& \hasfk (\pgfmathprintnumber[fixed, precision=2]{\phasfk}\%)
& \haslk (\pgfmathprintnumber[fixed, precision=2]{\phaslk}\%)
& \positivehas/\haslk (\pgfmathprintnumber[fixed, precision=2]{\ppositivehas}\%)\\

HSS& \hssfk (\pgfmathprintnumber[fixed, precision=2]{\phssfk}\%)
& \hsslk (\pgfmathprintnumber[fixed, precision=2]{\phsslk}\%)
& \positivehss/\hsslk (\pgfmathprintnumber[fixed, precision=2]{\ppositivehss}\%)\\

IOD& \iodfk (\pgfmathprintnumber[fixed, precision=2]{\piodfk}\%)
& \iodlk (\pgfmathprintnumber[fixed, precision=2]{\piodlk}\%)
& \positiveiod/\iodlk (\pgfmathprintnumber[fixed, precision=2]{\ppositiveiod}\%)\\

NLMR& \nlmrfk (\pgfmathprintnumber[fixed, precision=2]{\pnlmrfk}\%)
& \nlmrlk (\pgfmathprintnumber[fixed, precision=2]{\pnlmrlk}\%)
& \positivenlmr/\nlmrlk (\pgfmathprintnumber[fixed, precision=2]{\ppositivenlmr}\%)\\

UIO& \uiofk (\pgfmathprintnumber[fixed, precision=2]{\puiofk}\%)
& \uiolk (\pgfmathprintnumber[fixed, precision=2]{\puiolk}\%)
& \positiveuio/\uiolk (\pgfmathprintnumber[fixed, precision=2]{\ppositiveuio}\%)\\

\hline
\end{tabular}
\label{tab:evolution_score_kotlin}
\end{table*}

 \begin{figure}
\captionsetup{justification=centering}
    \centering
     \begin{subfigure}[b]{0.42\textwidth}
         \includegraphics[width=\textwidth]{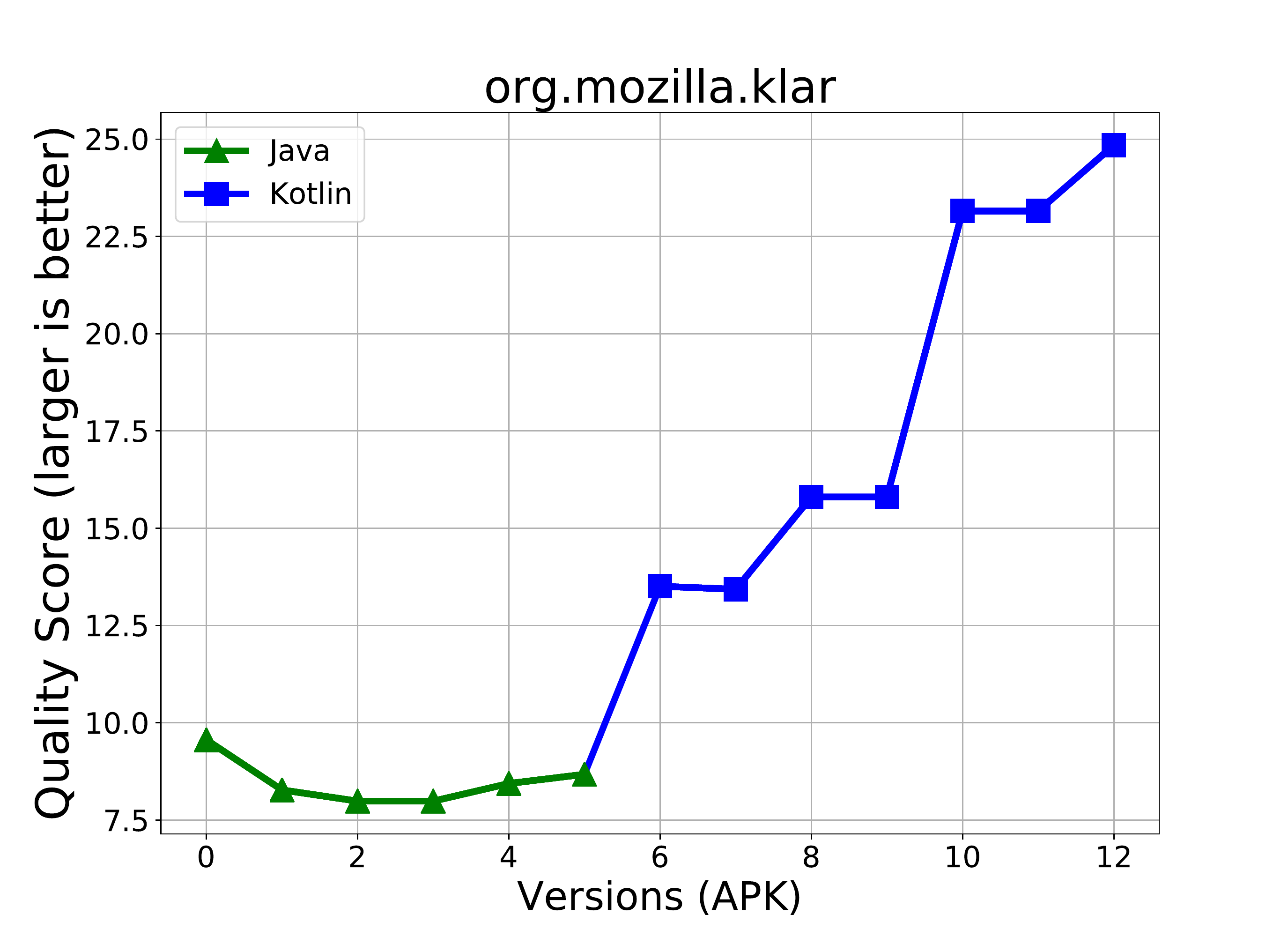}
       \caption{Improvement first and last Kotlin versions.}
        \label{fig:impr_aall}
    \end{subfigure}
    \begin{subfigure}[b]{0.42\textwidth}
          \includegraphics[width=\textwidth]{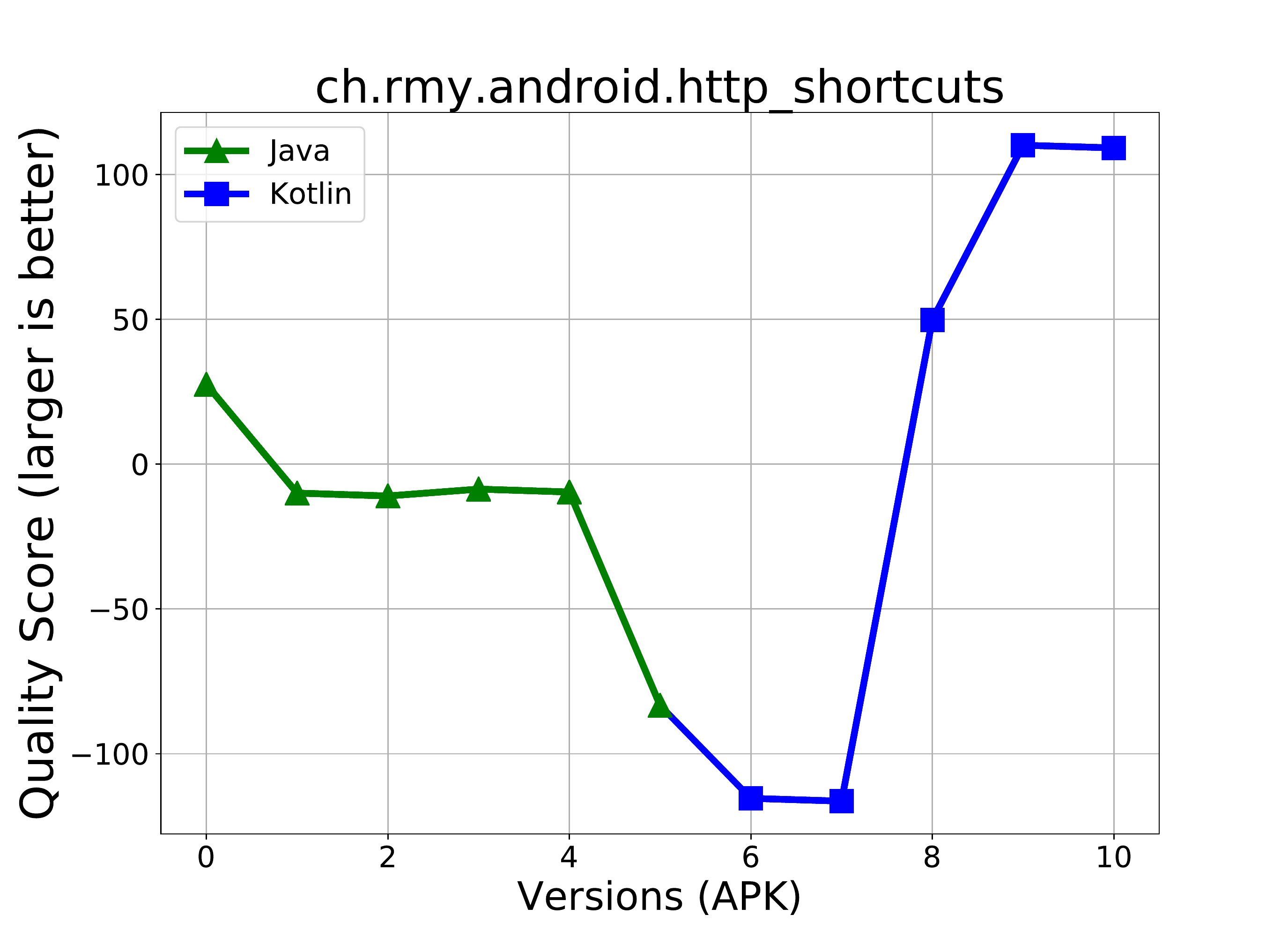}
        \caption{Improvement last Kotlin version.}
        \label{fig:impr_last}
    \end{subfigure}
    
      \begin{subfigure}[b]{0.42\textwidth}
        \includegraphics[width=\textwidth]{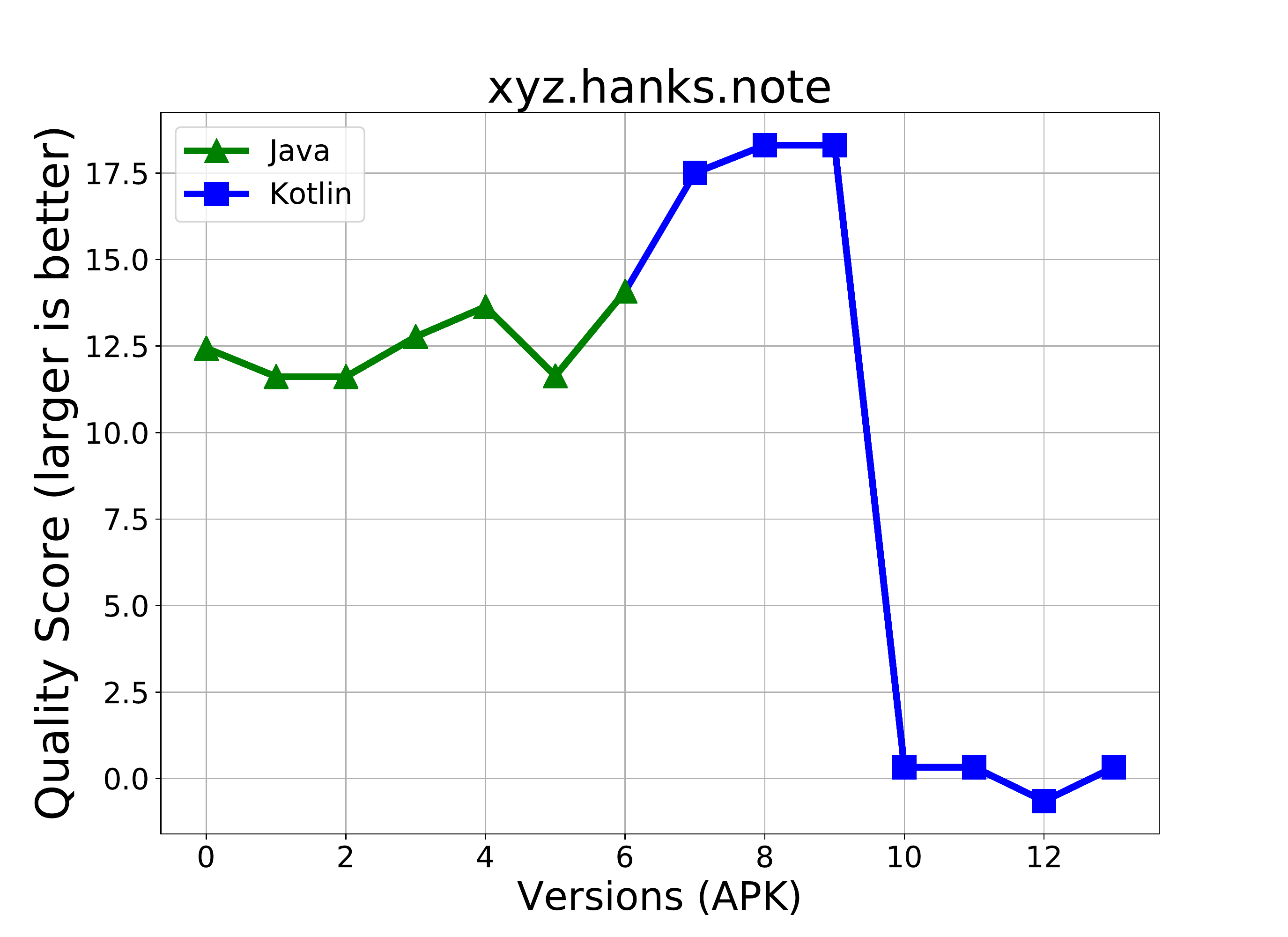}
        \caption{Improvement first Kotlin version.}
        \label{fig:impr_m1}
    \end{subfigure}
       \begin{subfigure}[b]{0.42\textwidth}
        \includegraphics[width=\textwidth]{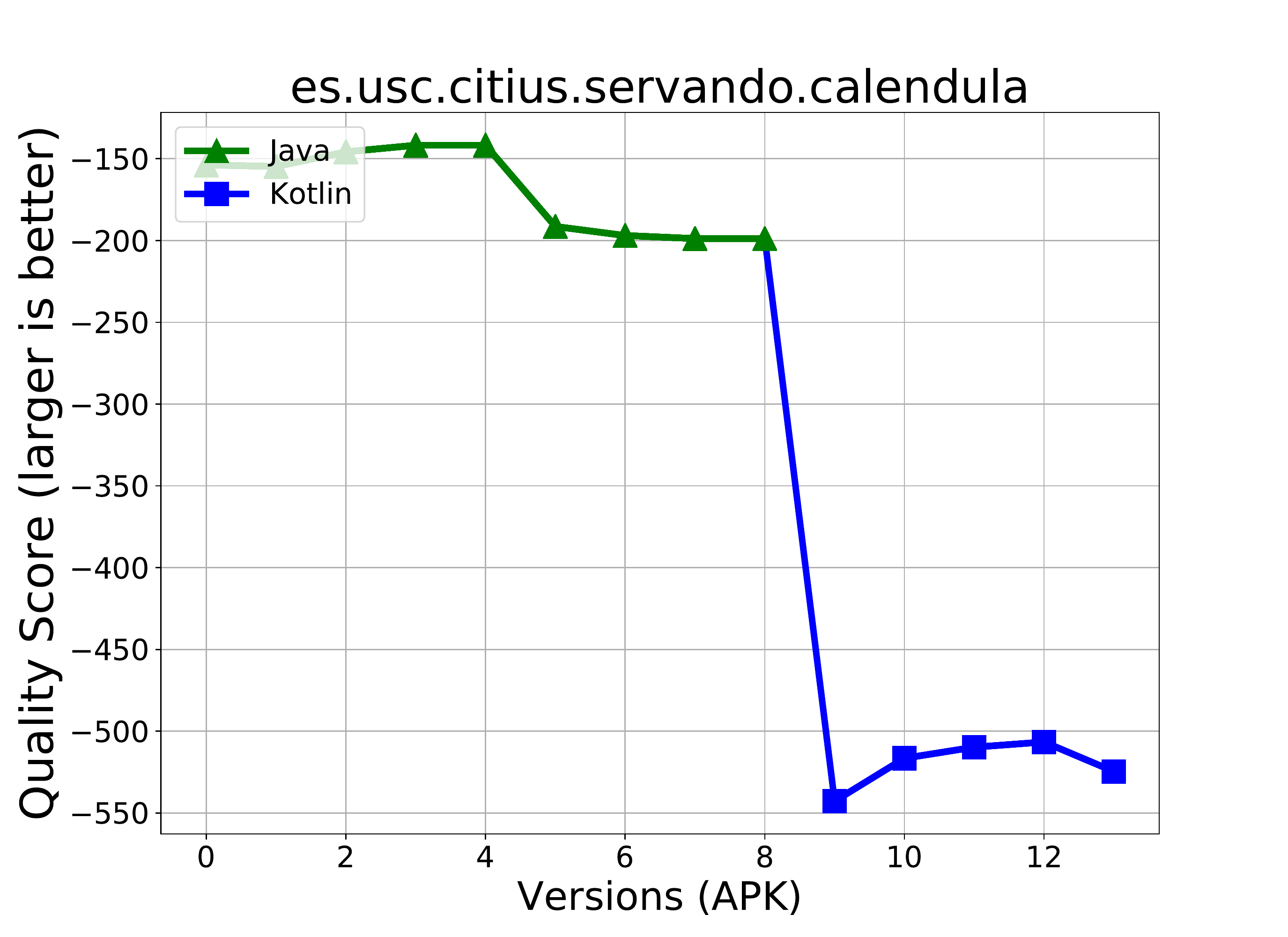}
        \caption{No improvement after introducing Kotlin.}
        \label{fig:no_impr}
    \end{subfigure}
       
    \caption{Evolution of quality scores based on CC smell along the version history.}
    \label{fig:score_evolution}
\end{figure}

The score quality is calculated using the output of Paprika, which takes as input \emph{apks}.
In our experiment, we found in total 57 applications, that initially had only Java code and later introduced Kotlin. 
Considering these applications, Table~\ref{tab:evolution_score_kotlin} shows, for each type of smell, the number of Kotlin applications whose quality score increases after the introduction of Kotlin code.
The increase of one quality score associated with one type of smell implies fewer instances of that smell and, consequently, a better quality of the application.

The results show that for the 10 smells, for at least the 50\% of the applications that introduced Kotlin code,
their quality scores increased between the last Java version and the first version that introduced Kotlin (see Table~\ref{tab:evolution_trend_kotlin} column `First Kotlin').
That means, for such applications the introduction of Kotlin code impacted the quality scores positively.
Note that 50\% of values is the lower bound value because smell LM improves precisely the 50\% of applications. However, all the other smells improve more than 50\%.
For instance, Table 7 shows that, for 8 out of 10 smells, the percentage of applications with quality improvement is larger than the 62\% and for 4 smells is larger than 76\%.
That means that for only one smell (LM) the 50\% of apps do not improve the quality. For the other smells, the number of apps with improvement is larger than the number of apps that do not enhance.
Furthermore, for all smells, at least the 50\% applications had an improvement of the quality score between the last Java version and the most recent (i.e., the last) Kotlin version (see Table~\ref{tab:evolution_trend_kotlin} column `Last Kotlin').

For instance, let us focus on smell CC (complex-class) at the second row of Table~\ref{tab:evolution_score_kotlin}.  
As column ``First Kotlin" shows, for~\ccfk~out of~\kotlinimprovement~(\pgfmathprintnumber[fixed, precision=2]{\pccfk}\%) applications, the version $v_k$ that introduces Kotlin code has larger (i.e., better)  quality score associated to smell CC than the last version without Kotlin $v_{k-1}$ (i.e., the previous version of $v_k$).
Figure \ref{fig:impr_aall} shows the quality score associated to the smell CC of each version of ``Mozilla Klar'' app. 
The last version that does not contain Kotlin code corresponds to X=5 in that figure.
We observed that the first version that has Kotlin code (X=6) increases the quality score, as well as all the subsequent versions (X=[6..12]) do.

Furthermore, as the column ``Last Kotlin" shows, for~\cclk~applications (\pgfmathprintnumber[fixed, precision=2]{\pcclk}\%), the most recent version with Kotlin code has larger (i.e., better)  quality score associated to smell CC  than the version before the introduction of Kotlin. 
Again, ``Mozilla Klar'' is one of those applications: the last version (X=12) has a higher score than the last Java version (X=5).
Note that, for the CC smell, there is 1 application (36 - 35) whose quality scores:
\begin{inparaenum}[\it a)]
\item decrease in the version whose introduces Kotlin, but 
\item increase in the most recent Kotlin version.
\end{inparaenum}
The evolution of the quality score of one of those applications, named ``HTTP Shortcuts'', is displayed in sub-figure \ref{fig:impr_last}.

We also observed in Table \ref{tab:evolution_score_kotlin} that for 5 out of 10 smells,
the number of applications with quality improvements after the first Kotlin version (column ``First Kotlin') is larger than the number of applications with quality improvements over the last Kotlin version (column ``Last Kotlin'). 
This means that the quality scores from some Kotlin applications decrease between the first and the last Kotlin versions.
For instance, sub-figure \ref{fig:impr_m1} shows the evolution of quality score of ``Hanks Note'' application: the first Kotlin version increases the score w.r.t the last Java version. 
However, during the subsequent versions, the quality scores drop, even lower than the score from the last Java version.  
Finally, sub-figure \ref{fig:no_impr} shows the evolution of the quality score of one application, named ``Calendula'', whose quality score constantly decreases after the introduction of Kotlin code. 
This finding shows that also the quality of Kotlin applications can be degraded along the app evolution.

\begin{figure}
\newcommand{\subfsize}{0.32}
\captionsetup{justification=centering}
    \centering
     \begin{subfigure}[b]{\subfsize\textwidth}
         \includegraphics[width=\textwidth]{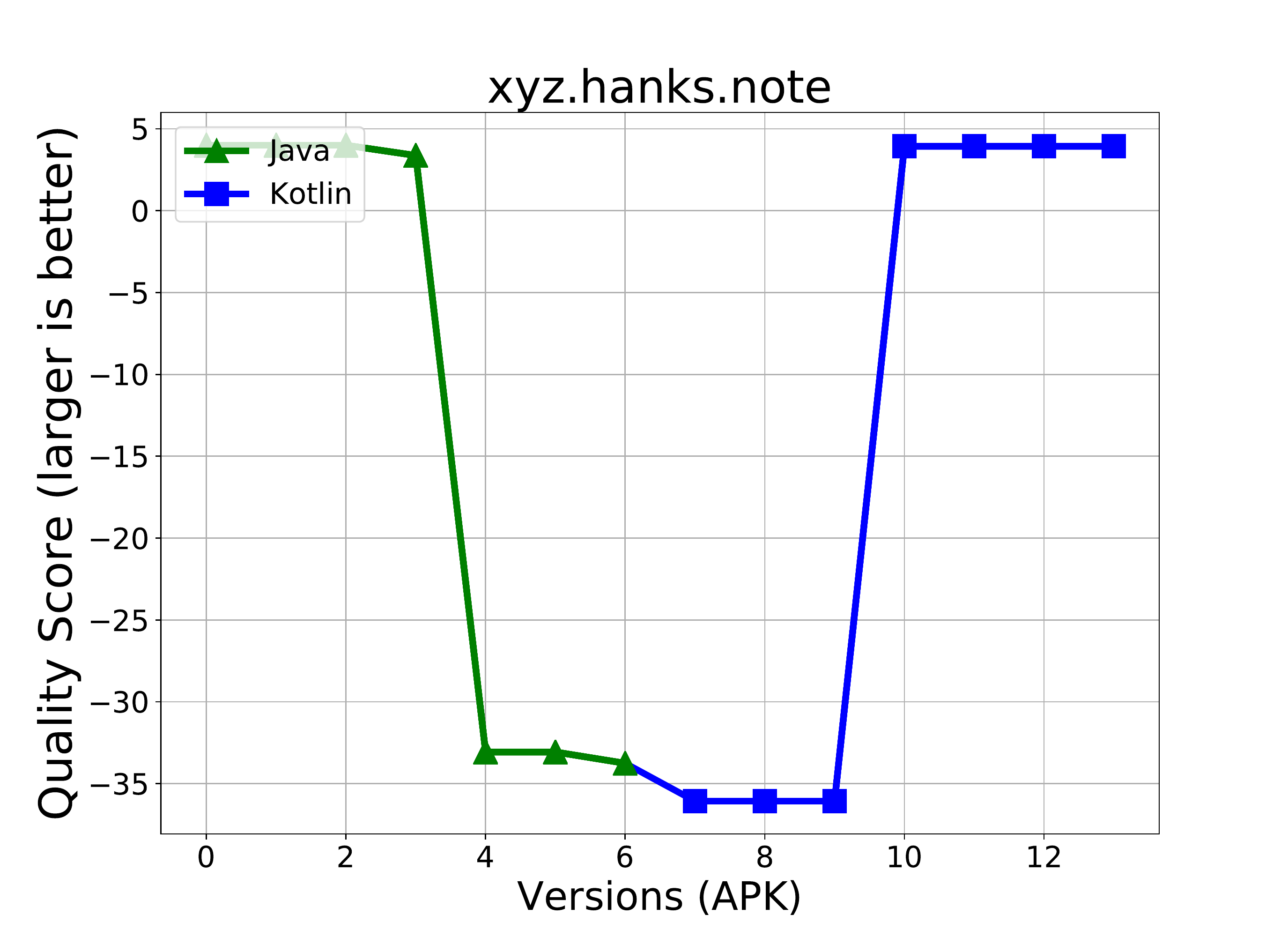}
        \caption{Positive change on quality trend (smell NLMR). Trend before: `Decline`, trend after: `Rise`.}
        \label{fig:trend_change_positive}
    \end{subfigure}
     \begin{subfigure}[b]{\subfsize\textwidth}
    \includegraphics[width=\textwidth]{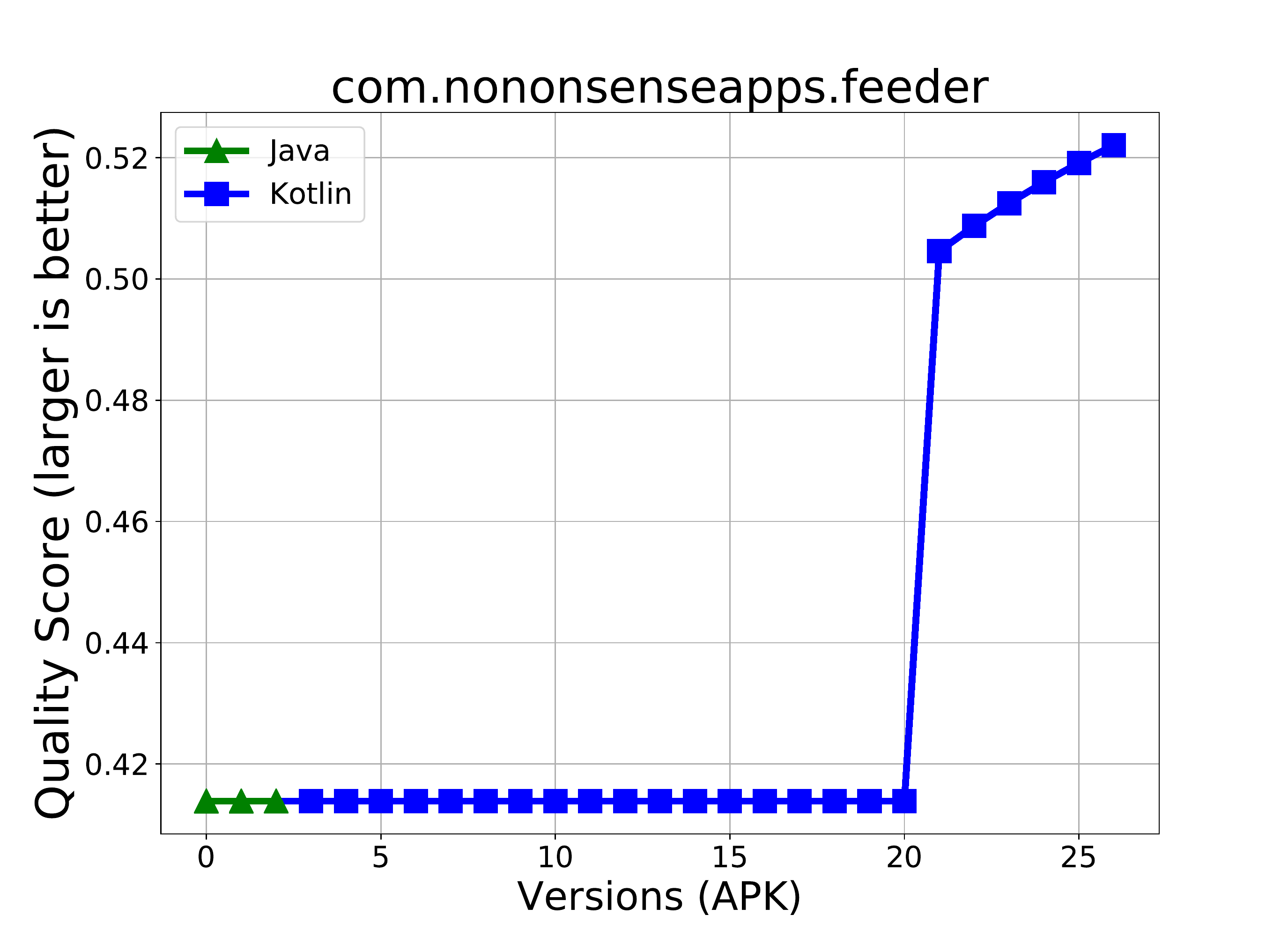}
       \caption{No change on quality trend (smell HAS). Trend before: `Stability`, trend after: `Stability` (even later is `Rise`).}
        \label{fig:trend_change_stable}
    \end{subfigure}
    \begin{subfigure}[b]{\subfsize\textwidth}
         \includegraphics[width=\textwidth]{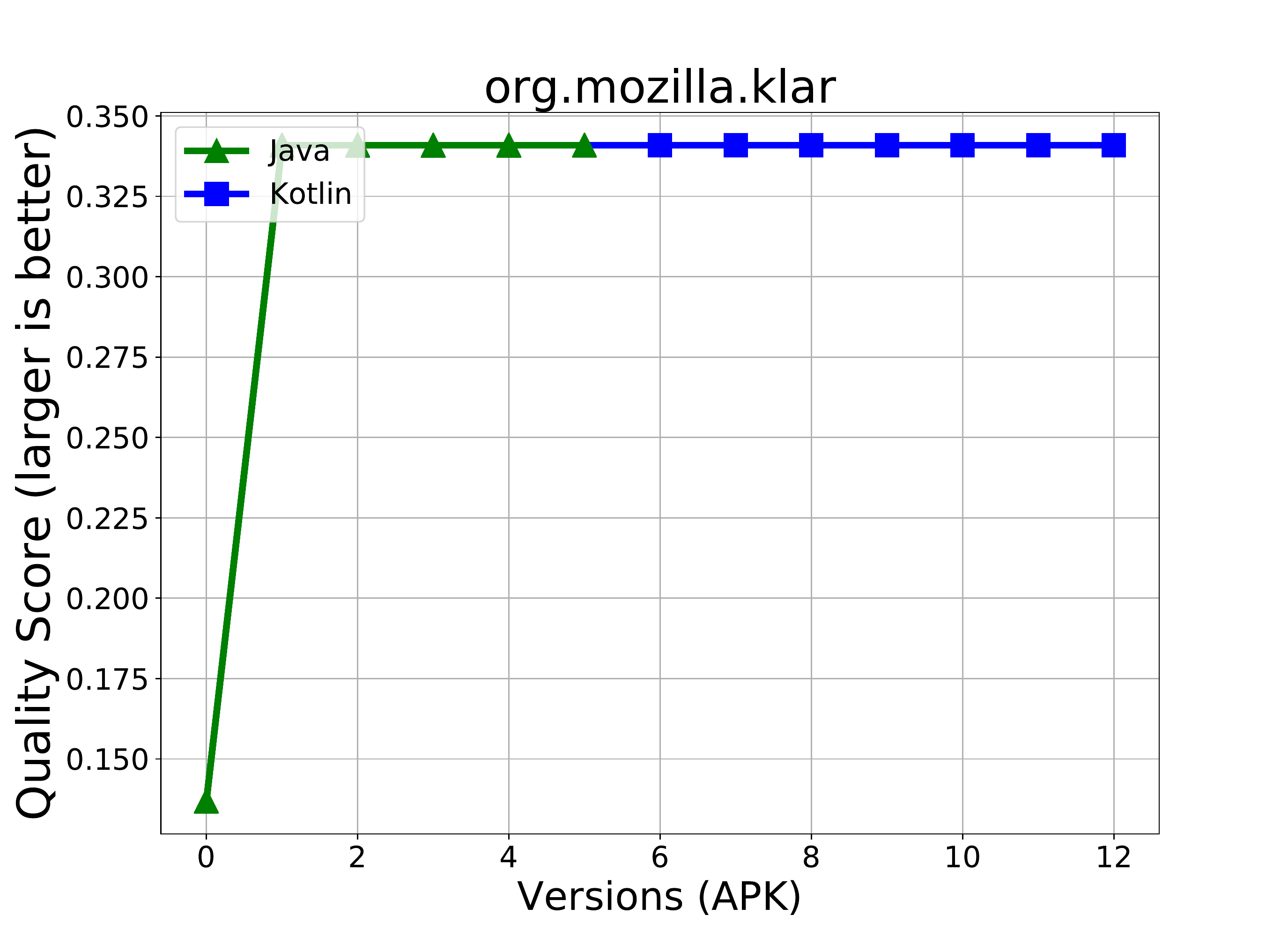}
        \caption{No change on quality trend (smell HBR). Trend before: `Rise`, trend after: `Rise`.}
        \label{fig:trend_change_rise}
    \end{subfigure}
    \caption{Three cases of classification of quality trend evolution \emph{before} and \emph{after} the introduction of Kotlin code.}
    \label{fig:score_trend_chage}
\end{figure}

\begin{tcolorbox}
Response to RQ 5:  \emph{\rqscorekotlin}

The introduction of Kotlin code in Android applications initially written in Java produces a rise in the quality scores from, at least, the 50\% of the Android applications.
More precisely, for 8 and 4 smells (out of 10), the first commits with Kotlin code produce a rise in the quality in, at least, the 62\% and 76\% of the studied applications, respectively. 
\end{tcolorbox}

Finally, the last column from Table \ref{tab:evolution_score_kotlin} shows the number of applications where the introduction of Kotlin has changed the quality evolution trend from `Decline' or `Stability' to `Rise' (Section \ref{sec:methodology:changes_evolve_trend}). We called those \emph{Positive changes} on quality evolution changes.
For the object-oriented smells, the percentage of applications that presented an improvement of quality between the last version of Java and the last of Kotlin varies between the 35.7\% and 42.8\%. For the Android smells, the number of applications with positive change is lower: between 6.6\% and 22.2\%.

Figure \ref{fig:score_trend_chage} shows three cases.
The first one, displayed in sub-figure \ref{fig:trend_change_positive}, corresponds to a positive change in the quality evolution trend.
Before the introduction of Kotlin, the quality scores were constantly declining.
The introduction of Kotlin has positively changed the evolution trend: after that, the quality scores constantly rise.
The second case, sub-figure \ref{fig:trend_change_stable}, shows that the introduction of Kotlin does not change the trend: the quality score before and after the introduction is stable. Note that, over the end of the evolution, the quality score suddenly rise. However, we did not associate this rise with the introduction of Kotlin, which was done much before. 
Finally, the third case, sub-figure \ref{fig:trend_change_rise}, does not present a change on the trend: the quality score was rising before the introduction of Kotlin and continues rising after that.

In conclusion, the study about the changes of quality evolution trend showed that some applications:
\begin{inparaenum}[\it a)]
\item stopped having a constant degradation of the quality of the app written in Java, and 
\item presented an improvement of quality that even rises after the introduction of Kotlin code.
\end{inparaenum}

\section{Threats to Validity}
\label{sec:threatsvalidity}

\subsection{Internal}

\paragraph{Classification of Android applications.}
In section \ref{sec:method:filteringapps} we defined a procedure for classifying Android applications in `Kotlin' and `Java' based on three heuristics that inspected both source code and apk.
By applying those heuristics, we assure the absence of false-positives in our dataset, i.e., applications classified as `Kotlin' but without Kotlin code along with their life-cycle.
However, it could exist some false negatives, i.e., applications that:
\begin{inparaenum}[\it 1)]
\item has versions with Kotlin but heuristic \hone{} does not classify them as Kotlin, and
\item the last version from the code repository does not have Kotlin code any more (by definition, not detected by the heuristic \htree{}).
\end{inparaenum}
The heuristic \htwo~could detect such applications in case that \hone{} fails, but it is expensive (time-consuming) to execute \htwo~over the complete FAMAZOA dataset considering our current infrastructure (remember that \htwo~analyzes each commit from the application's source code).

\paragraph{Modelisation of software quality models.}
There is a risk that the training dataset could not be representative. Thus, the quality model produces incorrect estimation.
We consider the same training dataset that previous work has used for creating quality models and for detecting  smells on Android applications \citep{Hecht2015,  Hecht2015a, Hecht2016}.
Furthermore, the use of that dataset allows having a training dataset and a validation dataset without any intersection, avoiding the generation of an overfitted model.

\paragraph{The set of smells studied.}
To compare fairly the presence of code smells in Java and Kotlin applications, we selected Paprika tool, which works at the level of bytecode. 
However, there are other code smells that are not considered in our work, that could impact in general quality of code. 
Nonetheless, according to \citet{Mannan2016}, 3 out 4 object-oriented code smells considered in this work are in the ranking of the most studied code smells in Android applications.

Furthermore, Kotlin provides new features and different syntax that could introduce new types of code smells. Nevertheless, to the best of our knowledge, there are no studies in the literature that investigate the impact of adoption of Kotlin, and consequently its features, at the source code quality.

\subsection{External}

\paragraph{Validity of Paprika.}
It could exist the risk that Paprika has 
\begin{inparaenum}[\it a)]
\item false positives, i.e.,  it detects smells instances that are not correct, and
\item false negatives, i.e., it does not detect smell instances.
\end{inparaenum}
However, Paprika has been exhaustively evaluated though different experiments \citep{hecht_thesis}.

\paragraph{Representativeness of FAMAZOA.}

Our experiments focus on studying both source and byte code of mobile applications.
For that reason, we decided to study the open-source application available on GitHub. 
To our knowledge combining F-Droid, AndroidTimeMachine, and AndroZoo, makes our dataset the largest Android repository that has both binary and source code of each app.
However, we cannot generalize our findings over applications that are not open-source.

\paragraph{Missing versions (apks).}
There is a risk that F-droid and AndroZoo do not contain all the released versions of an Android application. 
Consequently, this missing data could affect our analysis of the application quality, which is based on the analysis of all apks available of F-droid.

\paragraph{Comparison between Kotlin and Java code.} To compare the amount of Kotlin and Java code on each application, we used the tool CLOC.
Consequently, our results are dependent on the preciseness of that tool.
A bug on the tool could affect the analysis of the evolution of both languages during the applications' lifecycle.

\paragraph{Developers' background and team formation.}
In our work, we analyzed the source code of different Android applications. However, we did not consider information about developers, as experience with a particular language, neither information about the team of developers. It is possible that more experienced developers produce better code than beginners. Nevertheless, all applications that we analyzed are published F-droid or Google Play, then it represents a current snapshot of Android development.

\section{Discussion}
\label{sec:discussion}


Since the announcement of Kotlin as official Android programming language during Google I/O 2017, which is an annual Google Event for developers and consumers, on May 17, 2017,
mobile developers have available a new modern JVM language that they can use for developing Android applications.
This language has new features, such as lambdas (not available for the development of Android applications using Java) which aims at facilitating the development task.
In this paper, we first focus on studying the degree of adoption of this new language.
We found that around 11.26\% of the studied open-source applications have used Kotlin code.
This finding shows, that beyond the support announce is very recent, Kotlin has obtained important share faced of the well stable Java language and the development infrastructure for creating Android apps using Java.
Even more, we found that for some applications 6.1\% developers migrates from Java to Kotlin in few consecutive commits.

One of the advantages of Kotlin mentioned by the documentation is its interoperability with Java: a developer can write an app using both Java and Kotlin.
We found in this paper that 10.38\% applications from our dataset have used Kotlin and Java together. 

Despite the claiming of the Kotlin developers community in favor of Kotlin, no work has studied the impact of Kotlin adoption. Then, this work aims to support developers decision by measuring the quality of Android applications written in Kotlin and Java.
We found that the introduction of Koltin in applications initially written in Java produces a rise in the quality score from, at least, 50\% the applications. Since no work has compared the presence of code smells in Kotlin and Java applications and no work has not analyzed the impact of introducing Kotlin in mobile Android applications written in Java, our results bring new data for the studies about code smells in mobile Android applications.

\section{Conclusion and Future work}
\label{sec:conclusion}

During the last years, different development approaches have emerged for developing mobile applications. In this context,  Google has announced that Kotlin became officially supported language for Android development in 2017. Almost one year since the announcement, we conducted an empirical study to verify whether the Android applications developed using this new program language have better quality than the applications written using the traditional approach for developing Android applications, that means, to use Java language.

To realize our empirical study, we first created our study dataset of open-source Android applications, named FAMAZOA, by mining 3 existing Android datasets (F-Droid, AndroidTimeMachine and AndroZoo). 
The dataset is composed of application which:
\begin{inparaenum}[\it a)]
\item code repository is publicly available,  and
\item exist a version (apk) released after the 2017.
\end{inparaenum}
In total, FAMAZOA contains \numprint{\udataset}~applications and \numprint{\dataset} apks.
We then defined and applied 3 heuristics to classify those applications according to the amount of Kotlin code that they have.
Our study found that ~\ukotlindown~out~\numprint{\udataset}~applications that have at least, one version between the years 2017 and 2018 written using Kotlin.

Our study first focused on analyzing the evolution of the amount of Kotlin and Java code.
We found that for the 63.9\% of applications that include Kotlin code, the amount of code written in that language increases along the application evolution and, at the same time, the amount of Java code decreases or remains constant. 
Furthermore, for 25.8\% of the studied applications, Kotlin replaces Java completely.

Then, our study focused on the quality of Android application by analyzed the presence of code smells in applications from our dataset.
We found that 3 out of 4 object-oriented (OO) smells (LM, CC and BLOB) are present in, at least, the 93\% of both Java and Kotlin application. 
In percentage, 3 out of 4 OO smells (LM, CC and BLOB) are more frequent in Kotlin applications.
However, we found that Java applications have more entities affected by 5 out of 10 code smells, including all OO smells. 
For the remaining smells, we found that there was no difference with statistically.
Finally, we found that the introduction of Kotlin code on an Android application written in Java produced a rise of the quality on, at least, the 50\% of the studied applications.

As future work,  we plan to empirically verify if the advantages claimed by Kotlin develops community (Section \ref{sec:advantagesclaimed}) are true in the context of mobile development and to investigate more code smells and anti-patterns related, for instance, with the performance and energy consumption.

\bibliographystyle{spbasic}  
\bibliography{references}

\begin{thebibliography}{58}
\providecommand{\natexlab}[1]{#1}
\providecommand{\url}[1]{{#1}}
\providecommand{\urlprefix}{URL }
\expandafter\ifx\csname urlstyle\endcsname\relax
  \providecommand{\doi}[1]{DOI~\discretionary{}{}{}#1}\else
  \providecommand{\doi}{DOI~\discretionary{}{}{}\begingroup
  \urlstyle{rm}\Url}\fi
\providecommand{\eprint}[2][]{\url{#2}}

\bibitem[{Allix et~al.(2016)Allix, Bissyand{\'{e}}, Klein, and {Le
  Traon}}]{Allix2016}
Allix K, Bissyand{\'{e}} TF, Klein J, {Le Traon} Y (2016) {AndroZoo: Collecting
  Millions of Android Apps for the Research Community Kevin}. Proceedings of
  the 13th International Workshop on Mining Software Repositories - MSR '16 pp
  468--471, \doi{10.1145/2901739.2903508},
  \urlprefix\url{http://dl.acm.org/citation.cfm?doid=2901739.2903508}

\bibitem[{AndroidDoc(2018{\natexlab{a}})}]{android_dev_hmu}
AndroidDoc (2018{\natexlab{a}}) Array map | android developers.
  \urlprefix\url{https://developer.android.com/reference/android/support/v4/util/ArrayMap},
  [Online; accessed 17-July-2018]

\bibitem[{AndroidDoc(2018{\natexlab{b}})}]{android_dev_mim}
AndroidDoc (2018{\natexlab{b}}) Performance tips | android developers.
  \urlprefix\url{https://developer.android.com/training/articles/perf-tips#PreferStatic},
  [Online; accessed 17-July-2018]

\bibitem[{Aniche et~al.(2017)Aniche, Bavota, Treude, {Van Deursen}, and
  Gerosa}]{Aniche2017}
Aniche M, Bavota G, Treude C, {Van Deursen} A, Gerosa MA (2017) {A validated
  set of smells in model-view-controller architectures}. Proceedings - 2016
  IEEE International Conference on Software Maintenance and Evolution, ICSME
  2016 pp 233--243, \doi{10.1109/ICSME.2016.12}

\bibitem[{Brown et~al.(1998)Brown, Malveau, McCormick, and Mowbray}]{brown1998}
Brown WH, Malveau RC, McCormick HW, Mowbray TJ (1998) AntiPatterns: refactoring
  software, architectures, and projects in crisis. John Wiley \& Sons, Inc.

\bibitem[{Carette et~al.(2017)Carette, Younes, Hecht, Moha, and
  Rouvoy}]{Carette2017}
Carette A, Younes MAA, Hecht G, Moha N, Rouvoy R (2017) {Investigating the
  energy impact of Android smells}. SANER 2017 - 24th IEEE International
  Conference on Software Analysis, Evolution, and Reengineering pp 115--126,
  \doi{10.1109/SANER.2017.7884614}

\bibitem[{Chen et~al.(2014)Chen, Shang, Jiang, Hassan, Nasser, and
  Flora}]{chen2014}
Chen TH, Shang W, Jiang ZM, Hassan AE, Nasser M, Flora P (2014) Detecting
  performance anti-patterns for applications developed using object-relational
  mapping. In: Proceedings of the 36th International Conference on Software
  Engineering, ACM, pp 1001--1012

\bibitem[{Chidamber and Kemerer(1994)}]{Chidamber1994}
Chidamber SR, Kemerer CF (1994) {A metrics suite for object oriented design}.
  IEEE Transactions on Software Engineering 20(6):476--493,
  \doi{10.1109/32.295895}

\bibitem[{Cliff(2014)}]{cliff2014}
Cliff N (2014) Ordinal methods for behavioral data analysis. Psychology Press

\bibitem[{Cruz and Abreu(2017)}]{Cruz2017}
Cruz L, Abreu R (2017) {Performance-Based Guidelines for Energy Efficient
  Mobile Applications}. 2017 IEEE/ACM 4th International Conference on Mobile
  Software Engineering and Systems (MOBILESoft) pp 46--57,
  \doi{10.1109/MOBILESoft.2017.19},
  \urlprefix\url{http://ieeexplore.ieee.org/document/7972717/}

\bibitem[{Cruz and Abreu(2018)}]{Cruz2018}
Cruz L, Abreu R (2018) {Using Automatic Refactoring to Improve Energy
  Efficiency of Android Apps}. In Proceedings of the CIbSE XXI Ibero-American
  Conference on Software Engineering
  \urlprefix\url{http://arxiv.org/abs/1803.05889}, \eprint{1803.05889}

\bibitem[{Fowler et~al.(1999)Fowler, Beck, Brant, Opdyke, and
  Roberts}]{fowler1999}
Fowler M, Beck K, Brant J, Opdyke W, Roberts D (1999) Refactoring: improving
  the design of existing code. Addison-Wesley Professional

\bibitem[{{Geiger} and {Malavolta}(2018)}]{Geiger2018}
{Geiger} FX, {Malavolta} I (2018) {Datasets of Android Applications: a
  Literature Review}. ArXiv e-prints \eprint{1809.10069}

\bibitem[{Geiger et~al.(2018)Geiger, Malavolta, Pascarella, Palomba, Di~Nucci,
  and Bacchelli}]{Geiger2018:data}
Geiger FX, Malavolta I, Pascarella L, Palomba F, Di~Nucci D, Bacchelli A (2018)
  A graph-based dataset of commit history of real-world android apps. In:
  Proceedings of the 15th International Conference on Mining Software
  Repositories, ACM, New York, NY, USA, MSR '18, pp 30--33,
  \doi{10.1145/3196398.3196460},
  \urlprefix\url{http://doi.acm.org/10.1145/3196398.3196460}

\bibitem[{Grano et~al.(2017)Grano, {Di Sorbo}, Mercaldo, Visaggio, Canfora, and
  Panichella}]{Grano2017}
Grano G, {Di Sorbo} A, Mercaldo F, Visaggio CA, Canfora G, Panichella S (2017)
  {Android apps and user feedback: a dataset for software evolution and quality
  improvement}. Proceedings of the 2nd ACM SIGSOFT International Workshop on
  App Market Analytics - WAMA 2017 pp 8--11, \doi{10.1145/3121264.3121266},
  \urlprefix\url{http://dl.acm.org/citation.cfm?doid=3121264.3121266}

\bibitem[{Haase(2015)}]{medium_hmu}
Haase C (2015) Developing for android ii – google developers – medium.
  \urlprefix\url{https://medium.com/google-developers/developing-for-android-ii-bb9a51f8c8b9},
  [Online; accessed 17-July-2018]

\bibitem[{Habchi et~al.(2017)Habchi, Hecht, Rouvoy, and Moha}]{Habchi2017}
Habchi S, Hecht G, Rouvoy R, Moha N (2017) {Code Smells in iOS Apps: How Do
  They Compare to Android?} Proceedings - 2017 IEEE/ACM 4th International
  Conference on Mobile Software Engineering and Systems, MOBILESoft 2017 pp
  110--121, \doi{10.1109/MOBILESoft.2017.11}

\bibitem[{Hecht(2016)}]{hecht_thesis}
Hecht G (2016) {Detection and analysis of impact of code smells in mobile
  applications}. Phd thesis, {Universit{\'e} Lille 1 : Sciences et Technologies
  ; Universit{\'e} du Qu{\'e}bec {\`a} Montr{\'e}al},
  \urlprefix\url{https://tel.archives-ouvertes.fr/tel-01418158}

\bibitem[{Hecht et~al.(2015{\natexlab{a}})Hecht, Benomar, Rouvoy, Moha, and
  Duchien}]{Hecht2015}
Hecht G, Benomar O, Rouvoy R, Moha N, Duchien L (2015{\natexlab{a}}) Tracking
  the software quality of android applications along their evolution. In:
  Proceedings of the 2015 30th IEEE/ACM International Conference on Automated
  Software Engineering (ASE), IEEE Computer Society, Washington, DC, USA, ASE
  '15, pp 236--247, \doi{10.1109/ASE.2015.46},
  \urlprefix\url{https://doi.org/10.1109/ASE.2015.46}

\bibitem[{Hecht et~al.(2015{\natexlab{b}})Hecht, Rouvoy, Moha, and
  Duchien}]{Hecht2015a}
Hecht G, Rouvoy R, Moha N, Duchien L (2015{\natexlab{b}}) {Detecting
  Antipatterns in Android Apps}. Proceedings - 2nd ACM International Conference
  on Mobile Software Engineering and Systems, MOBILESoft 2015 pp 148--149,
  \doi{10.1109/MobileSoft.2015.38}

\bibitem[{Hecht et~al.(2016)Hecht, Moha, and Rouvoy}]{Hecht2016}
Hecht G, Moha N, Rouvoy R (2016) {An empirical study of the performance impacts
  of Android code smells}. Proceedings of the International Workshop on Mobile
  Software Engineering and Systems - MOBILESoft '16 pp 59--69,
  \doi{10.1145/2897073.2897100},
  \urlprefix\url{http://dl.acm.org/citation.cfm?doid=2897073.2897100},
  \eprint{arXiv:1508.06655v1}

\bibitem[{IDC(2017)}]{mobileshare}
IDC (2017) Smartphone os market share, 2017 q1.
  \urlprefix\url{https://www.idc.com/promo/smartphone-market-share/}

\bibitem[{Kessentini and Ouni(2017)}]{Kessentini2017}
Kessentini M, Ouni A (2017) {Detecting Android Smells Using Multi-Objective
  Genetic Programming}. Proceedings - 2017 IEEE/ACM 4th International
  Conference on Mobile Software Engineering and Systems, MOBILESoft 2017 pp
  122--132, \doi{10.1109/MOBILESoft.2017.29}

\bibitem[{Khalid et~al.(2016)Khalid, Nagappan, and Hassan}]{Khalid2016}
Khalid H, Nagappan M, Hassan AE (2016) {Examining the relationship between
  FindBugs warnings and app ratings}. IEEE Software 33(4):34--39,
  \doi{10.1109/MS.2015.29}

\bibitem[{KotlinDoc(2018)}]{kotlin_doc_static}
KotlinDoc (2018) Classes and inheritance - kotlin programming language.
  \urlprefix\url{https://kotlinlang.org/docs/reference/classes.html#companion-objects},
  [Online; accessed 17-July-2018]

\bibitem[{La(2017)}]{kotlinshare}
La J (2017) Update on kotlin for android.
  \urlprefix\url{https://android-developers.googleblog.com/2017/11/update-on-kotlin-for-android.html}

\bibitem[{Lehman(1980)}]{lehman1980programs}
Lehman MM (1980) Programs, life cycles, and laws of software evolution.
  Proceedings of the IEEE 68(9):1060--1076

\bibitem[{Li et~al.(2017{\natexlab{a}})Li, Guo, Shen, Li, and
  Huang}]{Li2017evo}
Li D, Guo B, Shen Y, Li J, Huang Y (2017{\natexlab{a}}) {The evolution of
  open-source mobile applications: An empirical study}. Journal of Software:
  Evolution and Process 29(7):1--18, \doi{10.1002/smr.1855}

\bibitem[{Li et~al.(2017{\natexlab{b}})Li, Bissyand, Papadakis, Rasthofer,
  Bartel, Octeau, Klein, and Traon}]{Li2017c}
Li L, Bissyand TF, Papadakis M, Rasthofer S, Bartel A, Octeau D, Klein J, Traon
  L (2017{\natexlab{b}}) Static analysis of android apps: A systematic
  literature review. Inf Softw Technol 88(C):67--95,
  \doi{10.1016/j.infsof.2017.04.001},
  \urlprefix\url{https://doi.org/10.1016/j.infsof.2017.04.001}

\bibitem[{Lockwood(2013)}]{androidpatterns_lic}
Lockwood A (2013) How to leak a context: Handlers \& inner classes.
  \urlprefix\url{https://www.androiddesignpatterns.com/2013/01/inner-class-handler-memory-leak.html},
  [Online; accessed 17-July-2018]

\bibitem[{Macbeth et~al.(2011)Macbeth, Razumiejczyk, and Ledesma}]{Macbeth2011}
Macbeth G, Razumiejczyk E, Ledesma RaD (2011) {Cliff's Delta Calculator: A
  non-parametric effect size program for two groups of observations}.
  {Universitas Psychologica} 10:545 -- 555,
  \urlprefix\url{http://www.scielo.org.co/scielo.php?script=sci_arttext&pid=S1657-92672011000200018\&nrm=iso}

\bibitem[{Malavolta et~al.(2015{\natexlab{a}})Malavolta, Ruberto, Soru, and
  Terragni}]{Malavolta2015EndUsers}
Malavolta I, Ruberto S, Soru T, Terragni V (2015{\natexlab{a}}) End users'
  perception of hybrid mobile apps in the google play store. In: 2015 IEEE
  International Conference on Mobile Services, pp 25--32,
  \doi{10.1109/MobServ.2015.14}

\bibitem[{Malavolta et~al.(2015{\natexlab{b}})Malavolta, Ruberto, Soru, and
  Terragni}]{Malavolta2015Hybrid}
Malavolta I, Ruberto S, Soru T, Terragni V (2015{\natexlab{b}}) Hybrid mobile
  apps in the google play store: An exploratory investigation. In: Proceedings
  of the Second ACM International Conference on Mobile Software Engineering and
  Systems, IEEE Press, Piscataway, NJ, USA, MOBILESoft '15, pp 56--59,
  \urlprefix\url{http://dl.acm.org/citation.cfm?id=2825041.2825051}

\bibitem[{Mannan et~al.(2016)Mannan, Ahmed, Almurshed, Dig, and
  Jensen}]{Mannan2016}
Mannan UA, Ahmed I, Almurshed RAM, Dig D, Jensen C (2016) {Understanding code
  smells in Android applications}. International Workshop on Mobile Software
  Engineering and Systems (MOBILESoft '16) pp 225--234,
  \doi{10.1145/2897073.2897094},
  \urlprefix\url{http://dl.acm.org/citation.cfm?doid=2897073.2897094}

\bibitem[{Mariotti(2013{\natexlab{a}})}]{blog_hbr}
Mariotti G (2013{\natexlab{a}}) Antipattern: freezing a ui with broadcast
  receiver.
  \urlprefix\url{http://gmariotti.blogspot.com/2013/02/antipattern-freezing-ui-with-broadcast.html},
  [Online; accessed 17-July-2018]

\bibitem[{Mariotti(2013{\natexlab{b}})}]{blog_hss}
Mariotti G (2013{\natexlab{b}}) Antipattern: freezing the ui with a service and
  an intentservice.
  \urlprefix\url{http://gmariotti.blogspot.com/2013/03/antipattern-freezing-ui-with-service.html},
  [Online; accessed 17-July-2018]

\bibitem[{Mariotti(2013{\natexlab{c}})}]{blog_has}
Mariotti G (2013{\natexlab{c}}) Antipattern: freezing the ui with an asynctask.
  \urlprefix\url{http://gmariotti.blogspot.com/2013/02/antipattern-freezing-ui-with-asynctask.html},
  [Online; accessed 17-July-2018]

\bibitem[{Martinez and Lecomte(2017)}]{Martinez:2017:TQI}
Martinez M, Lecomte S (2017) Towards the quality improvement of cross-platform
  mobile applications. In: Proceedings of the 4th International Conference on
  Mobile Software Engineering and Systems, IEEE Press, Piscataway, NJ, USA,
  MOBILESoft '17, pp 184--188, \doi{10.1109/MOBILESoft.2017.30},
  \urlprefix\url{https://doi.org/10.1109/MOBILESoft.2017.30}

\bibitem[{Mazinanian et~al.(2014)Mazinanian, Tsantalis, and
  Mesbah}]{mazinanian2014}
Mazinanian D, Tsantalis N, Mesbah A (2014) Discovering refactoring
  opportunities in cascading style sheets. In: Proceedings of the 22nd ACM
  SIGSOFT International Symposium on Foundations of Software Engineering, ACM,
  pp 496--506

\bibitem[{McAnlis(2015)}]{google_dev_uio}
McAnlis C (2015) Android performance patterns: Overdraw, cliprect, quickreject.
  \urlprefix\url{https://www.youtube.com/watch?v=vkTn3Ule4Ps}, [Online;
  accessed 17-July-2018]

\bibitem[{McCabe(1976)}]{mccabe1976}
McCabe TJ (1976) A complexity measure. IEEE Transactions on software
  Engineering (4):308--320

\bibitem[{Morales et~al.(2016)Morales, Saborido, Khomh, Chicano, and
  Antoniol}]{Morales2016}
Morales R, Saborido R, Khomh F, Chicano F, Antoniol G (2016) {Anti-patterns and
  the energy efficiency of Android applications}. ArXiv e-prints
  \eprint{1610.05711}

\bibitem[{Morales et~al.(2017)Morales, Saborido, Khomh, Chicano, and
  Antoniol}]{Morales2017}
Morales R, Saborido R, Khomh F, Chicano F, Antoniol G (2017) {EARMO: An
  Energy-Aware Refactoring Approach for Mobile Apps}. IEEE Transactions on
  Software Engineering X(X):1--31, \doi{10.1109/TSE.2017.2757486}

\bibitem[{Nagappan and Shihab(2016)}]{Nagappan2016}
Nagappan M, Shihab E (2016) {Future Trends in Software Engineering Research for
  Mobile Apps}. 2016 IEEE 23rd International Conference on Software Analysis,
  Evolution, and Reengineering (SANER) pp 21--32, \doi{10.1109/SANER.2016.88},
  \urlprefix\url{http://ieeexplore.ieee.org/document/7476770/}

\bibitem[{Ni-Lewis(2015{\natexlab{a}})}]{google_dev_iod}
Ni-Lewis I (2015{\natexlab{a}}) Avoiding allocations in ondraw() (100 days of
  google dev). \urlprefix\url{https://www.youtube.com/watch?v=HAK5acHQ53E},
  [Online; accessed 17-July-2018]

\bibitem[{Ni-Lewis(2015{\natexlab{b}})}]{google_dev_uha_iwr}
Ni-Lewis I (2015{\natexlab{b}}) Custom views and performance (100 days of
  google dev). \urlprefix\url{https://youtu.be/zK2i7ivzK7M?t=4m57s}, [Online;
  accessed 17-July-2018]

\bibitem[{Palomba et~al.(2015)Palomba, Bavota, {Di Penta}, Oliveto, Poshyvanyk,
  and {De Lucia}}]{Palomba2015}
Palomba F, Bavota G, {Di Penta} M, Oliveto R, Poshyvanyk D, {De Lucia} A (2015)
  {Mining version histories for detecting code smells}. IEEE Transactions on
  Software Engineering 41(5):462--489, \doi{10.1109/TSE.2014.2372760}

\bibitem[{Palomba et~al.(2017)Palomba, {Di Nucci}, Panichella, Zaidman, and {De
  Lucia}}]{Palomba2017}
Palomba F, {Di Nucci} D, Panichella A, Zaidman A, {De Lucia} A (2017)
  {Lightweight detection of Android-specific code smells: The aDoctor project}.
  SANER 2017 - 24th IEEE International Conference on Software Analysis,
  Evolution, and Reengineering pp 487--491, \doi{10.1109/SANER.2017.7884659},
  \urlprefix\url{https://dibt.unimol.it/staff/fpalomba/documents/C18.pdf}

\bibitem[{Palomba et~al.(2018)Palomba, Bavota, Penta, Fasano, Oliveto, and
  Lucia}]{Palomba2018}
Palomba F, Bavota G, Penta MD, Fasano F, Oliveto R, Lucia AD (2018) {On the
  diffuseness and the impact on maintainability of code smells: a large scale
  empirical investigation}. Empirical Software Engineering 23(3):1188--1221,
  \doi{10.1007/s10664-017-9535-z}

\bibitem[{Reimann et~al.(2014{\natexlab{a}})Reimann, Brylski, and
  A{\ss}mann}]{smell_catalogue}
Reimann J, Brylski M, A{\ss}mann U (2014{\natexlab{a}}) Android smells
  catalogue. \urlprefix\url{http://www.modelrefactoring.org/smell_catalog},
  [Online; accessed 17-July-2018]

\bibitem[{Reimann et~al.(2014{\natexlab{b}})Reimann, Brylski, and
  A{\ss}mann}]{reimann2014tool}
Reimann J, Brylski M, A{\ss}mann U (2014{\natexlab{b}}) A tool-supported
  quality smell catalogue for android developers. In: Proc. of the conference
  Modellierung 2014 in the Workshop Modellbasierte und modellgetriebene
  Softwaremodernisierung--MMSM, vol 2014

\bibitem[{Romano et~al.(2006)Romano, Kromrey, Coraggio, Skowronek, and
  Devine}]{romano2006}
Romano J, Kromrey JD, Coraggio J, Skowronek J, Devine L (2006) Exploring
  methods for evaluating group differences on the nsse and other surveys: Are
  the t-test and cohen’sd indices the most appropriate choices. In: annual
  meeting of the Southern Association for Institutional Research, Citeseer

\bibitem[{Saborido et~al.(2018)Saborido, Morales, Khomh, Gu{\'{e}}h{\'{e}}neuc,
  and Antoniol}]{Saborido2018}
Saborido R, Morales R, Khomh F, Gu{\'{e}}h{\'{e}}neuc YG, Antoniol G (2018)
  {Getting the most from map data structures in Android}. Empirical Software
  Engineering pp 1--36, \doi{10.1007/s10664-018-9607-8}

\bibitem[{Sharma and Spinellis(2018)}]{Sharma2018}
Sharma T, Spinellis D (2018) {A survey on software smells}. Journal of Systems
  and Software 138:158--173, \doi{10.1016/j.jss.2017.12.034},
  \urlprefix\url{https://doi.org/10.1016/j.jss.2017.12.034}

\bibitem[{Tufano et~al.(2015)Tufano, Palomba, Bavota, Oliveto, Di~Penta,
  De~Lucia, and Poshyvanyk}]{Tufano2015}
Tufano M, Palomba F, Bavota G, Oliveto R, Di~Penta M, De~Lucia A, Poshyvanyk D
  (2015) When and why your code starts to smell bad. In: Proceedings of the
  37th International Conference on Software Engineering - Volume 1, IEEE Press,
  Piscataway, NJ, USA, ICSE '15, pp 403--414,
  \urlprefix\url{http://dl.acm.org/citation.cfm?id=2818754.2818805}

\bibitem[{Verloop(2013)}]{verloop2013}
Verloop D (2013) {Code smells in the mobile applications domain}. Master
  thesis, {TU Delft, Delft University of Technology},
  \urlprefix\url{https://repository.tudelft.nl/islandora/object/uuid:bcba7e5b-e898-4e59-b636-234ad3fdc432?collection=education}

\bibitem[{Vinther(2017)}]{kotlin_adv}
Vinther M (2017) Why you should totally switch to kotlin.
  \urlprefix\url{https://medium.com/@magnus.chatt/why-you-should-totally-switch-to-kotlin-c7bbde9e10d5}

\bibitem[{Zadeh(1974)}]{zadeh1974}
Zadeh LA (1974) Fuzzy logic and its application to approximate reasoning. In:
  IFIP Congress, vol 591

\end{thebibliography}

\end{document}